\documentclass[11pt]{article}
\usepackage[utf8]{inputenc}
\usepackage{fancyhdr}
\usepackage{xspace,xcolor}
\usepackage{systeme}
\usepackage{authblk}
\usepackage[hypertexnames=false]{hyperref}
\usepackage{verbatim}
\usepackage{amsmath}
\numberwithin{equation}{section}

\usepackage{newtxtext}

\def\be{\begin{equation}}
\def\ee{\end{equation}}
\def\bes{\begin{equation*}}
\def\ees{\end{equation*}}
\def\bea{\begin{equation} \begin{aligned}}
\def\eea{\end{aligned} \end{equation}}
\def\beas{\begin{equation*} \begin{aligned}}
\def\eeas{\end{aligned} \end{equation*}}

\def\cX{\mathcal X}
\def\cY{\mathcal Y}

\newcommand{\X}{\mathbf{X}}
\newcommand{\x}{\mathbf{x}}

\newcommand{\W}{\mathbf{W}}

\def\bi{\begin{itemize}}
\def\ei{\end{itemize}}
\definecolor{rred}{rgb}{0.7,0,0.1}

\definecolor{greenrb}{rgb}{0.2,0.6,0.2}

\usepackage{TFG}
\graphicspath{{Figs/}}
\date{}

\title{Reduced-Order Models for Coupled Dynamical Systems:\\ 
 Data-driven Methods and the Koopman Operator}

\author[1,2]{Manuel Santos Guti\'errez\footnote{Corresponding author: \url{m.santos@pgr.reading.ac.uk}}}
\author[1,2]{Valerio Lucarini}
\author[3,4]{Micka\"el D. Chekroun}
\author[4,5,6]{Michael Ghil}
\affil[1]{Department of Mathematics and Statistics, University of Reading, Reading RG6 6AX, U.K.}
\affil[2]{Centre for the Mathematics of Planet Earth, University of Reading, Reading RG6 6AX, U.K.}
\affil[3]{Department of Earth and Planetary Sciences, Weizmann Institute, 
Rehovot 76100, Israel}
\affil[4]{Department of Atmospheric \& Oceanic Sciences, University of California at Los Angeles, Los Angeles, CA 90095, USA}
\affil[5]{Geosciences Department and Laboratoire de M\'et\'eorologie Dynamique (CNRS and IPSL), Ecole Normale Sup\'erieure and PSL University, Paris 75231, France}
\affil[6]{Institute of Applied Physics of the Russian Academy of Sciences, Nizhny Novgorod 603950, Russia}

\begin{document}
\maketitle

\begin{abstract}
Providing efficient and accurate parametrizations for model reduction is a key goal in many areas of science and technology. Here we present a strong link between data-driven and theoretical approaches to achieving this 
goal. Formal perturbation expansions of the Koopman operator allow us to derive general stochastic parametrizations of weakly coupled dynamical systems. Such parametrizations yield a set of stochastic integro-differential equations with explicit noise and memory kernel formulas to describe the effects of unresolved variables. We show that the perturbation expansions involved need not be truncated when the coupling is additive. 
The unwieldy integro-differential equations can be recast as a simpler multilevel Markovian model, and we establish an intuitive connection with a 
generalized Langevin equation. This connection helps setting up a parallelism between the top-down, equations-based methodology herein and the well-established empirical model reduction (EMR) methodology that has been shown to provide efficient dynamical closures to partially observed systems. Hence, our findings support, on the one hand, the physical basis and robustness of the EMR methodology and, on the other hand, illustrate the practical relevance of the perturbative expansion used for deriving the parametrizations. 
\end{abstract}

\tableofcontents

\begin{quotation}
\noindent	\bf{Parametrizations aim to reduce the complexity of high-dimensional dynamical systems. Here, a theory-based and a data-driven approach 
for the parametrization of coupled systems are compared, showing that both yield the same stochastic multilevel structure. The results provide a very strong support to the use of empirical methods in model reduction and 
clarify the practical relevance of the proposed theoretical framework.}
\end{quotation}
\section{Introduction and Motivation}\label{INTRO}

Multiscale systems are typically characterized by the presence of significant variability over a large range of spatial and temporal scales. This multiscale character is due to a combination of the following factors: the nature of the external forcings, the inhomogeneity of the properties of 
the system's various components, the complexity of the coupling mechanisms between the components, and the variety of instabilities, dissipative processes, and feedbacks acting at different scales. In many cases, both the theoretical understanding of such systems and the formulation of numerical models for simulating their properties are based on focusing upon a reduced range of large spatial and long temporal scales of interest, and upon devising an efficient way to capture effectively the impact of the faster dynamical processes acting predominantly in the neglected smaller spatial scales \cite{E.Lu.2011, Pavliotis2008}.

Given a high-dimensional dynamical system, we are thus interested in reformulating it in such a way that only the variables of interest are resolved.  A first-guess would be to ignore altogether the unresolved variables 
and consider uniquely the filtered evolution laws valid for the targeted range of scales. However, this is well-known to be inadequate, because nonlinearity guarantees that the unresolved variables have an impact on the 
resolved ones, both in terms of the detailed dynamics and of its statistical properties. Therefore, the problem of constructing accurate and efficient reduced-order models --- or, equivalently, of defining the coarse-grained dynamics --- is an essential and fundamental aspect of studying multiscale systems, both theoretically and through numerical simulations.

For the sake of concreteness, let us consider, the case of climate science. Is It is well-nigh impossible, therefore, given our current scientific 
knowledge and our available or even foreseeable technological capabilities, to create a numerical model able to directly simulate the climate system in all details for all the relevant time scales, which span a range of 
over 15 orders of magnitude \cite{Ghil2015, GhilLucarini2020, Peixoto1992}. Hence, one has to focus on a specific range of scales through suitably 
developed, approximate evolution equations that provide the basis for the 
numerical  modeling. Such equations are derived from the fundamental laws 
of climate dynamics through systematic asymptotic expansions that are based on imposing an approximate balance between the forces acting on geophysical flows. These balance relations lead to removing small-scale, fast processes that are assumed to play a minor role at the scales of interest by filtering out the corresponding waves  \cite{Holton, klein2010, vallis_atmospheric_2006}. 

In  climate science, parametrization schemes have been traditionally formulated in such a way that one expresses the net impact on the scales of interest of processes occurring within the unresolved scales via deterministic functions of the resolved variables, as in the pioneering work on the parametrization of convective processes by Arakawa and Schubert \cite{AS.74}. More recently, it has been recognized, mostly on empirical grounds, that parametrizations should involve  stochastic and non-Markovian components \cite{Berner2017, Franzke.ea.2015, palmer_stochastic_2009}.  Machine learning methods have been proposed as the next frontier of data-driven parametrizations \cite{Gentine2018, Ogorman2018, Rasp2018, weyn2020}, able to deliver a new generation of Earth system models \cite{Schneider2017}; see, though, the caveats discussed by \cite{Hosni2018, Faranda2019}.

%\subsection{The {\mg Projection Operator Formalism of Mori \cite{mori_transport_1965} and Zwanzig \cite{Zwanzig.2001}}} \label{ssec:MZ}
\subsection{The Projection Operator Formalism of Mori and Zwanzig}\label{ssec:MZ}
Let us reformulate the problem of constructing parameterizations as a projection of the dynamics onto the set of resolved variables. By working at 
the level of observables, Mori \cite{mori_transport_1965} and Zwanzig \cite{zwanzig_memory_1961} showed that the evolution laws for the projected dynamics incorporate a deterministic term that would be obtained by neglecting altogether the impact of the unresolved variables, to which a stochastic and a non-Markovian correction had to be added. A. J. Chorin and coauthors \cite{Chorin.ea.2002, Chorin.Lu.2015} played an important role in 
developing further these ideas and applying them to several important problems. We briefly recapitulate below the Mori-Zwanzig projection operator 
approach.

Formally, let $\Phi$ denote a generic observable defined on a state space 
viewed as the product of two finite-dimensional spaces $\mathcal{X}\times 
\mathcal{Y}$, with variables $\xx \in \cX$ and $\yy \in \cY$ being the resolved and unresolved variables, respectively. Next, let us define $\mathbb{P}$ to be a projector onto functions depending only on the target variables in $\mathcal{X}$, with the complementary projector on the unresolved variables being defined by $\mathbb{Q}=\mbox{Id} - \mathbb{P}$. %% The variable $\xx$ (resp.~$\yy$) in $\cX$ (resp.~$\cY$) will be sometimes referred to as the resolved (resp.~unresolved) variable.

Given a smooth flow $\psi_t$ on $\mathcal{X}\times \mathcal{Y}$, we consider its action on smooth observables $\Phi = \Phi(\xx, \yy)$ defined by 

\be
U_t \Phi (\xx,\yy)=\Phi (\psi_t(\xx,\yy)), \; (\xx, \yy) \in \cX\times \cY, \; t\geq 0.
\ee
The operator $U_t$ is the Koopman operator and the family $\{U_t\}_{t\geq 
0}$ of Koopman operators indexed by time forms a semigroup; see Sec.~\ref{weak-coupling limit parametrization} below for further details. The Koopman operator describes how functions on the phase space change under the action of the flow; its time evolution obeys the following equation:
\be
\partial _t \big(U_t\Phi\big)=\mathcal{L} \big(U_t \Phi\big). \label{koopman}
\ee
The linear operator $\mathcal{L}$ gives the instantaneous rate of change of $\Phi$ under the action of the flow $\psi_t$. Representing $U_t$ formally as the exponential of $\LLL$, $U_t=e^{t\LLL}$, is both useful and fully justified by operator semigroup theory \cite{Pazy2012}. With this representation, $\LLL$ satisfies the following identities 
\begin{subequations}
\begin{align}
\partial _t \big(U_t\Phi\big) & = \mathcal{L} e^{t\LLL}\Phi = e^{t\LLL}\mathcal{L}\Phi = e^{t\LLL}\left(\mathbb{P}+\mathbb{Q}\right)\mathcal{L}\Phi   \\
&= e^{t\LLL}\mathbb{P}\LLL\Phi + e^{t\mathbb{Q}\LLL}\mathbb{Q}\LLL\Phi  
+ \int_{0}^{t}e^{(t-s)\LLL}\mathbb{P}\LLL e^{s\mathbb{Q}\LLL}\mathbb{Q}\LLL \Phi  \dd s, \label{mori zwanzig 2}
\end{align}
\end{subequations}
where we have employed the Dyson identity \cite{Dyson1949,engel2006short} 
to obtain Eq.~\eqref{mori zwanzig 2}, as will be discussed in Sect.~\ref{WLapproximation}  %% \ref{ssec:MZ}. 
The first term in Eq.~\eqref{mori zwanzig 2} is the contribution of the resolved variables $\xx$ alone to the instantaneous rate of change of $\Phi$. The second term models the fluctuating effects of the unresolved $\yy$-variable by itself, while the third and last term represents, via an integral, the time-delayed influence upon $\xx$ of its interactions with $\yy$. 
 
 This formal calculation suggests that any closed model for the $\xx$-variables should incorporate a fluctuating term to account for the $\yy$ contributions and a memory or integral term for the $\yy$--$\xx$ interactions. Unfortunately, the Mori-Zwanzig equation \eqref{mori zwanzig 2}--- also known as the Generalized Langevin Equation (GLE) \cite{kondrashovdata2015, pavliotisbook2014} --- does not provide explicit analytic formulas to 
determine each of the three summands in the right-hand side (RHS) of Eq.~\eqref{mori zwanzig 2}. Hence, we need efficient ways to approximate such 
an equation.

In the limit of perfect time scale separation between the $\xx$- and  $\yy$-variables, the non-Markovian term drops out and the fluctuating term can be represented as a --- possibly multiplicative --- white-noise term, thus recovering the basic results obtained via homogenization theory \cite{Gottwald2013, Melbourne2011, Pavliotis2008}. When no such separation exists, however, one has to resort to finding an integral kernel beyond the 
abstract formulation of Eq.~\eqref{mori zwanzig 2}; see, for instance, the theoretical ansatz based on perturbation expansion presented in refs. \cite{Wouters2012, Wouters2013} and discussed later in the paper, and  refs.~\cite{kondrashovdata2015, Vissio2018a} for concrete applications.

In parallel with the theoretical approaches to approximate the Mori-Zwanzig equation \eqref{mori zwanzig 2}, data-driven methods have been proposed to model fluctuations and memory effects arising as a result of projecting a large state space onto (much) smaller subspaces. To this end, the work in \cite{kondrashovdata2015} provides a rigorous connection between the Mori-Zwanzig equation \eqref{mori zwanzig 2} and multilevel regression 
models that were initially introduced for climatological purposes in \cite{Kravtsov2005}. More recently, ref.~\cite{Lin.Lu.2021} proposed Nonlinear Auto-Regressive Moving Average with eXogenous input (NARMAX) models as a data-driven methodology that is comparable with the Mori-Zwanzig formalism and applied such models to the deterministic Kuramoto-Sivashinsky equation and to a stochastic Burgers equation. The complementarity of theory-based and data-driven model reduction methods in the absence of time scale separation is very well documented in \cite{Lin.Lu.2021} as well. A highly complementary recent contribution aimed at finding a common thread between data-driven methods and the Mori-Zwanzig theoretical framework is \cite{Lin2021}. 

Efforts at using ingeniously selected basis functions as a stepping stone in data-driven methods for model reduction, effective simulation with partial data, and even prediction are multiple and flourishing. Thus,
%It is well known that 
the eigenvalues and eigenfunctions of the Koopman and transfer operators have been used to capture the modes of variability of the underlying flow, regardless of the latter being deterministic or stochastic; see, respectively, \cite{B00} or \cite{chekroun2019c}. %% vAmong the data-driven methods, it is important to mention the
Dynamic mode decomposition \cite[DMD]{Schmid2010} allows one to reconstruct from observations the eigenvectors and eigenvalues of the Koopman operator for observables of interest even in high-dimensional dynamical systems  \cite{Mezic2005, Rowley2009}. The latter approach is complementary to the one presented herein because we shall use the  eigenvectors of the Koopman operator to build the projected dynamics for the observables of interest, which can then be re-written as a multilevel Markovian 
stochastic model.

Examples of other types of selection of dynamically interesting and effective bases are multichannel singular-spectrum analysis analysis \cite[MSSA]{Alessio.2015, Broom.ea.1987, Ghil.SSA.2002} and data-adaptive harmonic decomposition \cite[DAH-MSLM]{Chekroun.DK.2017, Kondrashov.ea.2018}. 
Both methodologies have been used extensively and successfully in the simulation, as well as the prediction, of complex phenomena \cite{ Ghil.SSA.2002, Keppenne.Ghil.1992, Kondrashov.ea.2018}.

Our main theoretical result, Theorem~\ref{proposition: 1} below, establishes how such spectral elements determine in turn the constitutive elements of data-driven non-Markovian closure of partially observed complex systems, when rewritten as a multilevel Markovian stochastic model. This theorem highlights, in particular, new bridges with Koopman modes and the DMD 
decomposition \cite{Mezic2005, Rowley2009, Schmid2010}, as well with other kinds of projections onto spectral bases \cite{Chekroun.DK.2017, Ghil.SSA.2002, Hasselmann.POPs.1988, KCYG.2018, Penland.1996, Tu.ea.2014}. A more thorough discussion of the complementary approaches involved is beyond 
the scope of this paper.

% to estimate Koopman and transfer operator can readily be estimated from time series by means of selecting a suitable basis of observed functions Under suitable assumptions, one can reconstruct unitary approximating matrices with eigenvalues indicating the growth rate and frequency of constituent periodic signals. Such unitarity, though, is not satisfied as soon as the underlying process is irreversible, for instance, when adding noise or when artificial diffusion is introduced \cite{Tantet2018}. The work along the lines of \cite{Mezic2005, Rowley2009} provides a protocol for analysing spatial structures together with an orthogonal decomposition of the flow.

\subsection{Goals of This Paper}\label{ssec:goals}
 
Many approaches to constructing theoretically rigorous parametrizations have been devised. These can be broadly divided into top-down and data-driven approaches: Top-down approaches aim at deriving the parametrizations by applying suitable approximations to the equations describing the dynamics of the whole systems \cite[for instance]{Majda2013, Majda2001, Vissio2018a, Wouters_2019a, Wouters2012, Wouters2013}, while data-driven parametrizations are built by constructing a statistical-dynamical model of the 
impact of unresolved scales on the scales of interest. In fact, partial observations of a time-evolving system can be used to deduce the fluctuating and delayed effects of the unobserved processes, as shown in \cite{kondrashovdata2015, Kondrashov.Kravtsov.ea.2006, Kravtsov2005, wilks_effects_2005}.  

In this paper, we will discuss and compare the properties of the Wouters-Lucarini (WL) top-down parametrization \cite{Vissio2018a, Wouters2012, Wouters2013} and of the Empirical Model Reduction (EMR) data-driven parametrization \cite{kondrashovdata2015, KKRG.ENSO.2005, Kondrashov.Kravtsov.ea.2006, Kravtsov2005, KKG.2010}. We will also see when and how the integro-differential equation occurring in the WL parametrization can be recast into a set of Markovian stochastic differential equations (SDEs). In other words, we investigate the quasi-Markovianity of the latter parametrization \cite{pavliotisbook2014}.

The two aforementioned methodologies are conceptually and practically distinct, even though the ultimate goal is to provide a  computationally practical approximation for the Mori-Zwanzig or GLE integro-differential equation. In other words, both approaches --- top-down and data-driven --- provide fluctuations in the form of stochastic noise and memory effects determined by an integral kernel. On the one hand, the WL approach assumes prior knowledge about the decoupled hidden dynamics but no information about the statistical properties of the coupled system. The empirical approach, on the other hand, samples the observed variables evolving according 
to the latter. The structure of the multilevel stochastic models (MSMs) that generalize EMRs \cite{kondrashovdata2015} allows one, moreover, 
to derive explicit formulas for the fluctuating and memory correction terms that parametrize the influence of hidden processes.

\begin{figure}[h]
	\centering
	\includegraphics[width=\textwidth]{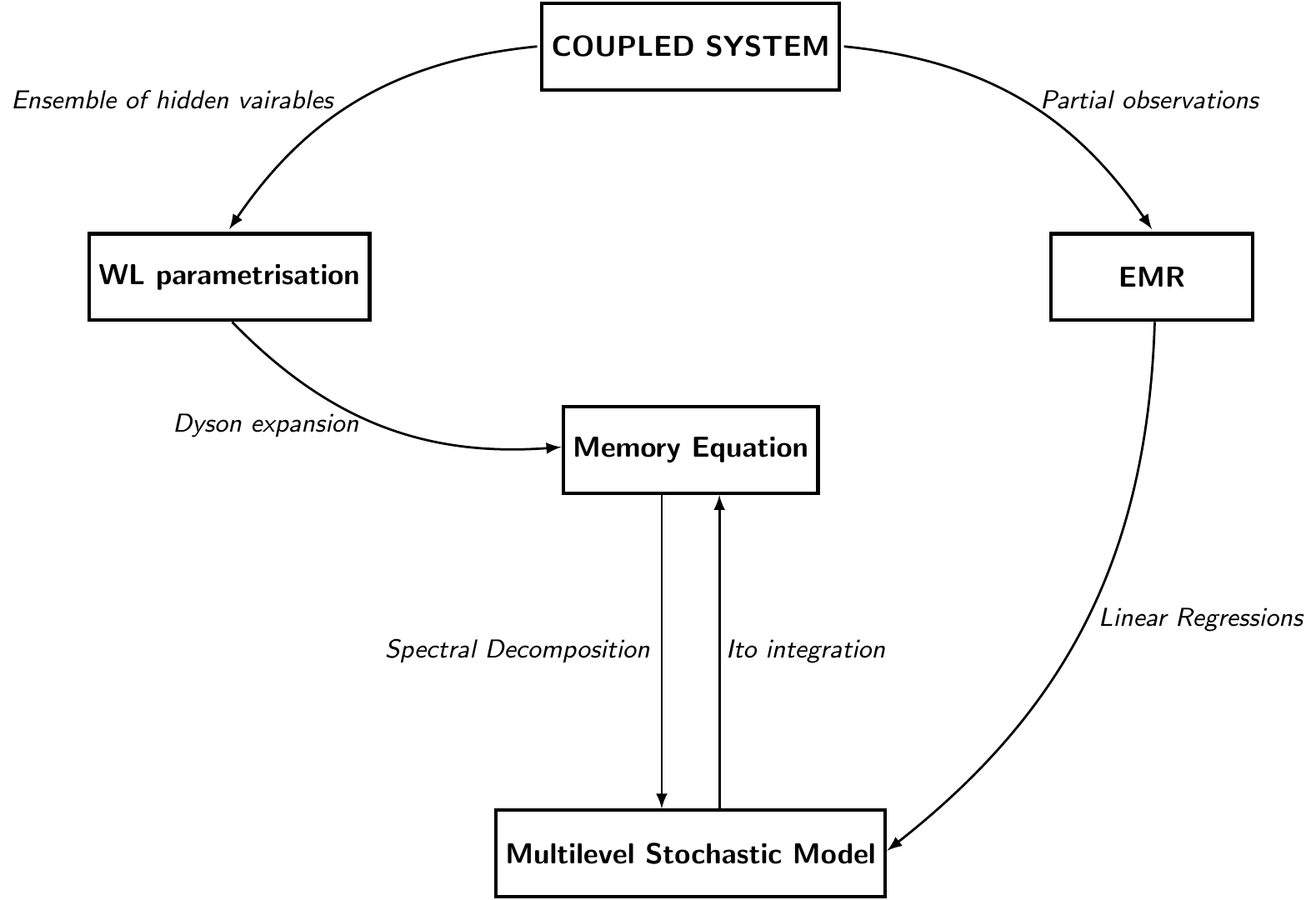}
	\caption{\label{schematic}Schematic view of the two complementary approaches 
	studied in this paper. 
	The arrows on the left-hand side indicate top-down, perturbative parametrizations; 
	on the right, they refer to bottom-up, empirical parametrizations.}
\end{figure}

The overall goal of this paper is to provide a conceptual and analytical link between these two approaches, aiming, on the one hand, to buttress the practical relevance of the WL perturbative approach and, on the other, 
to provide further insight into the well-documented robustness of the EMR 
method. Moreover, we will clarify how multilevel systems arise from both the \emph{top-down} (WL) and the \emph{bottom-up} (EMR) approaches. The paper explores the complete set of boxes and explains all the arrows in Figure~\ref{schematic}. The diagram in the figure shows that, starting from 
the top box, one can arrive at a memory equation, like \eqref{mori zwanzig 2}, via \emph{top-down} or \emph{bottom-up} methods, as indicated by the left and right sequence of arrows, respectively.

The paper is structured as follows. Section~\ref{weak-coupling limit parametrization} revisits the derivation of the WL parametrization method by applying the  Dyson expansion to the Koopman operator associated with two 
weakly coupled dynamical systems. We show that such an expansion need not 
be truncated for additively coupled models and consider more general coupling laws than those in \cite{Wouters2013}.

 Furthermore, we study the problem of finding Markovian representations of the memory equation in the WL approach based on the spectral decomposition of the Koopman semigroup. Specifically, Theorem~\ref{proposition: 1} shows, in the case of a scalar observable, how to recast the stochastic integro-differential equation arising in the WL parametrization as a multilevel Markovian stochastic system involving explicitly the spectral elements of the (uncoupled) Koopman operator, and we point out in Remark~\ref{Rem_genedim} how such a Markovianization extends to the multidimensional setting. Section~\ref{MSMEMR} provides new insights into the Markovian representation adopted in the MSM framework; these insights help one to determine, in particular, the number of levels required for EMR to converge. 
%% see Section~\ref{MSMEMR}.

Section~\ref{RESULTS} presents a comparison of the data-driven and top-down parametrization approaches using a simple conceptual stochastic climate model. Finally, we discuss the conclusions obtained from this investigation in Section~\ref{CONCLUSIONS}.  

Appendices are included in order to avoid diverting the attention of the reader from the main message of the paper. Appendix~\ref{App:proof} provides the proof of the key Theorem~\ref{proposition: 1}. Appendix~\ref{Markovianity}  briefly revisits the spectral decomposition of correlation functions and provides a criterion to quantify the loss of the semigroup property that is useful in identifying non-Markovian effects. Appendix~\ref{ITO} discusses the stochastic It\^o integration of the elementary form of 
an MSM. Appendix~\ref{L84L63} shows how EMR approaches can capture the dynamical properties of partially observed systems; in it, we consider a simple climate model, obtained by coupling the Lorenz atmospheric model \cite[(L84)]{Lorenz1984a} and the Lorenz convection model \cite[(L63)]{lorenzdeterministic1963}.

\section{Revisiting the Weak-Coupling--Limit Parametrization} \label{weak-coupling limit parametrization}
To study dynamical systems in which one can separate the variables into two groups, with weak coupling between the two, one often resorts to so-called parametrizations of the effects of one group on the other. In the weak-coupling limit, the coupling itself can be treated as a perturbation of the main dynamics \cite{lucariniwouters, Wouters2012, Wouters2013}. Granted such an assumption and some degree of structural stability of the system, one can apply response theory to derive explicit stochastic and memory terms to describe the impact of the variables we want to neglect on the variables of interest, in the Mori-Zwanzig spirit. Note that, to do so, no assumption on time-scale separation between the two groups of variables is necessary. This point is particularly relevant in fields like the climate sciences, where no clear time-scale separation is observed, so that asymptotic expansions of the kind used in homogenization theory are of 
limited utility.

\subsection{Deriving the WL Approximation for the Mori-Zwanzig Formalism}\label{WLapproximation}

Here, using a perturbative approach, we review the derivation of the parametrization presented in refs.~\cite{lucariniwouters, Wouters2012, Wouters2013}. Formally, we want to couple two dynamical systems generated independently by two vector fields $\FFF : \mathcal{X}\subseteq \mathbb{R}^{d_1} \longrightarrow \mathcal{X}$ and $\GGG : \mathcal{Y}\subseteq \mathbb{R}^{d_2}\longrightarrow \mathcal{Y}$ with possibly $d_1 \neq d_2$ and typically $d_1\ll d_2$. %% Then, by imposing some coupling law between the two systems, 
We study a broad class of systems of the form:
%%%%%%%%%%%%%%%%%%%%%%%
\begin{subequations}\label{Main_sys}
\begin{align} 
         \dot{\mathbf{x}}(t) &= \FFF(\mathbf{x}(t)) + \epsilon \CCC_{\xx}^{\xx}\left(\xx(t) \right) : \CCC_{\yy}^{\xx}\left(\yy(t) \right), 
	\label{mle1}\\
	\dot{\mathbf{y}}(t) &= \GGG(\mathbf{y}(t)) + \epsilon \CCC_{\xx}^{\yy}\left(\xx(t) \right):\CCC_{\yy}^{\yy}\left(\yy(t) \right). 
	\label{mle2}
\end{align}
\end{subequations}
%%%%%%%%%%%%%%%%%%%
%%%%%%%%%%%%%%%%%%%%%%%%%%%%%
The operation indicated by the colon $\xx:\yy$ denotes the Hadamard product that multiplies vectors or matrices component-wise. Here, four new vector fields have been introduced to model the coupling law, namely $\CCC_{\xx}^{\xx}:\mathcal{X} \longrightarrow \mathcal{X}$, $\CCC_{\xx}^{\yy}: \mathcal{X} \longrightarrow \mathcal{Y}$, $\CCC_{\yy}^{\xx}:\mathcal{Y} \longrightarrow \mathcal{X}$ and $\CCC_{\yy}^{\yy}:\mathcal{Y} \longrightarrow \mathcal{Y}$. 

The real parameter $\epsilon$  controls the strength of the coupling between the two groups of variables, $\xx(t)$ and $\yy(t)$, so that the $\xx$-  and $\yy$-variables are uncoupled for $\epsilon = 0$. We assume that 
the vector fields $ \FFF$ and $\GGG$, as well as the coupling laws %% $\CCC_{\xx}^{\xx}$, $\CCC_{\xx}^{\yy}$
in Eqs.~(\ref{mle1},\ref{mle2}) are such that system \eqref{Main_sys} possesses a global attractor. Furthermore, we assume throughout this article 
that this global attractor supports an physical invariant probability measure $\mu$ that describes the distribution of trajectories onto the global attractor.

The WL parametrization views the coupling as an $\epsilon$-perturbation of the otherwise independent $\xx$- and $\yy$-processes, with $\xx$ the observed and $\yy$ the hidden variables. One next assumes that the impacts of perturbations applied to these processes can be addressed using response theory \cite{ruelle_nonequilibrium_1998,ruelle_review_2009}, so that response formulas can be used to derive an effective equation for the $\xx$-variables. 

Taking the Mori-Zwanzig \cite{mori_transport_1965, zwanzig_memory_1961} point of view, we wish to calculate the evolution of observables that depend on the observed variables $\xx$ alone, $\Phi = \Phi (\xx(t))$. The idea, following \cite{Wouters2013}, is to perform a perturbative expansion 
of the differential operator $\LLL$ governing the evolution of $\Phi(\xx(t))$ under the action of the flow associated with Eq.~\eqref{Main_sys}. Denoting by $u(\xx,\yy,t)$ the time evolution of a smooth observable $\Phi$ in $\mathcal{C}^\infty(\cX\times \cY)$,  the first step of this %% {\mg 
Dyson-like operator 
expansion reads as follows:
\be\label{liouville equation}
\partial _t u = \LLL u =(\LLL_0 +\epsilon \LLL_1) u.
\ee
Here $ \LLL_0$ and $\LLL_1$ account for the advective effects of the uncoupled and coupling terms, respectively, that compose the RHS of Eq.~\eqref{Main_sys}, namely 
\begin{subequations} \label{eq:WL_LLL}
\begin{align} 
	\LLL_0 &= \begin{bmatrix}
	\FFF (\xx) \\ \GGG (\yy)
	\end{bmatrix}\cdot \begin{bmatrix}
	\nabla_{\xx} \\ \nabla_{\yy}
	\end{bmatrix}, \label{WL_LLL_0} \\
	 \LLL_1 &=  \begin{bmatrix}
	 \CCC_{\xx}^{\xx}\left(\xx \right) : \CCC_{\yy}^{\xx}\left(\yy \right) \\ \CCC_{\xx}^{\xx}\left(\xx \right) : \CCC_{\yy}^{\xx}\left(\yy \right)
	\end{bmatrix}\cdot \begin{bmatrix}
	\nabla_{\xx} \\ \nabla_{\yy}
	\end{bmatrix};\label{WL_LLL_1}
\end{align}
\end{subequations}
 in Eqs.~\eqref{eq:WL_LLL}, $\nabla_{\xx}$ and $\nabla_{\yy}$ denote the vector differential operators with respect to the variables $\xx$ and $\yy$.

Recalling Eq.~\eqref{koopman}, the solution operator of Eq.~\eqref{liouville equation} is the Koopman operator. Formally, its dual acts on densities and it is the so-called \emph{transfer operator} \cite{B00, lasota}. Equation~\eqref{liouville equation} is thus a transport equation, where the physical quantity or observable is advected by the vector field on the RHS of Eq.~\eqref{Main_sys}.

Note that the operator formalism presented here in the deterministic dynamical systems setting --- and the associated semigroup theory --- extends 
to Markov diffusion processes driven by a stochastic forcing \cite{chekroun2019c}.  In the latter case, the transport equation~\eqref{liouville equation} becomes the so-called backward Kolmogorov equation that describes 
the evolution of the expected value of observables. Loosely speaking, the 
corresponding extension amounts to adding a Laplacian-like operator to the advection operator $\LLL$ \cite{chekroun2019c, pavliotisbook2014}. See \cite{Hairer2010} for the appropriate context in testing the applicability of response theory to the independent  $\xx$  and $\yy$ processes, when 
$\epsilon=0$.

More precisely, one associates to the solution $u(\xx,\yy,t)$ of Eq.~\eqref{liouville equation} unfolding from an observable $\Phi = \Phi(\xx,\yy)$ at time $t=0$, a family of linear Koopman operators indexed by time 
$\{U_t\}_{t\geq 0}$ such that $u(\xx,\yy,t)=U_t \Phi (\xx, \yy)$, for any $t\geq 0$ and $(\xx, \yy)$ in $\cX \times \cY$. These operators are defined --- as mentioned already in connection with introducing the GLE~\eqref{mori zwanzig 2} --- as exponentials of the operator $\LLL$, i.e., $U_t=e^{t\LLL}$. This notation is formal as the operator $\LLL$ is unbounded; it is, however, useable as $\{U_t\}_{t\geq 0}$ satisfies  the semigroup property, i.e.~$U_{t+s}=U_t U_s$, $t,s\geq 0$, as for a standard exponential. Over the appropriate function space\footnote{ Such a space can be chosen, for instance, as $D_p=\{ \Phi \in L^{p}_{\mu} (\cX \times \cY) \, | \, A\Phi=\lim_{t\rightarrow 0} \ t^{-1}(U_t \Phi -\Phi) \textrm{ exists}\}$ for some $ p \in [1,\infty]$, with $\mu$ denoting a relevant 
invariant measure of the system~\eqref{Main_sys} supported by the global attractor, while the limit is taken in the sense of strong convergence \cite{engel2006short}.} of observables $\Phi$, this family actually forms 
a strongly continuous contracting semigroup \cite{engel2006short}.

The action of the flow on an observable $\Phi$ becomes thus more transparent thanks to the operator $U_t$, according to the equation
\begin{equation}\label{Def_Koopman}
U_t \Phi(\xx_0, \yy_0) = e^{t\LLL}\Phi (\xx_0, \yy_0) = \Phi (\xx(t;\xx_0),\yy(t;\yy_0)),
\end{equation}
where $(\xx(t;\xx_0),\yy(t;\yy_0))$ denotes the system's solution at time 
$t$ emanating from the initial state $(\xx_0, \yy_0)$ at time $t=0$.  In what follows, we omit the subscript $0$ in $(\xx_0, \yy_0)$ but still take it as an initial state. 

The semigroup $\{U_t\}_{t\geq 0}$ is known as the Koopman semigroup and for each $t$, $U_t$ is the Koopman operator mentioned above; see also \cite[Sec.~4]{kondrashovdata2015}. When the coupling parameter $\epsilon$ in system~\eqref{Main_sys} is small, one can use formal perturbation expansions of the Koopman semigroup to better isolate and assess the coupling effects at the level of observables. To do so, we follow here the perturbation expansion first introduced by Freeman J. Dyson in the context of quantum electrodynamics \cite{Dyson1949} and later formulated rigorously in mathematical terms in \cite{Gill2017}. Formally, this expansion reads as follows:
\begin{subequations}\label{pertexp}
\begin{align} 
%\Phi(\xx(t),\yy(t)) &=
%%%%%MC--->VC & MSG: useless to recall the relation between  \Phi(\xx(t),\yy(t))  and e^{t\LLL}\Phi(\xx,\yy) here
U_t\Phi (\xx,\yy) = e^{t\LLL}\Phi(\xx,\yy) & = e^{t\LLL_0+t\epsilon \LLL_1}  \Phi(\xx,\yy) \\ \label{pertexp1} &= e^{t\LLL_0}\Phi(\xx,\yy)+\epsilon\int_{0}^te^{s\LLL}\LLL_1e^{(t-s)\LLL_0}\Phi(\xx,\yy)\dd s\\ \label{pertexp2}  
&= e^{t\LLL_0}\Phi(\xx,\yy)+\epsilon\int_{0}^te^{(t-s)\LLL_0}\LLL_1e^{s\LLL}\Phi(\xx,\yy)\dd s,
\end{align}
\end{subequations}
and it yields the following expansion of the Koopman operator in $\epsilon$:
\be\label{pertexp3}
U_t\Phi (\xx,\yy) = e^{t\LLL_0}\Phi(\xx,\yy)+\epsilon\int_{0}^te^{(t-s)\LLL_0}\LLL_1e^{s\LLL_0}\Phi(\xx,\yy)\dd s + \mathcal{O}\left(\epsilon^2\right).
\ee
This identity shows that the evolution of a generic observable can be described as an  $\epsilon$-perturbation of its decoupled evolution according to ${\LLL_0}$. %% in the uncoupled regime. 
We note that these expansions are purely formal and in particular it is not  clear in which sense this expansion might converge. For a bounded perturbation operator $\LLL_1$, it would be straightforward to prove boundedness of the resulting perturbed semigroup. However, $\LLL_1$ here is a differential linear operator, for which direct estimates are more laborious. Leaving alone the functional analysis framework that would make such an expansion rigorously convergent, we shall use nevertheless the expansion~\eqref{pertexp3} throughout this article. %% and in particular to.

The objective now is, using this operator expansion, to derive an effective reduced-order model for the evolution of the $\xx$-variable without having to resolve the $\yy$-process. We start observing the system at $t=0$, but assume that it has already attained a steady state. Since we are only concerned with observables depending solely on the $\xx$-variables, we formulate now an evolution equation for such observables. To do so, we 
consider first the Liouville equation for a generic $\yy$-independent observable $\Phi$, this is, $\Phi(\xx,\yy)=\Phi(\xx)$, for every $\xx$ and 
$\yy$. 

For such an observable, at the time we start observing the coupled system, Eq.~\eqref{liouville equation} reduces to
\be\label{init liouville}
\partial _t \big(U_t \Phi\big) |_{t=0}=\left[ \FFF(\xx) + \epsilon \CCC_{\xx}^{\xx}\left(\xx \right):\CCC_{\yy}^{\xx}\left(\yy \right)  \right]\cdot \nabla_{\xx}\Phi,
\ee
where $\cdot$ denotes an inner product in $\mathcal{X}$. Equation~\eqref{init liouville} illustrates the trivial fact that the time evolution in Eq.~\eqref{Main_sys} of an $\xx$-dependent physical quantity is also affected by the $\yy$-variables. 

Following \cite{Wouters2012,Wouters2013}, the decoupled equations are assumed to have been evolving for some time prior to the coupling. Hence, we 
have to formally parametrize the evolution of the $\CCC_{\yy}^{\xx}\left(\yy \right)$-contribution to the vector field  which is, ultimately, a vector-valued observable. 

We do so by introducing an extended version of the Koopman operators that 
act on vectors component-wise, rather than just on real-valued observables. Consider $\mathbf{v}: \mathcal{X}\times \mathcal{Y}\longrightarrow \mathbb{R}^d$, for some positive integer $d$, and define the action of the Koopman operator $e^{t\LLL}$ on $\mathbf{v}$ as:
\be\label{v_Koop}
	\left[e^{t\LLL}\mathbf{v}(\xx,\yy)\right]_i = e^{t\LLL}\left[\mathbf{v}(\xx,\yy)\right]_i
\ee
for every $i=1,\ldots,d$. The definition~\eqref{v_Koop} will allow us to use the semigroup notation for observables of possibly different dimensions, all of which take their inputs in the phase space $\mathcal{X} \times \mathcal{Y}$ . Ultimately, this is a component-wise evaluation of our extended Koopman operator family, and its generator can be obtained analogously. As mentioned above, we have to model the effects of the coupling vector field $\CCC _{\yy}^{\xx}(\yy)$, whose state at time $t=0$ is the 
product of the evolution from time $-t$ to $0$. We then have, with the dynamics starting at time $-t$ and initial state $(\xx_0,\yy_0)$,
\begin{equation}
	\CCC_{\yy}^{\xx}\left(\yy \right) = e^{t\LLL}\CCC_{\yy}^{\xx}\left(\yy_0,-t \right)=e^{t\LLL}\CCC_{\yy}^{\xx}\left(\yy_0 \right). 
\end{equation}
%% where we have applied the assumption that the dynamics started at time 

Now, by using the perturbative expansions in Eqs.~(\ref{pertexp1})-(\ref{pertexp3}), we obtain:
\begin{subequations}
\begin{align}
   \CCC_{\yy}^{\xx}\left(\yy \right) &= e^{t\LLL}\CCC_{\yy}^{\xx}\left(\yy_0 \right)=e^{t\LLL_0+t\epsilon \LLL_1}	\CCC_{\yy}^{\xx}\left(\yy_0 \right) \\ \label{exact expansion1} 
   &= e^{t\LLL_0}\CCC_{\yy}^{\xx}\left(\yy_0 \right)+\epsilon\int_{0}^te^{s\LLL}\LLL_1e^{(t-s)\LLL_0}\CCC_{\yy}^{\xx}\left(\yy_0 \right)\dd s\\ \label{exact expansion2} 
   &= e^{t\LLL_0}\CCC_{\yy}^{\xx}\left(\yy_0 \right)+\epsilon\int_{0}^te^{(t-s)\LLL_0}\LLL_1e^{s\LLL}\CCC_{\yy}^{\xx}\left(\yy_0 \right)\dd s \\\label{lw approx}
   &= e^{t\LLL_0}\CCC_{\yy}^{\xx}\left(\yy_0 \right)+\epsilon\int_{0}^te^{(t-s)\LLL_0}\LLL_1e^{s\LLL_0}\CCC_{\yy}^{\xx}\left(\yy_0 \right)\dd s + \mathcal{O}\left(\epsilon^2\right).
\end{align}
\end{subequations}
Plugging the indentity in Eq.~\eqref{exact expansion2} into Eq.~(\ref{init liouville}, we find the following expression:
\begin{equation}\label{time zero liouville}
		\partial _t \big(U_t \Phi\big) |_{t=0} = \left[ \FFF(\xx) + \epsilon \CCC_{\xx}^{\xx}\left(\xx \right):\Big\{ e^{t\LLL_0}\CCC_{\yy}^{\xx}\left(\yy_0 \right)+\epsilon\int_{0}^te^{s\LLL}\LLL_1e^{(t-s)\LLL_0}\CCC_{\yy}^{\xx}\left(\yy_0 \right)\dd s \Big\}  \right]\cdot \nabla_{\xx} \Phi.
\end{equation}
This equation is an exact reformulation of the problem induced by Eq.~(\ref{init liouville}). This reformulation demonstrates that memory effects enter at second order in powers of the coupling parameter. Notice, though, that even if Eq.~\eqref{time zero liouville} reduces the dimensionality 
of the problem from $d_1 + d_2$ to $d_1$, it does not constitute an approximation for the evolution of $\Phi$ as an observable of $\xx$ alone, since it depends on the evolution of the $\yy$-variables in the coupled regime by means of the action of $e^{s\LLL}$ onto $\LLL_1$.

Therefore, we need to perform a further approximation by considering Eq.~(\ref{lw approx}) instead, which leads to:
\begin{equation}\label{dysonapprox}
\partial _t \big(U_t \Phi\big) |_{t=0} \simeq \left[ \FFF(\xx) + \epsilon \CCC_{\xx}^{\xx}\left(\xx \right):\Big\{ e^{t\LLL_0}\CCC_{\yy}^{\xx}\left(\yy_0 \right)+\epsilon\int_{0}^te^{(t-s)\LLL_0}\LLL_1e^{s\LLL_0}\CCC_{\yy}^{\xx}\left(\yy_0 \right)\dd s \Big\}  \right]\cdot \nabla_{\xx} \Phi ,
\end{equation}
where the terms of order $\epsilon ^3$ have been dropped. Equation~\eqref{dysonapprox}  is our equivalent of the Dyson approximation in quantum electrodynamics; it approximates the evolution of the $\xx$-variables with no need for the evolution of the $\yy$-variables in the coupled regime.

This result amounts to saying that --- by observing only the statistical properties of the decoupled dynamics of the $\yy$-process, obtained with $\epsilon=0$ --- one can 
construct a Markovian contribution
\bes 
\CCC_{\xx}^{\xx}\left(\xx \right):\Big\{e^{t\LLL_0}\CCC_{\yy}^{\xx}\left(\yy_0 \right)\Big\},
\ees
  and a non-Markovian contribution 
\bes
\CCC_{\xx}^{\xx}\left(\xx \right):\Big\{ \int_{0}^te^{(t-s)\LLL_0}\LLL_1e^{s\LLL_0}\CCC_{\yy}^{\xx}\left(\yy_0 \right)\dd s \Big\},
\ees
 to the dynamics of the $\xx$-variables that is able to describe the impact the the coupling. 

Expanding the kernel $\tilde{\mathcal{K}}$ of the memory contribution, we 
get: %% \cm{Why does ${\mathcal{K}}$ have a} \verb+\tilde+ \cm{here and not in Eq.~\eqref{kernel}?}
\begin{subequations}
\begin{align}
\tilde{\mathcal{K}}(t,s,\xx_0,\yy_0):&=e^{(t-s)\LLL_0}\LLL_1e^{s\LLL_0}\CCC_{\yy}^{\xx}\left(\yy_0 \right) \\ \label{kernel2} 
&= e^{(t-s)\LLL_0}\left(\left[ \CCC_{\xx}^{\xx}(\xx_0):\CCC_{\yy}^{\xx}(\yy_0) \right]\cdot \nabla_{\xx} + \left[\CCC_{\xx}^{\yy}(\xx_0):\CCC_{\yy}^{\yy}(\yy_0)  \right]\cdot \nabla_{\yy}\right)e^{s\LLL_0}\CCC_{\yy}^{\xx}\left(\yy_0 \right)\\ 
&= \left[e^{(t-s)\LLL_0} \left(\CCC_{\xx}^{\yy}(\xx_0):\CCC_{\yy}^{\yy}(\yy_0)  \right)\right]\cdot \nabla_{\yy}e^{s\LLL_0}\CCC_{\yy}^{\xx}\left(\yy_0 \right).
\end{align}
\end{subequations}
Note that the leading-order Koopman operator $e^{s\LLL_0}$ models the evolution of the observables in the uncoupled regime. Since there is no prior knowledge on initializing the coupled system at time $-t$, the initial state $\yy_0$ in the hidden variables should be drawn from an ensemble, according to a probability density function. At this stage, there is freedom in the choice of such a prior. However, since we are assuming that the 
coupled system was initialized at time $-t$, it is natural to draw $\yy_0$ according to the invariant measure $\nu$ associated with the dynamical system generated by the vector field $\GGG$ from Eq.~\eqref{Main_sys}.

We wish to sample initial conditions from the coupled steady state, but do not assume any prior knowledge of the coupled statistics. As discussed in \cite{Wouters2012,Wouters2013}, we can take advantage of response theory to address this situation. Indeed, for any sufficiently smooth observable $\Psi$, we have:
$$ 	\langle \Psi \rangle_{\epsilon} = \langle  \Psi \rangle_{\epsilon=0} + \sum_{k=1}^{\infty}\epsilon ^{k}\delta_k [\Psi],$$
where $\langle \Psi \rangle_{\epsilon}$ is the expectation value of $\Psi$ in the coupled system~\eqref{Main_sys}, %% given by Eqs. \ref{mle1}-\ref{mle2}, 
$\langle \Psi \rangle_{\epsilon=0}$ is the expectation value of $\Psi$ according to the statistics generated by the uncoupled $\yy$ process obtained by setting $\epsilon=0$ in Eq.~\eqref{mle2}, and $\epsilon ^{k}\delta_k [\Psi]$ is the $k^{\mathrm{th}}$-order response. In what follows, we remove the subscripts for the averages when $\epsilon = 0$. Therefore, we have that the expected value of the coupling function reads as:
\begin{equation}
	\bigg\langle  \CCC_{\yy}^{\xx} \bigg \rangle_{\epsilon} = \bigg\langle 
 \CCC_{\yy}^{\xx}\bigg \rangle + \sum_{k=1}^{\infty}\epsilon ^{k}\delta_k [\CCC^{\xx}_{\yy}].
\end{equation}

Likewise, we can calculate the average of such function at time $t$:
\begin{equation}
	\bigg\langle  e^{t\LLL_0}\CCC_{\yy}^{\xx} \bigg \rangle_{\epsilon} = \bigg\langle  e^{t\LLL_0}\CCC_{\yy}^{\xx} \bigg \rangle + \sum_{k=1}^{\infty}\epsilon ^{k}\delta_k [e^{t\LLL_0}\CCC^{\xx}_{\yy}].
\end{equation}
Now, by letting $\tilde{\eta}(t,\yy_0)= e^{t\LLL_0}\CCC_{\yy}^{\xx}(\yy_0)$, we find that in order for the approximate statistics to agree up to 
second order in $\epsilon$ with the exact one, we only need to impose the 
following conditions upon the first two moments of the parametrized noisy 
fluctuations (see also \cite{Wouters2012}):
\begin{subequations}\label{eq:moments}
\begin{align}
	\bigg \langle \tilde{\eta}(t,\yy_0) \bigg \rangle &= \int \nu\left(\dd 
\yy_0\right)\CCC_{\yy}^{\xx}\left(\yy_0 \right), \label{moments1} \\
	\bigg \langle \tilde{\eta}(t,\yy_0)\tilde{\eta}^{\top}(0,\yy_0) \bigg \rangle &= \int \nu(\dd \yy_0) e^{t\LLL_0}\CCC_{\yy}^{\xx}\left(\yy_0 \right)\left(\CCC_{\yy}^{\xx}\left(\yy_0 \right)\right)^{\top}; \label{moments2}
\end{align}
\end{subequations}
here $(\cdot)^{\top}$ stands for the transpose of a vector or a matrix. It follows that any stochastic noise $\eta(t)$ that satisfies the two conditions above will be suitable for parametrizing the fluctuations in the $\yy$-dynamics tied to the lack of knowledge in the initial state. Each of 
the entries in the correlation matrix given by Eq.~\eqref{moments2} is the correlation function between the components of the vector field $\CCC^{\xx}_{\yy}$ and these will become explicit provided a suitable spectral decomposition is at hand. Such a decomposition will be provided later in Sect.~\ref{ssec:Koopman-eigen}, although the reader is referred at this point to Appedix~\ref{Markovianity} for clarity.

In the memory term, though, we neglect $\epsilon$-corrections to its statistics since memory effects are of order $\epsilon ^2$ already. Thus, we have, by averaging the kernel $\tilde{\mathcal{K}}(t,s,\xx_0,\yy_0)$ in Eq.~\eqref{kernel2} with respect to the $\yy$-variables,
\begin{subequations}\label{kernel}
\begin{align}
	\mathcal{K}(t,s,\xx):&=\int \nu(\dd \yy_0) \left[e^{(t-s)\LLL_0} \left(\CCC_{\xx}^{\yy}(\xx_0):\CCC_{\yy}^{\yy}(\yy_0)  \right)\right]\cdot \nabla_{\yy}e^{s\LLL_0}\CCC_{\yy}^{\xx}\left(\yy_0 \right) \\
	& = \int \nu(\dd \yy_0)  \left[e^{(t-s)\LLL_0} \CCC_{\xx}^{\yy}(\xx_0): e^{(t-s)\LLL_0}\CCC_{\yy}^{\yy}(\yy_0)\right]  \cdot \nabla_{\yy}e^{s\LLL_0}\CCC_{\yy}^{\xx}\left(\yy_0 \right) \\ 
	& = \int \nu(\dd \yy_0)\left[ \CCC_{\xx}^{\yy}(\xx(t-s)): e^{(t-s)\LLL_0}\CCC_{\yy}^{\yy}(\yy_0)\right]  \cdot \nabla_{\yy}e^{s\LLL_0}\CCC_{\yy}^{\xx}\left(\yy_0 \right)
\end{align}
\end{subequations}
This way, the memory kernel only depends on the $\xx$-variables. Hence, we find a self-consistent evolution of the $\xx$-variables, subject to the 
influence of unobserved variables $\yy$, in the form of a stochastic integro-differential equation (SIDE) resembling the 
GLE \eqref{mori zwanzig 2}:
\begin{equation}\label{lw parametrization}
	\dot{\xx}(t)=\FFF(\xx) + \epsilon \CCC_{\xx}^{\xx}\left(\xx \right): \eta(t)+\epsilon^2\CCC_{\xx}^{\xx}(\xx)\cdot \int_{0}^t\mathcal{K}(t,s,\xx)\dd s,
\end{equation}
where $\eta(t)$ is a stochastic forcing that agrees with the mean and correlation properties stated in Eq.~\eqref{eq:moments}.

We emphasize that the solution $\xx(t)$ of the original system of ordinary differential equations~\eqref{Main_sys} does not satisfy Eq.~\eqref{lw parametrization}: it is just the proposed reduced-order model for the $\xx$-variables. The closure provided by expressing the corrections in the second and third term on the RHS of Eq.~\eqref{lw parametrization} as functions of $\xx$ alone is typically called a parametrization of the effect of the unobserved $\yy$-variables in the climate sciences \cite{GhilLucarini2020}.

Note that there is considerable freedom in the choosing the noise, since we only require agreement up to the second moment. However, a direct consequence of this weak-coupling parametrization is that realizations of the 
noise can be produced by directly integrating the decoupled hidden variables, or by representing it using simple autoregressive models \cite{Vissio2018b}. We are assuming here that the uncoupled dynamics leads to a noisy signal; this can be achieved either by the presence of stochastic forcing in the hidden variables \cite{Vissio2018b, Wouters2016} or by their uncoupled dynamics being chaotic \cite{Vissio2018a}.

To summarize, the weak-coupling limit allows one to develop a parametrization of the hidden variables for a system of coupled equations where no separation of time scales is assumed. Moreover, this approach provides explicit approximate expressions for the deterministic, stochastic, and non-Markovian terms in the Mori-Zwanzig formalism of Eq.~\eqref{mori zwanzig 2}. 

There are two sources of error in the parametrization proposed in Eq.~(\ref{lw parametrization}). First, the truncation performed in the Dyson expansion neglects higher-order effects, which are weighted by the third power of the small coupling parameter. Secondly, averaging over the statistics of the uncoupled dynamics can also introduce errors. Furthermore, the nature of the stochastic correction is not fully determined except for its lagged correlation.

The perturbation operator approach taken here is analogous to that of \cite{Wouters2013}, who only considered the independent or additive-coupling 
cases; the latter is expanded upon in Sect.~\ref{additive}. Here, though, 
we generalize further the parametrization formulas that can be obtained via perturbative expansion of linear operators. In fact, the present approach can also be extended to weakly coupled systems of the form:
\begin{subequations}
	\begin{align}
		\label{mle12}
		\dot{\mathbf{x}}(t) &= \FFF(\mathbf{x}(t)) + \epsilon \CCC^{\xx}\left(\yy(t) \right), \\
		\label{mle22}
		\dot{\mathbf{y}}(t) &= \GGG(\mathbf{y}(t)) + \epsilon \CCC^{\yy}\left(\xx(t),\yy(t) \right),
	\end{align}
\end{subequations}
where $\CCC^{\yy}$ encodes interactions that need not be separable between the $\xx$- and $\yy$-variables in the hidden layer of the model. Note that the full parametrization of arbitrary couplings was discussed by the two authors of \cite{Wouters2013} in previous work \cite{Wouters2012}, in 
which they used a response-theoretic approach.\\

% \noindent \cm{MG got to here in this version, on Friday, 26 March.}

\subsection{The Additive-Coupling Case}\label{additive}

The approximate Dyson expansion given in Eq.~\eqref{dysonapprox} is exact 
in the case of additive coupling. Such systems take the form:  
\begin{subequations}\label{add_mle}
	\begin{align}
	\label{add_mle12}
	\dot{\mathbf{x}}(t) &= \FFF(\mathbf{x}(t)) + \epsilon \CCC^{\xx}\left(\yy(t) \right), \\
	\label{add_mle22b}
	\dot{\mathbf{y}}(t) &= \GGG(\mathbf{y}(t)) + \epsilon \CCC^{\yy}\left(\xx(t) \right).
	\end{align}
\end{subequations}
Indeed, letting $\CCC ^{\yy}(\xx,\yy)=\CCC ^{\yy}(\xx)$ in Eq.~(\ref{mle22}) and using Eq.~(\ref{exact expansion1}) allows us to avoid the truncation of the Dyson expansion and yields the following expression for the memory term:
\begin{subequations}
\begin{align}
	\tilde{\mathcal{K}}(t,s,\xx,\yy_0)&=e^{s\LLL}\LLL_1e^{(t-s)\LLL_0}\CCC^{\xx}\left(\yy_0 \right) \\ &= e^{s\LLL}\left(\CCC^{\xx}(\yy_0) \cdot \nabla_{\xx} + \CCC^{\yy}(\xx) \cdot \nabla_{\yy}\right)e^{(t-s)\LLL_0} \CCC^{\xx}\left(\yy_0 \right)\\ &= e^{s\LLL} \left(\CCC^{\yy}(\xx) \right)\cdot \nabla_{\yy}e^{(t-s)\LLL_0} \CCC^{\xx}\left(\yy_0 \right),
\end{align}
\end{subequations}
which is exact.
Next, taking averages with respect to $\nu$, we obtain:
\begin{align}\label{wl parametrization additive}
\mathcal{K}(t,s,\xx)&=  \int \nu(\dd \yy_0)e^{s\LLL} \CCC^{\yy}(\xx) \cdot \nabla_{\yy}e^{(t-s)\LLL_0}\CCC^{\xx}\left(\yy_0 \right).
\end{align}
Hence, the parametrization in this additive-coupling case is exact, as no 
terms proportional to $\epsilon^k$, $k\geq3$ are present. The only assumption made is that the statistics in the $\yy$ variables have reached a steady state according to the unperturbed system. Finally, the full SIDE in 
this case takes the form:
\begin{equation}\label{lw parametrization independent}
	\dot{\xx}(t)=\FFF(\xx) + \epsilon \eta(t)+\epsilon^2\int_{0}^t\mathcal{K}(t,s,\xx)\dd s,
\end{equation}
where the stochastic process $\eta$ has the mean and correlation properties given by Eqs.~\eqref{eq:moments}. %%-\eqref{moments2}.
This equation is, thus, exactly the one obtained in \cite{Wouters2013}. 

Memory effects represented by integral terms seem unavoidable unless the memory kernels vanish quickly with respect to time. Infinite time scale separation between the two sets of variables leads, though, to the vanishing of the associated integral expressions \cite{Pavliotis2008}. Here, we are not assuming no such property in the coupled dynamical system under study; see Eqs.~\eqref{Main_sys} and \eqref{add_mle}. On the other hand, reduced phase spaces can help explain  the statistics of the dynamical system without resorting to delayed effects that entail the integrals in Eqs.~\eqref{lw parametrization} and \eqref{lw parametrization independent}. Following \cite{chekroun2019c}, we briefly review in Appendix \ref{Markovianity} a criterion based on Koopman operators --- and, more generally, Markov operators --- that enables one to decide whether memory effects can help explain 
the dynamics and statistics in reduced phase spaces.

\subsection{Markovian Representation through Leading Koopman Eigenfunctions}\label{ssec:Koopman-eigen}

 In the context of Langevin dynamics, there are known conditions on memory kernels that allow one to recast certain stochastic integro-differential equations into a Markovian SDE by means of extended variables; see \cite[Sec.~8.2]{pavliotisbook2014}. The stochastic processes that allow such a procedure are called quasi-Markovian \cite{pavliotisbook2014}.\footnote{Such a Markovianization procedure is actually not limited to stochastic processes and it relies on the same type of ideas in other contexts; see refs. \cite{Chek_al11_memo, chekroun_glatt-holtz, dafermos1970asymptotic} 
and \cite[Sec.~1.3]{kondrashovdata2015}.} 

This Markovianization theory can be formulated in the setting of near-equilibrium statistical mechanics, where one uses fluctuation-dissipation--like relations that link the decay properties of the memory kernel and the 
decorrelation rates of the  fluctuations. Here, we follow the approach in 
\cite{pavliotisbook2014} but without making any assumptions on the Hamiltonian behavior of the $\yy$-variables. We need, though, to make assumptions on the spectral properties of the generator of the $\yy$-dynamics, as explained below. 

We define the generator $\LLL^{\yy} _0$ of the Koopman semigroup associated with the $\yy$-dynamics by:
\begin{equation}\label{Def_L0}
\LLL_0^{\yy}\Phi(\yy)=\GGG(\yy)\cdot \nabla\Phi(\yy),
\end{equation}
for every real-valued observable $\Phi \in \mathcal{C}^{\infty}\left(\mathbb{R}^{d_2}\right)$ and we denote the associated Koopman operator at time $t$ by $U_t=e^{\LLL^{\yy} _0 t}$; the subscript $\mathbf{y}$ has been 
dropped from the $\nabla$ operator herewith, for notational clarity. Recall that the the spectrum of such operators provides useful insights into the statistical properties of the system; this topic is beyond the scope of the present paper but it is treated in detail in \cite{chekroun2019c}.

It suffices to show below that the spectrum of $U_t$ allows one to characterize the constitutive ingredients of the WL parameterization \eqref{lw parametrization} and \eqref{lw parametrization independent}, subject to natural assumptions. Even though we have clarified in Eq.~\eqref{v_Koop} how the Koopman operator acts on vector-valued observables of any dimension, we restrict now its action for simplicity to scalar real-valued observables, as in Eq.~\eqref{Def_L0}.  In this case, along the lines of the methodology of dynamic mode decomposition \cite{Mezic2005,Rowley2009,Schmid2010}, we can (formally) decompose the operator as:
\begin{equation}\label{Eq_L0_decomp}
e^{\LLL^{\yy}_0 t} = \sum _{j=1}^{  N} e^{t\lambda _j}\Pi_j + \mathcal{R}(t),
\end{equation}
where $\{\lambda_j\}_{j=1}^{  N}$ are the eigenvalues that form the point spectrum of $\LLL^{\yy} _0$ and $\Pi _j$ is the spectral projector onto the eigenspace spanned by the eigenfunction $\psi _j$. Here, $\mathcal{R}(t)$ is the residual operator associated with the essential spectrum of 
$\LLL^{\yy} _0$ and its norm is controlled by a decaying exponential. We assume furthermore that the spectrum of $\LLL^{\yy} _0$ lies in the complex left half-plane, and that in particular $\mathfrak{Re}\lambda _j \leq 0$ for any $j$. 

Such a spectral decomposition and its properties can be rigorously justified for a broad class of differential equations perturbed by small noise disturbances; see \cite[Theorem 1 and Appendix A.5]{chekroun2019c}. %% and \cite[Appendix A.5]{chekroun2019c}. 
 Based on these rigorous results, we assume, roughly speaking,  that these properties survive in a certain small-noise limit, and concentrate here 
on vector fields $\yy$ given by $\GGG$ in \eqref{Def_L0} for which a decomposition such as \eqref{Eq_L0_decomp} holds and a spectral gap does exist in the appropriate functional space.

 In the following lines, we examine the expression of the memory kernel $\mathcal{K}$ appearing in Eq.~\eqref{lw parametrization independent} using the eigendecomposition proposed in Eq.~\eqref{Eq_L0_decomp}. In particular, we study such an integral kernel $\mathcal{K}$ component-wise:
\begin{align}\label{memory kernel spectral decomposition}
	[\mathcal{K}(t,s,\xx)]_{i} &= \CCC^{\yy}(\xx(s))\cdot \bigg \langle \nabla\sum_{j=1}^Ne^{\lambda_j (t-s)}\alpha_{i,j} \psi _j (\yy) \bigg \rangle + \mathcal{R}(t-s)[\CCC ^{\xx}]_i \\&\approx \CCC^{\yy}(\xx(s))\cdot 
\bigg \langle \nabla\sum_{j=1}^Ne^{\lambda_j (t-s)}\alpha_{i,j} \psi _j 
(\yy)  \bigg \rangle \\ &=\CCC^{\yy}(\xx(s))\cdot  \sum_{j=1}^Ne^{\lambda_j (t-s)}\alpha_{i,j}\bigg \langle \nabla\psi _j (\yy)  \bigg \rangle,
\end{align}
for  $i=1,\ldots,d_1$, where $$\alpha_{i,j}= \left\langle \psi^{\ast}_j, [\CCC ^{\xx}]_i \right\rangle = \int \nu(\dd \yy)\overline{\psi_j^{\ast}(\yy)}[\CCC ^{\xx}(\yy)]_i,$$ and we have neglected the contribution 
coming from the essential spectrum. The $(\cdot)^{\ast}$ superscript is used to denote the dual eigenfunction. 

This decomposition highlights the fact that the leading eigenvalues of the operator governing the evolution of observables in the uncoupled $\yy$-dynamics set the time scale for the memory kernel. Furthermore, this spectral approximation implies that the correlation functions of the noise have the same decay properties, as will become apparent later in the proof of Theorem~\ref{proposition: 1}. It follows that the correspondence between the noise and integral time scales allows us to recast the SIDE in the 
WL equation~\eqref{lw parametrization independent} into a fully Markovian 
version with linearly driven hidden variables that are forced by the observed variables, through a functional dependence that can be nonlinear. More exactly, we have the following theorem.

\begin{theorem}\label{proposition: 1}
	Consider the system \eqref{add_mle} where Eq.~\eqref{add_mle12} is, instead, a scalar equation for a real-valued variable $x(t)$. 	
 	Let $\nu$ be the physical invariant measure associated with the equation
	\be\label{Eq_G}
	\dot{\yy}= \mathbf{G}(\yy),
	\ee
	i.e., with the flow determined by the vector field $\mathbf{G}$ in system \eqref{add_mle}, for $\epsilon=0$. Moreover, let  $\LLL_{0}^{\yy}$ be 
the (uncoupled) Koopman operator associated with \eqref{Eq_G} as defined in Eq.~\eqref{Def_L0}.

The point spectrum of $\LLL_{0}^{\yy}$ is assumed to be constituted of $N$ simple eigenvalues, whose corresponding eigenpairs
$\left\{ (\lambda_j,\psi_{j}), \; j=1,\ldots,N \right\}$ are ordered as 
follows: $0\geq \mathfrak{Re}\lambda_j\geq \mathfrak{Re}\lambda_{j+1}$ and $\lambda_j=\overline{\lambda_{j+1}}$ when $\mathfrak{Im}\lambda_j>0$, 
for $j$ in $\{1,\ldots, N\}$.

We assume that  $\CCC^{\xx}$ in \eqref{add_mle} lies in the $\mathrm{span}\left\{ \psi_{j}, \; j=1,\ldots,N \right\}$ and has $\nu$-mean zero.

Then, the WL equation \eqref{lw parametrization independent} associated with system \eqref{add_mle} admits a Markovianization of the form:
	\begin{subequations}\label{remarkovianisation1}
		\begin{align}
		\label{remarkovianisation3}	\dot{x}(t) & = \FFF (x(t)) + \epsilon \Lambda\cdot \mathbf{Z}(t), \\ 
		\label{remarkovianisation2} \dd\mathbf{Z}(t) & =
		\left(\epsilon \mathbf{R}(x(t))  +
		\mathrm{D}\mathbf{Z}(t)\right)\dd t + \Sigma \dd W_t,
		\end{align}
	\end{subequations}
	where $\Lambda$ and $\mathbf{Z}(t)$ lie in $\mathbb{C}^{N}$ for every $t$,  while the inner product $\Lambda \cdot\mathbf{Z}(t)$ is real. 
	Here, $\mathbf{R}$ is mapping  $\mathbb{R}$ into $\mathbb{C}^{N}$, $W_t$ 
is a (real-valued) $N$-dimensional Wiener process with covariance matrix $\Sigma$,  and $\mathrm{D}$ is an $N \times N$ matrix with complex entries, as specified below.
	
More precisely, 
	\begin{equation}\label{Eq_lambda}
		\Lambda=\left[  \alpha_1^{1/2}\beta_1^{1/2}, \ldots,\alpha_N^{1/2}\beta_N^{1/2} \right]^{\top},
		\end{equation}	
where 
		\begin{subequations}
			\begin{align}
		\label{eq:alpha}	\alpha_{j}&=\left \langle \psi^{\ast}_j , \CCC^{\xx} 
\right \rangle = \int \nu(\dd \yy)\overline{\psi_j^{\ast}(\yy)}\CCC^{\xx}(\yy), \\
		\label{eq:beta}	\beta_j&=\left \langle\CCC^{\xx}, \psi_j \right \rangle = \int \nu(\dd \yy)\CCC^{\xx}(\yy)\psi_j(\yy).
			\end{align}
		\end{subequations}
The $\mathbb{C}^{N}$-valued mapping $\mathbf{R}$ is defined as 
			\begin{equation}\label{Eq_Rexpression}
		\mathbf{R}\left(x \right)= \left(\CCC^{\yy}(x)\cdot\frac{\alpha_1^{1/2}}{\beta_1^{1/2}}\bigg \langle \nabla\psi _1 (\yy)  \bigg \rangle,\ldots,\CCC^{\yy}(x)\cdot\frac{\alpha_N^{1/2}}{\beta_N^{1/2}}\bigg \langle \nabla\psi _N (\yy)  \bigg \rangle\right),
		\end{equation}
where $\CCC^{\yy}$ is defined in  \eqref{add_mle22b}, and $\langle \cdot \rangle$ denotes averaging with respect to the invariant measure $\nu$.
	
Finally, $\mathrm{D}=\mathrm{diag}\left(\lambda_1,\ldots,\lambda_N\right)$ and the covariance matrix $\Sigma$ is given by
	\begin{align} \label{eq: def covariance matrix proposition}
	\Sigma=\left(-\left(\mathrm{D}+\mathrm{D}^{\ast}\right)\right)^{1/2}\mathrm{H},
	\end{align}
where $\mathrm{H}$ is an $N\times N$  matrix whose entries are defined as 
follows: If $\lambda_j$ is real, then $\mathrm{H}_{j,j}=1$, and if $\lambda_j=\overline{\lambda_{j+1}}$, 
	then
	\bea\label{Eq_H2}
	&\mathrm{H}_{j,j}=1,\\
	&\mathrm{H}_{j+1,j+1}=0,\\
	&\mathrm{H}_{j+1,j}=1,
	\eea
	while all other entries are zero.
\end{theorem}
%%%%%%%%%%%%%%%%%%%%%%%%%%%%%%%%%%%%%

The full proof appears in Appendix~\ref{App:proof}.

\begin{remark}\label{rem:fast}
Note that the Koopman operator of interest here is the one associated with the $\yy$-subspace $\mathcal{Y} \subseteq \mathbb{R}^{d_2}$ and not with the entire $(\xx, \yy)$-space $\mathcal{X}\times \mathcal{Y}\subseteq \mathbb{R}^{d_1+d_2}$. Other techniques, like the DMD mentioned in Sect.~\ref{ssec:MZ}, aim at extracting the modes of variability of the full system by means of studying the Koopman operator in the entire phase space through suitably defined observables. To this end, the latter methods employ projections of observables onto the eigenfunctions of the Koopman operator to obtain the so-called Koopman modes, which are susceptible of capturing the underlying dynamics. Notice that in Theorem 2.1, instead, we are using the Koopman eigenfunctions to identify the closure model, while projections only come into play in the definition of the coefficients $\alpha_{j}$ and $\beta_j$; see Eqs.~\eqref{eq:alpha} and \eqref{eq:beta}, respectively.
	
\end{remark}

\begin{remark}
\hspace*{.1em}  \vspace*{-0.1em}
\bi
\item[(i)] Assumptions on $\FFF$, $\mathbf{R}$ and $\Lambda$ that ensure that \eqref
{remarkovianisation1} possesses a global random attractor --- and thus a stable asymptotic behavior in the pullback sense ---  appear in \cite[Theorem 3.1 and Corollary 3.2]{kondrashovdata2015}. %%see also  \cite[Corollary 3.2]{kondrashovdata2015}. 

\item[(ii)] Note that Theorem~\ref{proposition: 1} can be viewed as a generalization of other Markovianization results for GLEs that appeared in the literature; see \cite{pavliotisbook2014}. For instance, %% if one considers 
the scalar GLE in $\mathbb{R}$ reduces to
\be\label{Eq_GLE0}
\dot{x}=F (x(t)) -\int_0^t K(t-s) x(s) \dd s +\eta(t),
\ee
where $\lambda$ is in $\mathbb{R}^{n}$, $M$ is a positive definite $n\times n$ matrix and $K(t-s)=\left(e^{M (t-s)} \lambda\right)\cdot \lambda$ 
determines the autocorrelation of the process $\eta(t)$.  In this setting, Eq.~\eqref{Eq_GLE0} is equivalent to the following SDE:
\bea
&\dot{x}=F(x(t))+ \lambda \cdot z\\
&\dd z  =(x\lambda -M z) \dd t + \Sigma \dd W_t,
\eea
with $\Sigma \Sigma^{\ast}=M+M^\ast$. Theorem~\ref{proposition: 1} allows for nonlinear dependence on $x$ in the $z$-equation, 
and thus for memory kernels that are more complicated  than in \eqref{Eq_GLE0}. Such a  generalization is of practical importance since the process $z$ can then have a more complex correlation dependence on the observed 
variable $x$ than the one afforded by linear memory terms. 
\ei
\end{remark}

\begin{remark}
 When $\LLL_{0}^{\yy}$ is self-adjoint --- in a suitable Hilbert space, as outlined in Appendix~\ref{Markovianity}, i.e., when $\LLL_{0}^{\yy}=\LLL_{0}^{\yy^{\ast}}$ --- the eigenvalues are real and the eigenvectors are mutually orthogonal. Self-adjointness thus  implies that there are no oscillations in the correlation functions of the noise or, equivalently, peaks in their power spectrum.  With respect to Theorem~\ref{proposition: 
1}, the matrix $\mathrm{H}$ in this case would be the identity, since the 
eigenvalues and eigenfunctions are real and the It\^o solutions of \ref{remarkovianisation2} are, hence, real as well.
\end{remark}

\begin{remark}\label{Rem_interpretation}
The resulting system given by Eq.~(\ref{remarkovianisation1}) is now fully Markovian and the only sources of error with respect to the original SIDE %% stochastic integro-differential equation 
\eqref{lw parametrization} lie (i) in the effects of the essential spectrum, which are neglected herein; and (ii) the assumptions about the coupling terms. Neglecting the essential spectrum is only valid for Koopman operators with a point spectrum capable of capturing the correlations in the 
decoupled $\yy$-system; the latter might only hold in the case of Markovian diffusion processes and not for deterministic ones. Also, the assumption that the coupling functions project solely on the point spectrum might 
not hold in general.

From a practical perspective, though, a suitable choice of dominant eigendirections can reduce the number of extra dimensions needed to integrate the system. Such a suitable choice boils down to neglecting particular eigendirections and this can be done according to two handy criteria: 
\bi
\item[(i)] The weight determined by the $\alpha_{j}$ and $\beta_j$ coefficients defined in Eqs.~(\ref{eq:alpha}, \ref{eq:beta}) is small; and 
\item[(ii)] The eigenvalues $\lambda_j$ of $\LLL_{0}^{\yy}$ satisfy $\mathfrak{Re} \lambda_{j_0} \ll \lambda^{\dagger}$, for some $j_0\in \{1,\ldots,N\}$, in which case $e^{\lambda_{j_0} t}$ decays rapidly as $t$ grows; 
here $\lambda^{\dagger} < 0$ and $|\lambda^{\dagger}|$ is some characteristic inverse time for the deterministic system $\FFF$. In addition, if $\alpha_{j_0}>\alpha_{j}$ for $j=1,\ldots,N$ and $j\neq j_0$, both the memory and the noise correlations die out fast. Hence, one can neglect the integral terms and perform a fully Markovian parametrization, which is possible in the presence of white noise.
\ei
\end{remark}

\begin{remark}\label{Rem_genedim}
Theorem~\ref{proposition: 1} is stated for $x$ scalar for the sake of simplicity, but this result extends to the $d_1$-dimensional  Eq.~\eqref{add_mle} for the (observed) variables $\xx$. In this Remark, we sketch the main elements that permit such a generalization. 

 Aside from the obvious generalization of the assumptions in Theorem~\ref{proposition: 1} to a multidimensional setting, the main  hypothesis consists of assuming that the now vector-valued coupling function $\CCC^{\xx}$ in Eq.~\eqref{add_mle} has components $\{\left[\CCC^{\xx}\right]_{i}: i=1,\ldots, d_1\}$ that project onto the $N$ simple eigenspaces of the decoupled Koopman operator introduced in Eq.~\eqref{Def_L0}. In this case, 
the construction of a multilevel Markovianization like Eq.~\eqref{remarkovianisation1} can be done in the following fashion:
	\begin{subequations}\label{eq: multilevel markovianization1}
		\begin{align}
		\dot\xx(t) &= \FFF(\xx(t))+\epsilon \Lambda\mathbf{Z}(t), \\
		\dd \mathbf{Z}(t) &= \left(\epsilon \mathbf{R}(\xx(t))+\mathcal{D}\mathbf{Z}(t) \right)\dd t + \mathcal{S} \dd W_t. 
		\end{align}
	\end{subequations}
	Here $W_t$ is a $d_1N$-dimensional Wiener process, $\mathbf{Z}(t)$ is a $d_1N$-dimensional vector, $\Lambda$ is a matrix of size $d_1\times d_1N$, $\mathbf{R}:\mathbb{R}^{d_1}\longrightarrow \mathbb{C}^{d_1N}$ and $\mathcal{D}$ and $\mathcal{S}$ are $d_1N\times d_1N$ block-diagonal matrices 
given by:
	\begin{equation}\label{Eq_multilevel matrices}
	\mathcal{D}=\begin{bmatrix}\mathrm{D}_1 &&\\ &\ddots &\\ && \mathrm{D}_N\end{bmatrix} \text{ and } \mathcal{S}=\begin{bmatrix}\Sigma_1 &&\\ &\ddots &\\ && \Sigma_N\end{bmatrix},
	\end{equation}
	where $\mathrm{D}_j$ and $\Sigma_j$ are $d_1\times d_1$ diagonal matrices with every (non-zero) element being equal to $\lambda_j$ or $\sqrt{-2\mathfrak{Re}\lambda_j}$, respectively. More importantly, the vectors $\mathbf{Z}(t)$ and $\mathbf{R}(\xx)$ are split into $N$ column vectors $\mathbf{z}_j(t)$ and $\mathbf{r}_j(\xx)$ of length $d_1$ with $j$ in $\{1,\ldots, N  \}$. This way, $\mathbf{Z}(t)=\left[ \mathbf{z}^{\top}_1(t),\ldots,\mathbf{z}^{\top}_{N}(t) \right]^{\top}$ and $\mathbf{R}(\xx)=\left[ 
\mathbf{r}^{\top}_1(t),\ldots,\mathbf{r}^{\top}_{N}(t) \right]^{\top}$. 
	
	Therefore Eq.~\eqref{eq: multilevel markovianization1} can be written as:
	\begin{subequations}\label{eq: multilevel markovianization2}
		\begin{align}
		\dot\xx(t) &= \FFF(\xx(t))+\epsilon \Lambda\mathbf{Z}(t), \\
		\dd \mathbf{z}_1(t) &= \left(\epsilon \mathbf{r}_1(\xx(t))+\mathrm{D}_1\mathbf{z}_1(t) \right)\dd t + \Sigma_1 
		\dd W^{(1)}_t, \\ & \vdots \nonumber \\ \label{eq: multilevel markovianization2 last level}	
		\dd \mathbf{z}_N(t) &= \left(\epsilon \mathbf{r}_N(\xx(t))+\mathrm{D}_N\mathbf{z}_N(t) \right)\dd t + \Sigma_N \dd W^{(N)}_t,
		\end{align}
	\end{subequations}
 where $W^{(j)}_t$ is a $d_1$-dimensional Wiener process. The vectors $\mathbf{r}_j(\xx)$ are given by:
	\begin{equation}
	\mathbf{r}_j(\xx)=\left[ \CCC^{\yy}(\xx)\cdot \gamma_{1,j}\bigg \langle \nabla\psi _1 (\yy)  \bigg \rangle,\ldots, \CCC^{\yy}(\xx)\cdot \gamma_{d_1,j}\bigg \langle \nabla\psi _1 (\yy)  \bigg \rangle \right]^{\top},
	\end{equation}
	where $\gamma_{i,j}$ are defined in terms of the parameters $(\alpha_{j}, \beta_j)$ introduced in Eqs.~(\ref{eq:alpha}, \ref{eq:beta}), respectively. Here we don't give the explicit expression of $\Lambda$, but its role is to provide suitable weights, in the spirit of Eq.~\eqref{Eq_lambda}, 
to the levels in Eq.~\eqref{eq: multilevel markovianization2} so that: (a) the correlation functions match those of the coupling function $\CCC^{\xx}$ in the uncoupled regime; and (b) the resulting term $\Lambda\mathbf{Z}(t)$ is real.
	
	The system Eq.~\eqref{eq: multilevel markovianization2} has the general structure one would obtain if the coupling function $\CCC^{\xx}$ projected along all the eigendirections in the point spectrum. This might not be true in general, but the drift matrix $\mathcal{D}$ can be rearranged so that only the relevant modes of variability be modeled --- following criteria (i) and (ii), as formulated in Remark \ref{Rem_interpretation} --- and still afford a reduction of the number of levels $N$. Notice that the $N$th level variables $\mathbf{z}_{N}$ described by Eq.~\eqref{eq: multilevel markovianization2 last level} decorrelate the fastest, %%compared to 
the rest, 
	since their exponential decorrelation rate is given by $|\mathfrak{Re}\lambda_N |\geq |\mathfrak{Re}\lambda_j|$, for all $j=1,\ldots,N-1$.

\end{remark}

The advantages of the Markovian system of Eqs.~\eqref{remarkovianisation1} and \eqref{eq: multilevel markovianization2} over the original WL equation~\eqref{lw parametrization independent} are twofold. First, we identify situations in which the WL equation can be Markovianized by introducing 
extended, hidden variables. This idea was already introduced in a preliminary application of the WL parametrization \cite{Wouters2016}, in which the authors resorted to a Markovian system to perform their simulations.  In fact, one of their examples is studied in the present framework; see  Sect.~\ref{preliminary}.

Second, memory equations contain nonlocal terms that are cumbersome and computationally expensive to integrate, as well as requiring much larger storage for the full history of the system's variables. The efficient Markovianization of evolution equations with memory terms is an active field of research in diverse areas of mathematics and the applied sciences; these areas include the study of bifurcations of delay differential equations \cite{CGLW16,CKL20}, the reduction of stochastic partial differential equations to stochastic invariant manifolds \cite{CLW15_vol1,CLW15_vol2}, and material sciences \cite{dafermos1970asymptotic}, among many others.

\subsection{Preliminary Example}\label{preliminary}

As seen earlier in Theorem~\ref{proposition: 1}, if the coupling function 
is resonant with the Koopman operator associated with the $\yy$-dynamics, 
one can identify the dominant exponential rates of decay of the memory term and the characteristic decorrelation time of the noise. As a consequence, one can Markovianize the parametrization and greatly facilitate the numerical integrations involved. 

To illustrate the above statement, we revisit here the preliminary application of the WL parametrization in the context of multiscale triads \cite{Wouters2016}. In that work, the authors implemented the parametrization for a collection of three-dimensional models that do exhibit time scale separation and compare here the corresponding outputs to those obtained via homogenization. The results are encouraging, since the parametrizations 
in \cite{Wouters2016} were obtained only from the decoupled hidden dynamics, in the lines of the present paper as well; see derivation of Eq.~\eqref{lw parametrization}.

One of the first multiscale triads studied in \cite{Wouters2016} is the following:
\begin{subequations}\label{eq:triad}
\begin{align} 
         \dot{x}(t) & = \epsilon B^{(0)}y_1y_2,  
	\label{eq:B0}\\
	\dot{y_1}(t) & = \epsilon B^{(1)}xy_2 - \gamma _1y_1 + \sigma _1\dd W^{(1)}_t,  \label{eq:B1}\\
	\dot{y_2}(t) & = \epsilon B^{(2)}xy_1 - \gamma _2y_2 + \sigma _2\dd W^{(2)}_t. \label{eq:B2}
\end{align}
\end{subequations}
Here we require that $\sum_jB^{(j)} = 0$, $\dd W^{(1)}_t$ and $\dd W^{(2)}_t$ are scalar Brownian increments and the parameter $\epsilon$ indicates both the time scale separation and the coupling strength. Notice that 
when the system is decoupled, i.e. when $\epsilon=0$, the fast dynamics 
evolve according to an Ornstein-Uhlenbeck (OU) process whose steady-state 
statistics are governed by Gaussian distributions with explicit mean and variance (see, e.g. \cite{pavliotisbook2014}). Hence, by virtue of the previous formulas or by following \cite{Wouters2016}, the WL parametrization yields the following scalar SIDE:
\begin{equation}\label{eq:scalar_memory}
	\dot{x}(t)=\epsilon \eta (t) + \epsilon ^2 \int _0^t\mathcal{K}(s,x(t-s))\dd s;
\end{equation}
here,
\begin{subequations}
\begin{align}
\left\langle \eta(t) \right\rangle & = 0, \\
\left\langle \eta(t+s)\eta(s) \right\rangle & = \left( B^{(0)}\right)^2e^{-(\gamma_1 + \gamma_2)t}\frac{\sigma_1^2}{2\gamma_1}\frac{\sigma_2^2}{2\gamma_2},\\ \label{eq: memory triad 1}
\mathcal{K}(s,x)&= \begin{bmatrix}
	x \\ x
\end{bmatrix}\cdot \bigg \langle \begin{bmatrix}
B^{(1)}y_2 \\B^{(2)} y_1
\end{bmatrix}: \nabla_{\yy} y_1(s)y_2(s) \bigg \rangle,
\end{align}
\end{subequations}
where the angular brackets refer to the ensemble averages according to the already mentioned Gaussian distributions arising from the decoupled model. Expanding these averages Eq.~\eqref{eq: memory triad 1} leads to:
\begin{subequations}
\begin{align}
\mathcal{K}(s,x)&= xe^{-(\gamma_1 + \gamma_2)}\bigg \langle B^{(1)}y^2_2 +  B^{(2)}y^2_1\bigg \rangle  \\ & = xB^{(0)}e^{-(\gamma_1 + \gamma_2)s}\left(B^{(1)}  \frac{\sigma_2^2}{2\gamma_2}+ B^{(2)} \frac{\sigma_1^2}{2\gamma_1}\right).
\end{align}
\end{subequations}

The time scales are indicated by the exponents in the formulas above and they are the same for the noise and the memory kernel. This equality suggests the possibility of Markovianizing the memory equation into the following two-dimensional system:
\begin{subequations}
\begin{align}
	\dot{z_1}(t)&=\epsilon B^{(0)}z_2, \\ 
	\dot{z_2}(t)&=-(\gamma_1 + \gamma_2)z_2 + \frac{\sigma_1\sigma_2}{2\gamma_1\gamma_2}    
	\{2(\gamma_1 + \gamma_2)\}^{1/2}
	\dd W_t +\epsilon \left( B^{(1)}\frac{\sigma_2^2}{2\gamma_2} + B^{(2)}\frac{\sigma_1^2}{2\gamma_1} \right)z_1.
\end{align}
\end{subequations}
Clearly, performing a numerical integration of this system is easier %less numerically difficult 
than for a memory equation like Eq.~\eqref{eq:scalar_memory}.

The results of Sec.~\ref{ssec:Koopman-eigen} allow us to carry out the dimension reduction of the multiscale triad by analyzing the spectral properties of the Koopman operator associated with the decoupled $\yy$-dynamics. Since the $\yy$-variables evolve stochastically, the Koopman operator becomes the \emph{backward-Kolmogorov} equation, which governs the evolution of the expectation values of the observables. Thus, for a generic observable $\Psi$ in the $\yy$ phase space, the evolution of its expectation 
value is given by:
\begin{equation}
	\partial _t \Psi (y_1,y_2)= \LLL_0^{\yy}\Psi(y_1,y_2) = \begin{bmatrix}
	-\gamma_1 y_1 \\ -\gamma_2y_2
	\end{bmatrix}\cdot \nabla \Psi (y_1,y_2) + \sigma^2_1\partial^2_{y_1}\Psi(y_1,y_2) + \sigma^2_2\partial^2_{y_2}\Psi(y_1,y_2).
\end{equation}
Now, let $\Psi(y_1,y_2)=y_1y_2$ be the coupling function of the triad system~\eqref{eq:triad}, for which we find that:
\begin{equation}
	\LLL^{\yy}_{0}\Psi(y_1,y_2) = -(\gamma_1 + \gamma_2)\Psi(y_1,y_2).
\end{equation}
The above equation is an eigenvalue problem, showing that this particular 
$\Psi$ is an eigenfunction of the Koopman operator associated with the eigenvalue $e^{-(\gamma_1 + \gamma_2)}$. This is no surprise, since $y_1$ and $y_2$ are respectively the Hermite polynomial eigenfunctions of the backward-Kolmogorov equation of the scalar OU process \cite{ornsteinuhlenbeck}. Hence, the product $y_1y_2$ is also an eigenfunction of the same equation for the joint process. Therefore, we can immediately re-Markovianize the parametrization according to Eqs.~(\ref{remarkovianisation1}), where $\mathrm{D}=\gamma_1 + \gamma_2$ and 
\begin{equation}
	\mathbf{R}(x(t))=\left( B^{(1)}\frac{\sigma_2^2}{2\gamma_2} + B^{(2)}\frac{\sigma_1^2}{2\gamma_1} \right)x(t).
\end{equation}

\section{Multilevel Stochastic Models and Empirical Model Reduction (EMR)}\label{MSMEMR}

\subsection{Multilevel Stochastic Models (MSMs)} \label{ssec:MSM}
MSMs are a general class of  SDEs that were introduced in \cite{kondrashovdata2015} and are, by their layered structure,  susceptible to provide a 
good approximation of the GLE \eqref{mori zwanzig 2} formulated by Mori and Zwanzig when a high-dimensional system is partially observed; see  \cite[Proposition 3.3 and Sec.~5]{kondrashovdata2015}. The MSM framework allows one to provide such approximations that are accompanied by useful dynamical properties, such as the existence of random attractors \cite[Theorem 3.1]{kondrashovdata2015}. The conditions on the high-dimensional system's coupling interactions between the resolved and hidden variables are also well understood \cite[Corollary 3.2]{kondrashovdata2015}. \footnote{Moreover, random attractors with fractal structures that survive highly degenerate noise \cite{Chekroun2011}---a situation that might occur when approximating deterministic chaotic dynamics by stochastic pathwise dynamics---can still be present in MSMs; see \cite[Sec.~7]{kondrashovdata2015}.}

As discussed in \cite{kondrashovdata2015}, MSMs arise in a variety of data-driven protocols for model reduction that typically use successive regressions from partial observations; see Sec.~\ref{Sec_EMR} below. The general form of an MSM is given by \cite[Eq.~({\bf MSM})]{kondrashovdata2015}; we only use herein its most basic version, which has the following structure:
\begin{subequations}\label{Eq_MSM}
\begin{align}
\dd \xx(t) & = \Big(\FFF (\xx(t)) + \epsilon \Pi \rr(t)\Big) \dd t, \label{msm1} \\
\dd \rr(t) & = \Big(\epsilon \mathrm{C}\xx(t) -\mathrm{D}\rr(t)\Big) \dd t + \Sigma \dd \W_t. \label{msm2}
\end{align}
\end{subequations}
Here the observed vector variable $\xx(t)$ lies in $\mathbb{R}^{d_1}$ and, for $\epsilon = 0$, the hidden variables $\rr(t) \in \mathbb{R}^{d_2}$ evolve in time independently. Otherwise, the dynamics of the $\xx$-variables is linearly coupled to that of the $\rr$-variables, which act upon \eqref{msm1} as a stochastic forcing, via the canonical projection $\Pi: \mathbb{R}^{d_2} \longrightarrow \mathbb{R}^{d_1}$, while $\W_t$ in \eqref{msm2} is a $d_1$-dimensional Wiener process. 

The matrix $\mathrm{C}$ in $\mathbb{R}^{d_2\times d_1}$ models the feedback of the $\xx$-process onto the $\rr$-variables. In the case of $\mathrm{C}\equiv 0$, $\rr$ whould evolve according to an OU process with drift matrix $\mathrm{D}$ and covariance matrix $\Sigma\Sigma ^{\ast}$. For the sake of simplicity, we restrict ourselves to the case $d_1=d_2$ so that 
the projection $\Pi$ reduces to the identity.

The more general MSM with nonlinear coupling considered in \cite{kondrashovdata2015} was shown to be equivalent to a SIDE with explicit expressions for the memory kernels and stochastic forcing being obtained; see \cite[Proposition 3.3]{kondrashovdata2015}. The noise term there results from successive convolutions of the homogeneous solutions of the lower levels of the system with an OU process. In particular, using the It\^o stochastic calculus, one readily obtains a SIDE that is equivalent to an MSM; see 
\cite[Sec.~3.2]{kondrashovdata2015} and Appendix \ref{ITO} below. 

We show next that the same SIDE can actually be obtained by using the operator formalism presented in Sec.~\ref{weak-coupling limit parametrization} above. One might object that an MSM is a stochastic system, due to the 
presence of white noise in the hidden layer, whereas the theory presented 
above applies to deterministic dynamics. However, as clarified below, the 
operator formalism applies equally well to the MSM case. 

In fact, given a smooth, $\CC^\infty$ observable
\be
\Phi :\mathbb{R}^{d_1}\times \mathbb{R}^{d_2} \rightarrow \mathbb{R}, \quad (\xx,\rr) \mapsto \Phi (\xx,\rr),
\ee 
its expected value along a stochastic trajectory $X_t=(\xx(t),\rr(t))^{\top}$ solving  Eq.~\eqref{Eq_MSM}, namely $\mathbb{E}(\Phi(X_t))$, defines a Markov semigroup $P_t$ by
\be
P_t \Phi =\mathbb{E}(\Phi(X_t)),
\ee
which solves the backward Kolmogorov equation \cite{chekroun2019c} associated with  Eq.~\eqref{Eq_MSM}:
\begin{equation}\label{backward Kolmogorov}
\partial _t \big (P_t\Phi \big) =\begin{bmatrix}\FFF(\xx)+\epsilon \rr\\ \epsilon \mathrm{C}\xx - \mathrm{D}\rr \end{bmatrix}\cdot \nabla P_t\Phi + \frac{1}{2}\begin{bmatrix} 0 \\  \Sigma\Sigma^{\ast}\nabla_{\rr}^2 P_t\Phi  \end{bmatrix};
\end{equation}
the only difference with respect to the transport equation \eqref{liouville equation} lies in the presence of a second-order differential operator 
induced by the white noise. 
%%%%%%%%%

We introduce the operators:
\begin{subequations} \label{eq:MSM_LLL}
\begin{align} 
&  \LLL_0=\begin{bmatrix} \FFF(\xx)\\ - \mathrm{D}\rr \end{bmatrix}\cdot \nabla +\begin{bmatrix} 0 \\     \Sigma\Sigma^{\ast}\nabla_{\rr}^2 \end{bmatrix}, \label{eq:LLL_0} \\
& \LLL_1=\begin{bmatrix} \rr\\ \mathrm{C}\xx \end{bmatrix}\cdot \nabla, 
\label{eq:LLL_1}
\end{align}
\end{subequations}
which play a role that is analogous to their deterministic relatives in Eq.~\eqref{eq:WL_LLL} of the previous section. Again, the operator $\LLL_1$ is viewed as a perturbation to the operator $\LLL_0$ due to the coupling. 

If one considers observables $\Phi=\Phi(\xx)$, Eq.~(\ref{backward Kolmogorov}) becomes at time $t=0$:
\begin{equation}
\partial _t \big (P_t\Phi \big)|_{t=0}=\begin{bmatrix}\FFF(\xx)+\epsilon \rr \end{bmatrix}\cdot \nabla _{\xx} \Phi,
\end{equation}
and we apply now, as in Sec.~\ref{weak-coupling limit parametrization}, the  Dyson perturbative expansion. By virtue of the formula \eqref{lw parametrization},
the parametrization leads to a reduced equation of the form:
\begin{equation}
\dot{\xx}(t) = \FFF(\xx(t)) + \epsilon \eta (t) + \epsilon^2 \int _0^t\mathcal{K}(s , \xx(t-s))\dd s,
\end{equation}
where the hidden variables in the decoupled regime are governed by an OU process with invariant measure $\mu_{\rr}$. The properties of the stochastic noise $\eta(t)$ are given by:
\begin{subequations}
\begin{align}\label{msm wl}
	\left\langle \eta(t)\eta^{\top}(0) \right\rangle  &= \int \dd \mu_{\rr}(\rr_0) e^{t\LLL_0} \rr_0\rr_0^{\top} \\&= \int \dd \mu_{\rr}(\rr_0) \mathbb{E}\left(\rr (t)|\rr_0 \right)\left(\mathbb{E}\left(\rr(0)|\rr_0 \right)\right)^{\top} \\
	&= \int \dd \mu_{\rr}(\rr_0) e^{-t\mathrm{D}} \rr_0\rr_0^{\top} 	
	\\
	&= \int \dd \mu_{\rr}(\rr_0) e^{-t\mathrm{D}} \rr_0\rr_0^{\top} = e^{-t\mathrm{D}}\Sigma\Sigma^{\ast},
\end{align}
\end{subequations}
where $\rr$ is a function analogous to the coupling function $\CCC^{\xx}_{\yy}$ in the previous section and the initial condition $\rr_0$ is assumed to be normally distributed with zero mean and variance $\Sigma\Sigma^{\ast}$. The memory kernel is given by:
\begin{subequations}
\begin{align}
	\mathcal{K}(s,\xx(t-s))&=\int \dd \mu_{\rr}(\rr_0) \mathrm{C}\xx(t-s)\cdot \nabla_{\rr_0}\mathbb{E}\left( \rr(s)|\rr_0 \right)\\&=\int \dd \mu_{\rr}(\rr_0) \mathrm{C}\xx(t-s)\cdot \nabla_{\rr_0}e^{-s\mathrm{D}}\rr_0
	\\&= \mathrm{C}\xx(t-s)\cdot e^{-s\mathrm{D}},
	\\&= e^{-s\mathrm{D}}\mathrm{C}\xx(t-s).
\end{align}
\end{subequations}
Using the intermediate steps above, the explicit parametrization becomes, 
finally:
\begin{equation}
\dot{\xx}(t) = \FFF(\xx(t)) + \epsilon \eta (t) + \epsilon^2 \int _0^te^{-s\mathrm{D}}\mathrm{C}\xx(t-s)\dd s.
\end{equation}

The integro-differential equation above is the same as Eq.~\eqref{eq:ITO} 
one obtains using the It\^o integration described in Appendix \ref{ITO}. This similarity of results occurs because we are considering the case of additive coupling, and the Dyson expansion can be truncated after the memory term proportional to $\epsilon^2$, cf. Sect.~\ref{additive}.

\subsection{Empirical Model Reduction (EMR)}\label{Sec_EMR}

As discussed in Sects.~\ref{INTRO} and \ref{weak-coupling limit parametrization}, and illustrated in Fig.~\ref{schematic}, the evolution of the resolved variables is forced by fluctuating terms and the effects of the previous state of the system. It is desirable, therefore, to construct a full model of the system even when only capable to partially observe it. The EMR methodology \cite{kondrashovdata2015, Kondrashov.Kravtsov.ea.2006, Kravtsov2005, KKG.2010} aims at achieving this goal; we discuss it below in the broader context of MSMs. Note that EMR provides a solution for the 
dynamical closure of partially observed systems and thus it differs from the methodology recently proposed in ref.~\cite{Brunton2016}, which requires one to fully observe the system for the data-driven discovery of 
its underlying equations to work.  %% of the system is obtained from fully observing it.

Having a set of reduced $d_1$-dimensional observations $\{\xx_i: i = 1, 
\ldots, n\}$ every $\dd t$ time units, one seeks to regress the tendencies $\{\dd \xx_i: i = 1, \ldots, n\}$ of the data onto a quadratic function of the form:
\begin{equation}\label{eq: EMR_vector_field}
	\FFF(\mathbf{x})=\mathbf{f} + \mathbf{b}\cdot \mathbf{x} +\mathbf{Q}(\xx),
\end{equation}
 where $\mathbf{b}$ in $\mathbb{R}^{d_1\times d_1}$ describes dissipative 
processes and $\mathbf{Q}$ is a quadratic form describing self-interaction between the $\xx$ variables. The $i$th component of the quadratic form is given by:
 \begin{equation}
 [\mathbf{Q}]_i=\xx^{\top}\mathbf{A}_i\xx,
 \end{equation}
where $\mathbf{A}_i$ in $\mathbb{R}^{d_1 \times d_1}$. The function $\FFF$ is expected to approximate the vector field driving the dynamics in the 
absence of hidden external influences. Of course, performing regressions yields an error called \emph{residual} $\{\rr_i: i = 1, \ldots, n\}$. Hence, the evolution of the $\xx$-variables satisfies the equation:
 \begin{equation}
 	\frac{\dd \xx}{\dd t}  = \FFF(\xx) + \rr.
 \end{equation}
 
At this point, one can study the properties of the residual time series $\{\rr_i\}_{i=1}^{n}$ and construct a model able to reproduce its main statistical features. However, we know that if it is possible to sample all the variables of the dynamical system of interest, one expects that the 
residuals are explained by the errors committed exclusively in the regression algorithm. If some sort of subsampling is done, whether spatial or temporal, the residuals are due also to the delayed influence of unresolved processes that are involved in the coupling. 

Allowing for the main level variables $\xx$ to be linearly coupled with the residual $\rr$, we are creating a model that is able to incorporate memory effects as well. Hence, for each component $i$ of $\xx$ we have:
\begin{subequations}\label{eq:EMR}
\begin{align}
	& \frac{\dd  [\mathbf{x}]_i }{\dd t} = [\mathbf{f}]_i + \mathbf{b}_i^{(0)}\cdot \mathbf{x}+ \mathbf{x}^{\top}\mathbf{A}_i\mathbf{x}  + [\rr^{(0)}]_i, \label{EMR_0}\\
	& \frac{\dd   [\rr^{(0)}]_i}{\dd t}  = \mathbf{b}_i^{(1)}\cdot [\mathbf{x},\mathbf{r}^{(0)}] + [\rr^{(1)}]_i, \label{EMR_1}\\
	& \frac{\dd   [\rr^{(1)}]_i }{\dd t} = \mathbf{b}_i^{(2)}\cdot [\mathbf{x},\mathbf{r}^{(0)},\mathbf{r}^{(1)}] + 
		[\rr^{(2)}]_i, \label{EMR_2}\\
       & \qquad \vdots  \label{EMR_k} \\
	& \frac{\dd   [\rr^{(l)}]_i}{\dd t} = \mathbf{b}_i^{(l+1)}\cdot [\mathbf{x},\mathbf{r}^{(0)},\mathbf{r}^{(1)},\ldots, ,\mathbf{r}^{(l)}] 	+ [\rr^{(l+1)}]_i, \label{EMR_l}
\end{align}
\end{subequations}
where we have introduced new matrices $\mathbf{b}^{(j)}\in \mathbb{R}^{d_1 \times (j+2)d_1}$ that model the linear coupling.  The residual at the last level $[\rr^{(l+1)}]_i$ is assumed to obey a Wiener process for which the correlation matrix is obtained from the last residual time series. The choice of stochastic process in the last step can only be done if the 
decorrelation of $[\rr^{(l+1)}]_i$ is sufficiently fast according to the time scale set by $\dd t$. This motivates the problem of choosing the optimal number $l$ of levels.

Several criteria have been established to determine the optimal number of 
levels $l$.
The basic idea is that the resulting $(l + 1)$-residual in Eq.~\eqref{EMR_l} should be well approximated by Gaussian white noise \cite{kondrashovdata2015, KKG.2010}. One has, therefore, to test whether the residual variables decorrelate at lag $\dd t$ and whether the lag-0 covariance matrix is invariant in the last levels. 

Therefore, regression on the tendency of the optimal level $\rr^{(l)}$ should yield:
\begin{equation}
	\rr^{(l+1)}-\rr^{(l)} \simeq -\rr^{(l)} + \gamma ^{(l)},
\end{equation}
where $\gamma^{(l)}$ is the residual of the previous regression and is approximately equal to $\rr^{(l+1)}$. Hence, $\gamma^{(l)}$ would become a lagged version of $\rr^{(l+1)}$. Subject to this assumption, it is possible to estimate the optimal value of the coefficient of determination $R^2$:
\begin{equation}\label{emrdiagnostic}
	R^2=1-\frac{\sum _{k}\gamma_l^2}{\sum_{k}\left(\rr^{(l+1)} - \rr^{(l)} 
\right)^2} \simeq 1 - \frac{\sum _l{\rr^{(l+1)}}^2}{\sum _l {\rr^{(l+1)}}^2 + {\rr^{(l)}}^2} \simeq 0.5.
\end{equation}
This means that, when the amount of unexplained variance of the last regression is $50\%$, one has reached the optimal number of levels. 
It is worth stressing that the empirical model~\eqref{eq:EMR} has the structure \eqref{Eq_MSM} of an MSM, as discussed in \cite{kondrashovdata2015}. It can, therewith, be integrated to transform it into an integro-differential equation with explicit formulas for the fluctuating noise and memory kernel, cf.~\cite[Proposition 3.3]{kondrashovdata2015}; see also Sec.~\ref{ssec:MSM} for such a transformation from another perspective.

 Finally, note that the aforementioned stopping criterion for EMR --- namely $R^2\simeq 0.5$, see \cite[Appendix A]{kondrashovdata2015}) --- is based on decorrelation times and it is also present in the multilevel WL equation \eqref{eq: multilevel markovianization2}. We noted, in fact, in Remark~\ref{Rem_genedim} of Sect.~\ref{ssec:Koopman-eigen} that the last level modeled by Eq.~\eqref{eq: multilevel markovianization2 last level} decorrelates the fastest with respect to the rest. Ultimately, making this points amounts to saying that a low number of levels is expected to arise 
in the EMR method, provided most of the eigenvalues $\lambda_j$ in Theorem~\ref{proposition: 1} are located far away from the imaginary axis, except for a very few of them. %% see also Remark~\ref{Rem_genedim}. 
 Conversely, if the Koopman eigenvalues cluster near the imaginary axis or do not exhibit a spectral gap located at a, suitably defined, small negative real part, many levels are expected to be needed to capture the hidden dynamics; see again Remark \ref{Rem_genedim}.
%%%%The small paragraph that was below on March 22 has been moved up, right after the equation%%%%%%

\section{Numerical Experiments} \label{RESULTS}

In Sects.~\ref{weak-coupling limit parametrization} and \ref{MSMEMR}, we have shown that both the WL top-down approach and the EMR data-driven method yield a set of multilevel equations for the variables of interest in a multi-scale system. 
In particular, both approaches give explicit formulas for the fluctuation 
term and memory kernel in the GLE~\eqref{mori zwanzig 2} of the Mori-Zwanzig formalism. Furthermore, their Markovian representation share that the 
hidden layers are linearly driven with a white noise background (see Eq.~\ref{remarkovianisation1} and Eq.~\eqref{Eq_MSM}). We now compare the two 
approaches to model reduction in a simple, conceptual stochastic climate model. 

Since the  modeling of geophysical flows is the primary motivation for this research, we consider a set of SDEs proposed in \cite[among others]{Franzke2007} as a physically consistent climate ``toy'' model. In such a model, the main $\xx$-variables are slow and weakly coupled to the fast $\yy$-variables. The latter correspond to weather fluctuations and carry, in 
fact, most of the system's variance. The model's governing equations are:
\begin{subequations} \label{stochastic model 1}
\begin{align}
	\dd x_1 & = \left\lbrace -x_2\left( L_{12} + a_1x_1 + a_2x_2\right) -d_1x_1 +F_1 +\epsilon\left(L_{13}y_1+c_{134}y_1y_2\right)\right\rbrace \dd 
t, 	\label{stochastic model 2} \\
	\dd x_2 & =  \left\lbrace x_1\left( L_{21} + a_1x_1 + a_2x_2 \right) -d_2x_2 +F_2 + 
	\epsilon L_{24}y_2  \right\rbrace \dd t, \label{stochastic model 3} \\
	\dd y_1 & = \left \lbrace \epsilon \left(-L_{13}x_1 + c_{341}y_2x_1\right) + F_3 - \frac{\gamma_1}{h}y_1  \right\rbrace \dd t +\frac{\sigma _1}{\sqrt{h}}\dd W^{(1)}_t, \label{stochastic model 4} \\
	\dd y_2 & = \left\lbrace -\epsilon \left(L_{24}x_2 + c_{413}y_1x_2\right) +F_4 - \frac{\gamma_2}{h}y_2 \right \rbrace \dd t + \frac{\sigma_2}{\sqrt{h}}\dd W^{(2)}_t,\label{stochastic model 5}
\end{align}
\end{subequations}
where $W^{(1)}_t$ and $W^{(2)}_t$ are two independent Wiener processes.

These equations describe the evolution of four real variables $\xx=(x_1,x_2)$ and $\yy=(y_1,y_2)$; their time scale separation is determined by the parameter $h$ and the coupling strength is controlled by $\epsilon$. The parameter values used herein are: $c_{134}=c_{341}=0.25$, $c_{413}=-0.5$, $L_{12}=L_{21}=1$, $L_{24}=-L_{13}=1$, $a_1=-a_2=1$, $d_1=0.2$, $d_2=0.1$, $F_1=-0.25$, $F_2=F_3=F_4=0$, $\gamma _1 = 2 $, $\gamma _2 = 1$ and $\sigma _1 = \sigma _2 = 1$. The time scale separation and the coupling strength are $h=0.1$ and $\epsilon= 0.5$, respectively.

\subsection{WL Approximation}

Notice that the hidden variables evolve according to a decoupled OU process. Taking advantage of this fact, we calculate the weak-coupling--limit parametrization of the model, according to the formulas presented in Sect.~\ref{weak-coupling limit parametrization} for the separable coupling functions given by:
\begin{subequations} \label{eq:separate}
\begin{align}
\CCC ^{\xx}(\xx , \yy)=\CCC _{\yy}^{\xx}(\yy) &= \begin{bmatrix}
L_{13}y_1 + c_{134}y_1y_2, \\ 
L_{24}y_2 
\end{bmatrix}, \label{eq:slow} \\ 
\CCC ^{\yy}(\xx , \yy)=\CCC ^{\yy}_{\xx}(\xx ):\CCC _{\yy}^{\xx}(\yy) &= \begin{bmatrix}
-L_{13}x_1 + c_{341}y_2x_1 \\ 
-L_{24}x_2 + c_{413}y_1x_2
\end{bmatrix}. \label{eq:fast}	
\end{align}
\end{subequations}
The coupling function $\CCC ^{\xx}$ in the slow equation~\eqref{eq:slow} is independent of the $\xx$-variables, indicating that the noise correction can be additively incorporated and implemented by examining the decoupled hidden process. Note that the functional form of  $\CCC ^{\yy}$ implies that the WL parametrization cannot be exact in $\epsilon$, as noted in 
Eqs.~\eqref{time zero liouville} and \eqref{dysonapprox}. This indicates that the WL reduced model will not only introduce an error in averaging over the decoupled steady state, but also that the Dyson expansion Eq.~\eqref{pertexp} has to be truncated at $\epsilon ^{3}$, rather than merely at $\epsilon ^{2}$ where no memory effects would be included.

According to the WL parametrization discussed in Sect.~\ref{weak-coupling 
limit parametrization}, the fluctuation terms correspond to the decoupled 
evolution of the coupling function $\CCC ^{\xx}_{\yy}$, concretely as in Eq.~\eqref{eq:moments}. This allows to directly compute the correlation function:
\begin{align}\label{autocorrelation function wl}
	\bigg \langle  \CCC ^{\xx}_{\yy}(\yy)\CCC ^{\xx}_{\yy}(\yy (t))^{\top} \bigg \rangle = \begin{bmatrix} L_{13}^2e^{-(\gamma_1/h)t}\frac{\sigma_1^2}{2\gamma_1} + c_{134}^2e^{-(\gamma_1+\gamma_2)t/h}\frac{\sigma_1^2\sigma_2^2}{2\gamma _1 \gamma_2}& 0 \\ 0 & L_{24}^2e^{-(\gamma_2/h)t}\end{bmatrix},
\end{align}
From Eq.~\eqref{autocorrelation function wl} we deduce that the noise covariance matrix for the given parameter values is given by:
\begin{equation}\label{covariance matrix wl}
	\bigg \langle  \CCC ^{\xx}_{\yy}(\yy)\CCC ^{\xx}_{\yy}(\yy)^{\top} \bigg 
\rangle = \begin{bmatrix}
	0.2578... & 0 \\ 0 & 0.5 
	\end{bmatrix}
\end{equation}

The memory kernel $\mathcal{K}$, which is a vector of two components $(\kappa _1, \kappa _2)^{\top}$, is given by:
\begin{align}
\mathcal{K}(s,\xx)=\begin{bmatrix} \kappa _1(s,\xx) \\ \kappa_2(s,\xx) \end{bmatrix} = \bigg \langle  \CCC^{\yy}(\xx , \yy)\cdot \nabla_{\yy} \CCC^{\xx}(\xx (s),\yy (s)) \bigg \rangle; 
\end{align}
here the brackets $\left \langle \cdot  \right \rangle$ indicate the averages for the uncoupled equilibrium in the $\yy$-variables, which happen to be a set of independent OU processes. Explicitly,
\begin{subequations} %% \label{eq:separate}
\begin{align}
\kappa_1(s,\xx) =&  \bigg \langle \left( -L_{13}x_1+c_{341}y_2x_1 \right)\partial _{y_1}\left( L_{13}y_1(s) + c_{134}y_1(s)y_2(s) \right) \bigg  
\rangle \\ 
&-\bigg \langle \left( L_{24}x_2 + c_{413}y_1x_2 \right) \partial _{y_2}\left( L_{13}y_1(s) + c_{134}y_1(s)y_2(s) \right) \bigg  \rangle \\ =& -L_{13}^2e^{-(\gamma _1/h)s}x_1 + c_{341}c_{134}e^{-\left(\gamma_1 + \gamma_2\right)s/h}\frac{\sigma _2^2}{2\gamma_2}x_1+ c_{134}c_{341}e^{-\left(\gamma_1 + \gamma_2\right)s/h}\frac{\sigma _1^2}{2\gamma_1}x_1, \\
%% and \begin{align}
\kappa_2(s,\xx) =&  \bigg \langle \left( -L_{13}x_1+c_{341}y_2x_1 \right)\partial _{y_1}\left( L_{24}y_2(s) \right) \bigg  \rangle
-\bigg \langle L_{24}x_2 \partial _{y_2}\left( L_{24}y_2(s) \right) \bigg 
 \rangle \\ =& -L_{24}^2e^{-\left(\gamma _2/h\right)s}x_2.
\end{align}
\end{subequations} 

The reduced-order model obtained herewith does give explicit formulas for 
the evaluation of the stochastic noise and the memory kernel, independently of the time scale separation $h$, although these formulas are rather complicated. Still, the scheme remains the same when changing parameter values, so it is flexible in studying different scenarios.

\subsection{EMR Model and Results}

\paragraph{Basic EMR algorithm implementation.}
Regarding the data-driven EMR protocol, we integrated the full model with 
a time step of $\dd_{\ell} t = 10^{-3}$ time units for a duration of $T_{\ell} = 10^{4}$ time units in order to learn the model parameters. Then, a separate run was performed in order to examine the ability of the inferred model to reproduce the general statistical features. This time, the EMR system was integrated together with the full model using a time step of $\dd_{\tau} t = 10^{-2}$ time units for a total of $T_{\tau} = 10^5$ time units. The equations were solved using a fourth-order Runge-Kutta and a Euler-Maruyama method for the deterministic and stochastic components, respectively. 

By sampling every time step, we learned an EMR model whose coefficients were explicitly found. The convergence criterion $R^2 \simeq 0.5$ was attained by adding two extra levels., for a total of three. The convergence was not affected by changes in the time scale separation parameter --- namely, $h=0.1$ and $h=1$ in the case at hand. Probably the value of $h$ 
was not that important here because of the low dimensionality and stochastic nature of the hidden process. However, convergence is likely to be altered in more complicated models, as illustrated in Appendix~\ref{L84L63}.

\begin{figure}[H]
	\centering
	\begin{subfigure}[b]{0.32\textwidth}
		\centering
		\includegraphics[width=\textwidth]{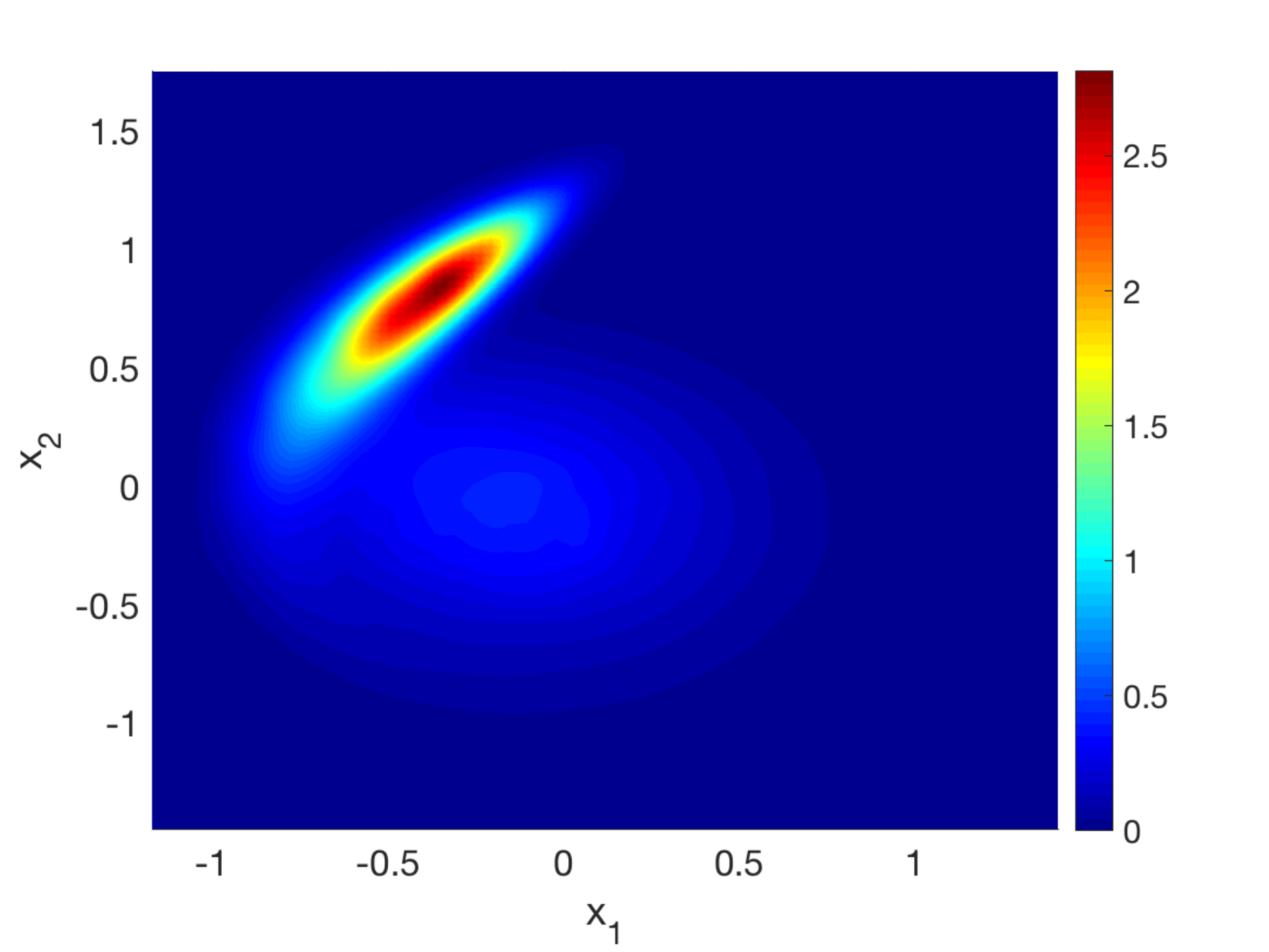}
		\caption{\label{climatologies 1 a}Full model}
	\end{subfigure}
	\hfill
	\begin{subfigure}[b]{0.32\textwidth}
		\centering
		\includegraphics[width=\textwidth]{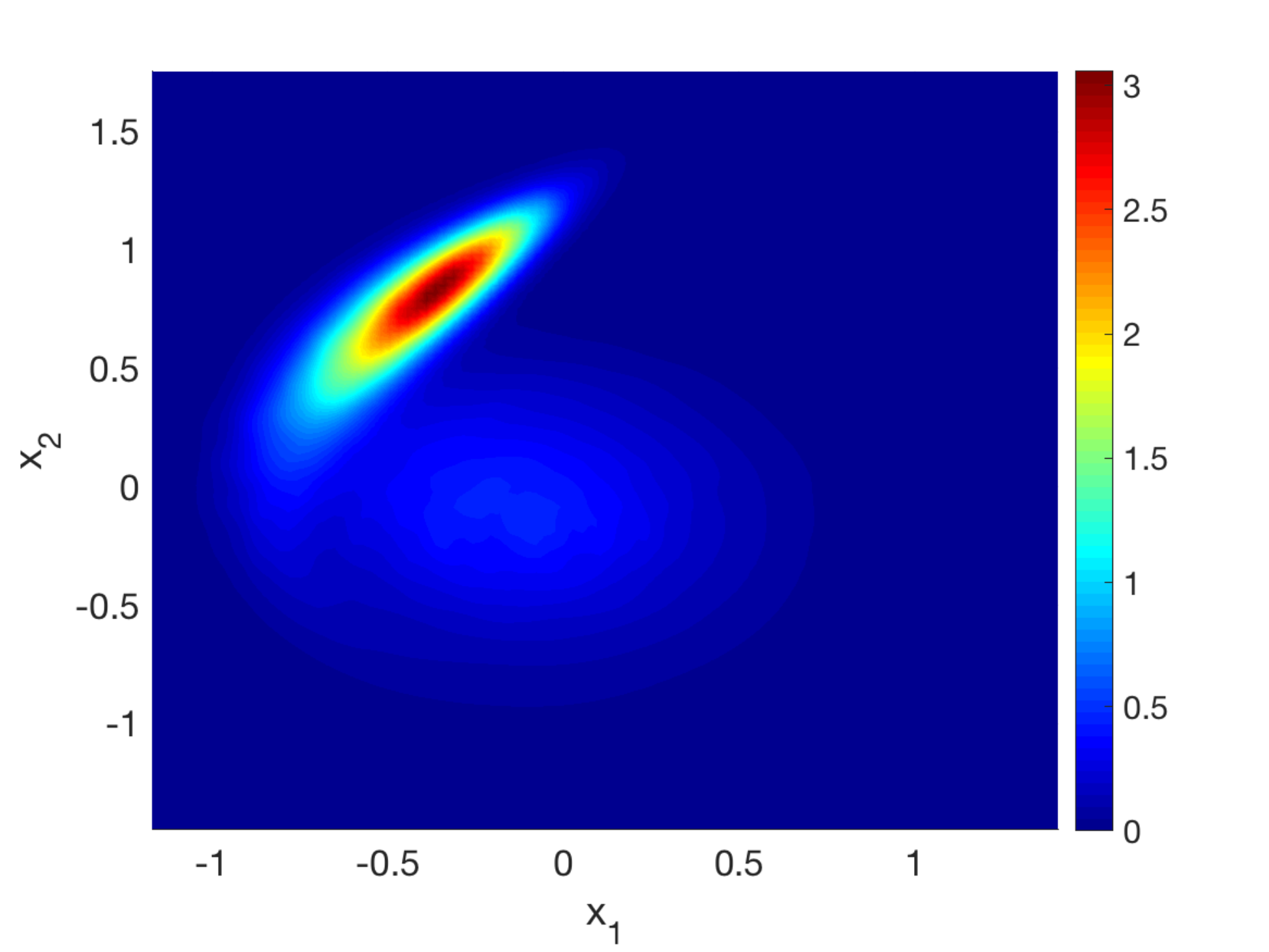}
		\caption{\label{climatologies 1 b}EMR model}
	\end{subfigure}
	\hfill
	\begin{subfigure}[b]{0.32\textwidth}
		\centering
		\includegraphics[width=\textwidth]{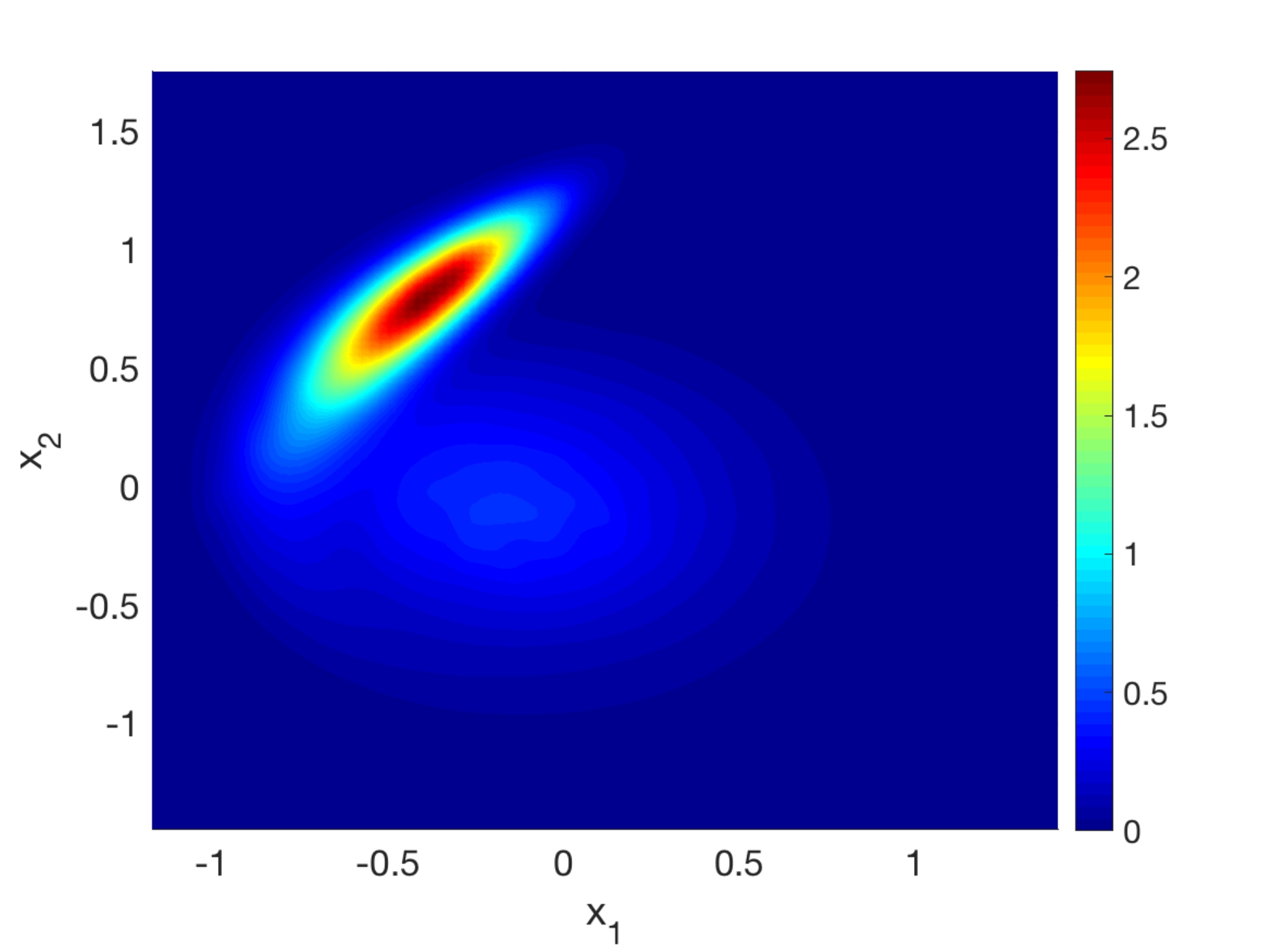}
		\caption{\label{climatologies 1 c}WL parametrization}
	\end{subfigure}
	\caption{\label{climatologies 1}Two-dimensional probability density functions (PDFs) of the stochastic
	model~\eqref{stochastic model 1} in the $(x_1, x_2)$-plane, as obtained with: (a) the full integration; (b) an integration of the EMR model; and (c) the WL parametrization. The time scale separation parameter used is $h = 0.1$. The PDFs shown here and in Fig.~\ref{histograms 2d epsilon 1} 
were obtained by using the Matlab R2019a kernel smoothing function \textit{ksdensity}.
}
\end{figure}

\begin{figure}[H]
	\centering
	\begin{subfigure}[b]{0.49\textwidth}
		\centering
		\includegraphics[width=\textwidth]{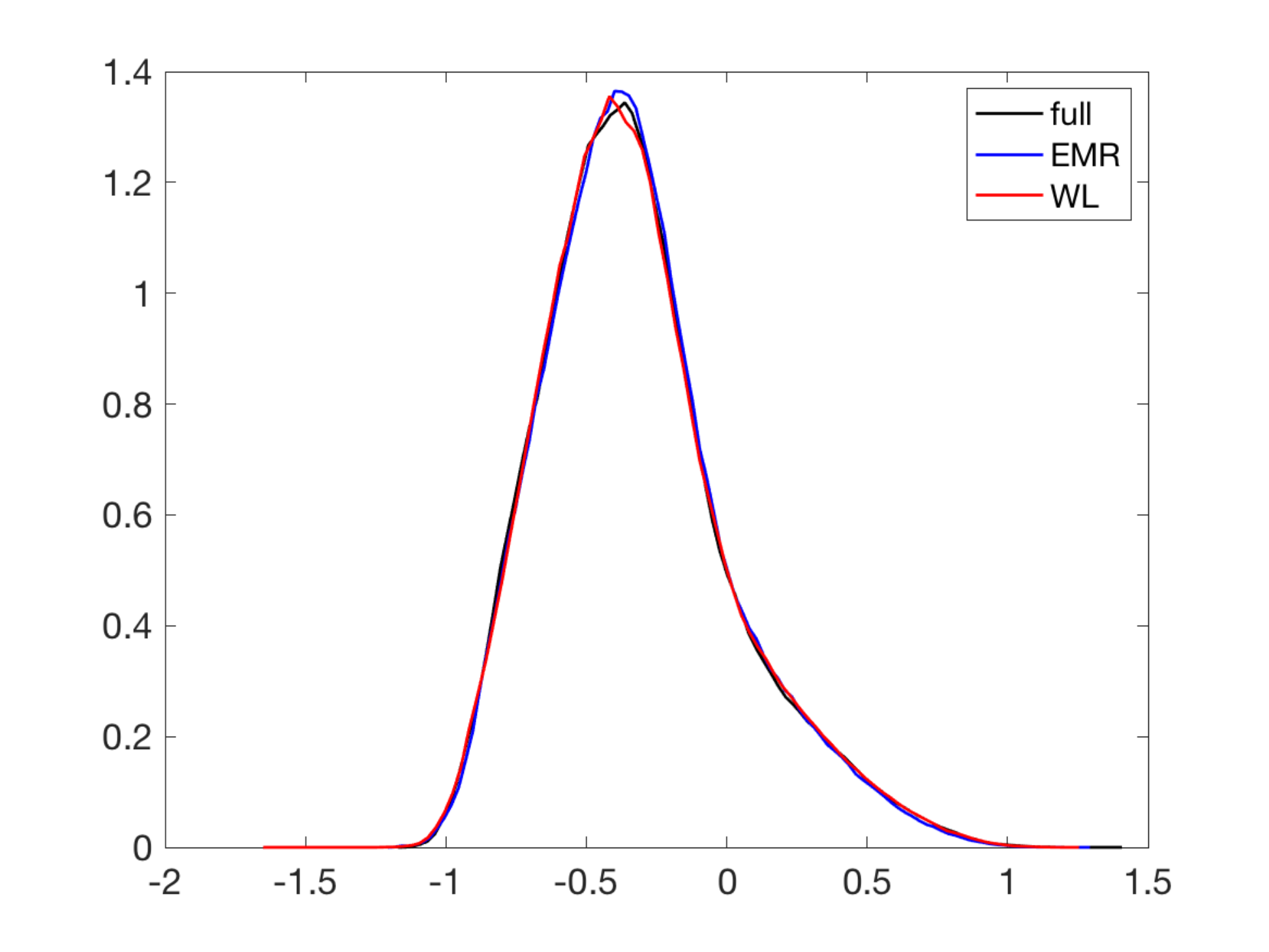}
		\caption{$x_1$-PDF}
	\end{subfigure}
	\begin{subfigure}[b]{0.49\textwidth}
		\centering
		\includegraphics[width=\textwidth]{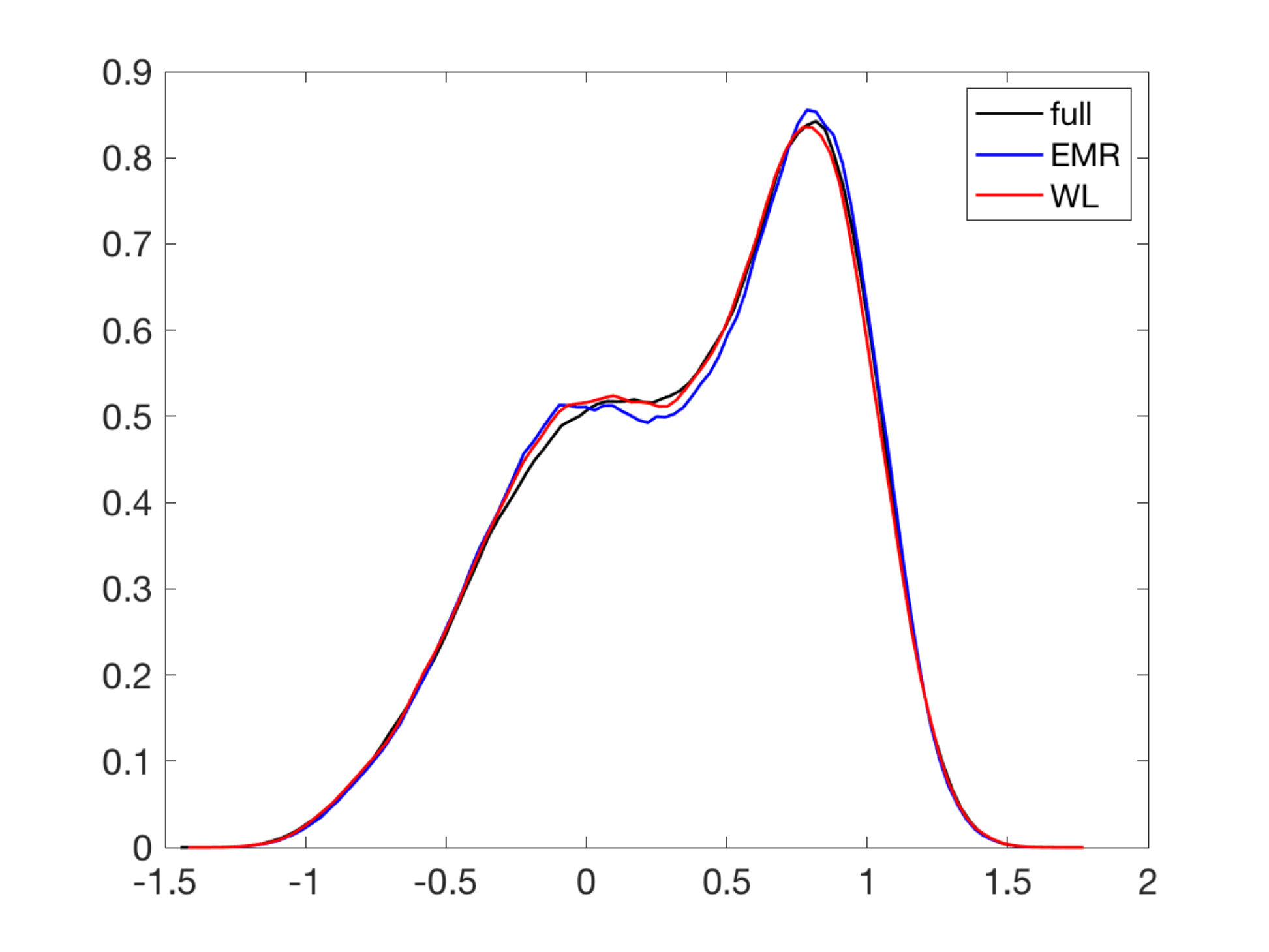}
		\caption{$x_2$-PDF}

	\end{subfigure}
	\caption{\label{marginals 1}PDFs of  (a) the $x_1$ variable; and (b) the 
$x_2$ variable.
	The separation parameter is $h = 0.1$ and colors used for each method are indicated by the legends inside 
	the panels.
	}
\end{figure}

The climatologies of the slow $\xx$-variables are obtained using data from the full model, the EMR model and the WL parametrization. The two-dimensional probability density functions (PDFs) of the stochastic model~\eqref{stochastic model 1} in the $(x_1, x_2)$-plane are shown in Fig.~\ref{climatologies 1}. These PDFs were calculated by employing the Matlab R2019a kernel smoothing function \textit{ksdensity}. Their respective marginals are shown in Fig.~\ref{marginals 1}. The agreement between the two methodologies when approximating the clearly non-Gaussian density arising 
from the full model is clearly excellent.

\begin{figure}[H]
	\centering
	\begin{subfigure}[b]{0.49\textwidth}
		\centering
		\includegraphics[width=\textwidth]{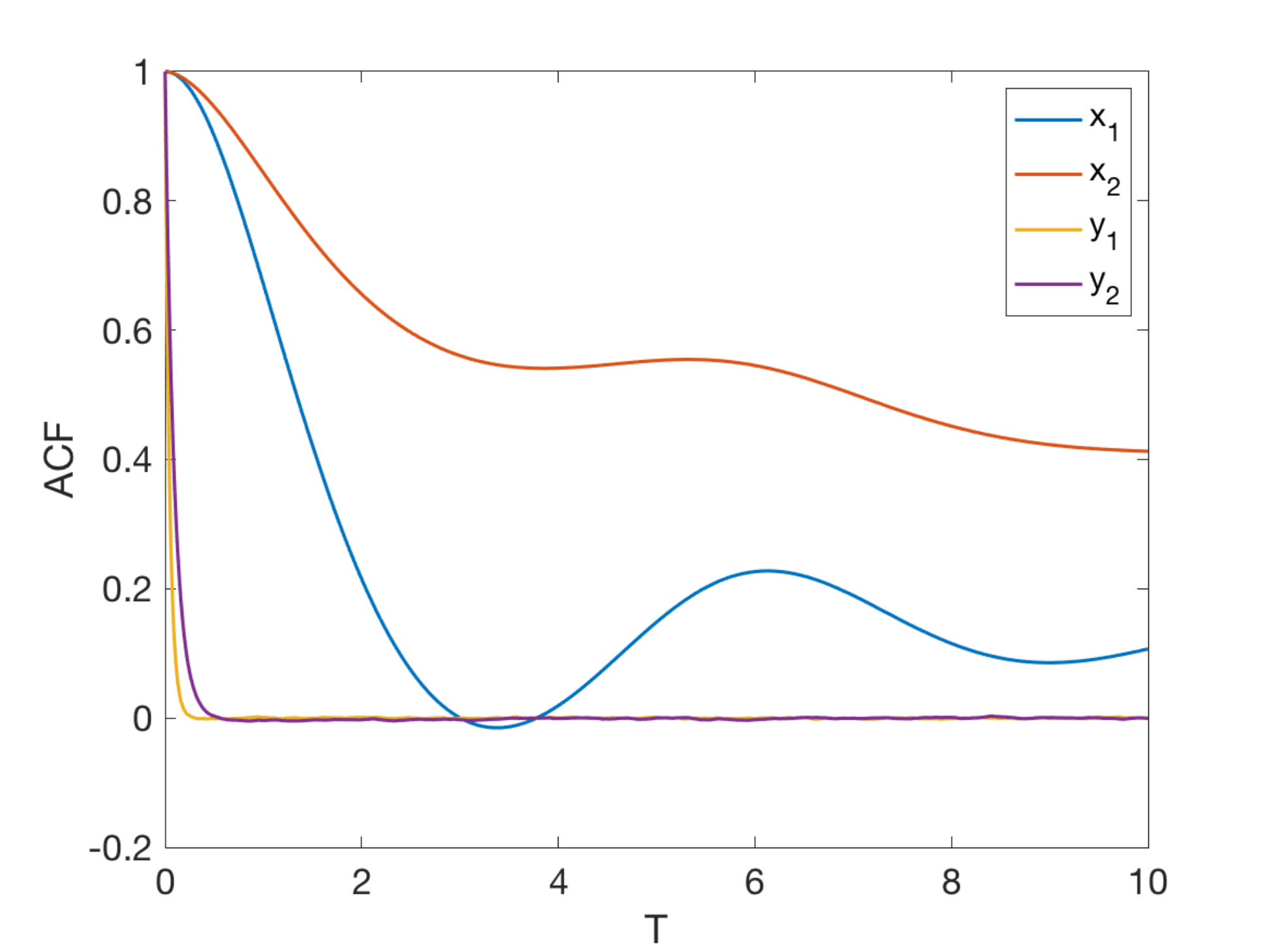}
		\caption{}
	\end{subfigure}
	\begin{subfigure}[b]{0.49\textwidth}
		\centering
		\includegraphics[width=\textwidth]{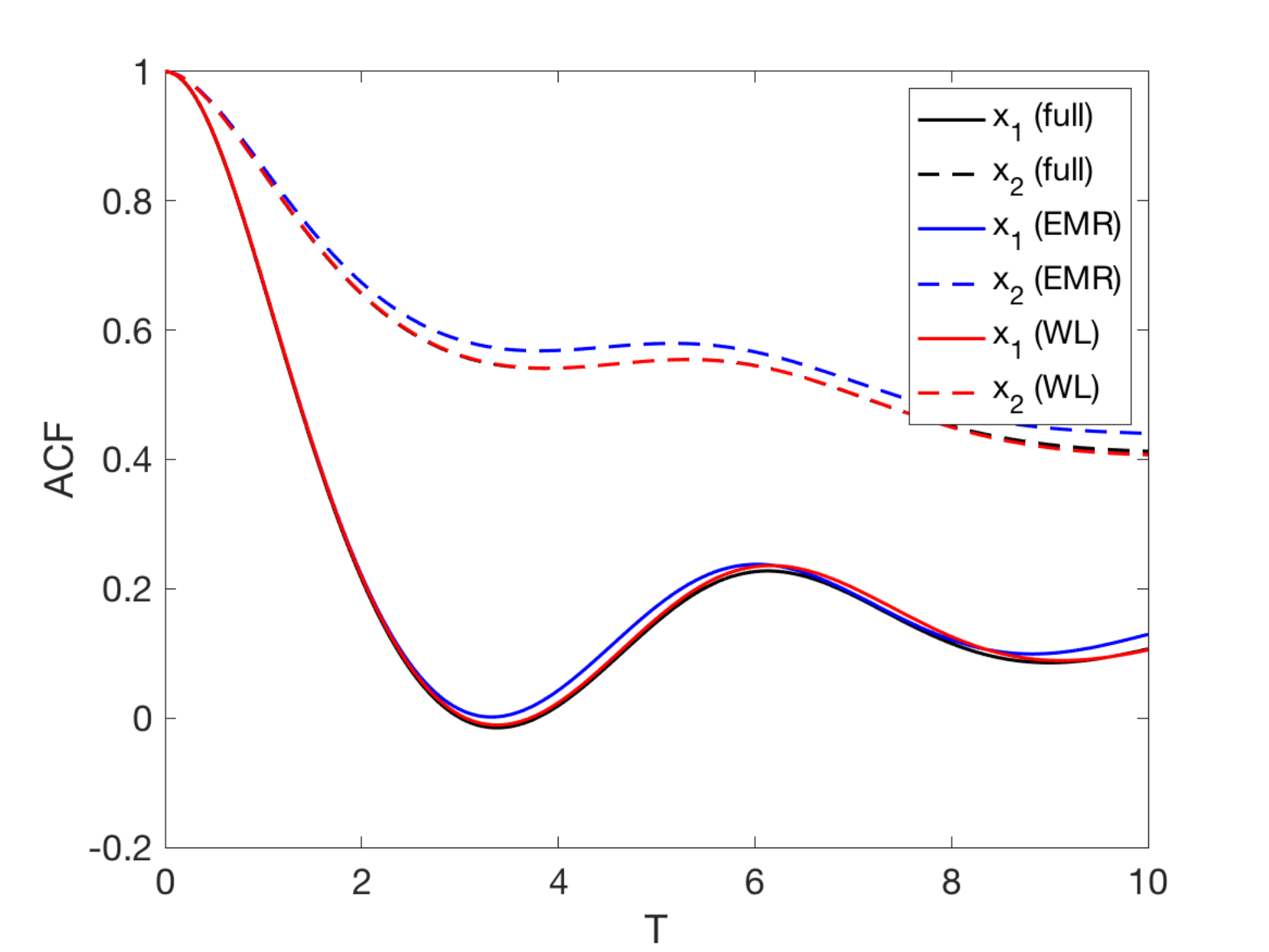}
		\caption{}
	\end{subfigure}
	\caption{\label{correlations 1}Autocorrelation functions for the four variables 
	 $x_1, x_2, y_1, y_2$ obtained (a) from the full model; %% (left); %% and the comparison of those obtained with
	and (b) the comparison of the corresponding results for $x_1, x_2$ with the full model, 
	the EMR model, and the WL parametrization. See legend for the choice of lines; $h = 0.1$.}
\end{figure}

The time scale separation between the $\xx$-variables and the $\yy$-variables is clearly depicted in the left panel of Fig.~\ref{correlations 1}, where the fast variables decorrelate almost instantly compared to the slow ones. The approximation of these autocorrelation functions is also obtained using the EMR and WL methods.

In general, cf.~\cite{Kravtsov2005}, the regressions performed in the main level~\eqref{EMR_0} of the EMR model allow one to effectively reconstruct the coefficients of a weakly coupled model; see Appendix~\ref{L84L63}. 
The EMR methodology, though, only allows for linear coupling between the slow $\xx$'s and the fast $\yy$'s. The nonlinear coupling between the slow and fast variables in system~\eqref{stochastic model 1} compromises the 
estimation of the main model parameters in Eq.~\eqref{EMR_0}, so that we cannot expect to recover the original, full model's behavior given by \eqref{stochastic model 1}. The EMR model coefficients at the first and second levels are as shown on tables \ref{tab:h1_row1} and \ref{tab:h1_row2}, respectively.

\begin{table}[H]\renewcommand{\arraystretch}{2}
\centering
\caption{Empirically estimated EMR model coefficients at the first level,
              Eq.~\eqref{EMR_0}, for $h = 0.1$. First column gives the coefficients for the constant forcing $\mathbf{f}^{(0)}$, the second and third columns indicate the linear component of the vector field $\mathbf{b}^{(0)}$ and the last three columns determine the quadratic form $\mathbf{A}$.}
\begin{tabular}{cccccc}
	\hline 
	$f$ & $x_1$ & $x_2$ & $x_1^2$ & $x_1x_2$ & $x_2^2$ \\ 
	\hline 
	-0.31404 & -0.50954 & -0.065313 & 0 & -1.0092 & 0.99704 \\ 
	-0.15356 & 0.12353 & 0.21979 & 1.0092 & -0.99704 & 0 \\ 
	\hline 
\end{tabular} \label{tab:h1_row1} 
\end{table}

\begin{table}[H]\renewcommand{\arraystretch}{2}
	\centering
	\caption{Empirically estimated EMR model coefficients at the second 
	level, Eq.~\eqref{EMR_1}, for $h = 0.1$. First column gives the coefficients for the constant forcing $\mathbf{f}^{(1)}$, the next two columns indicate the linear coupling to the main level (i.e. the first two columns of $\mathbf{b}^{(1)}$) and the last two columns determine the linear drift for the second level (i.e. the last two columns of $\mathbf{b}^{(1)}$).}
\begin{tabular}{ccccc}
	\hline 
	$f^{(1)}$ & $x_1$ & $x_2$ & $r^{(1)}_1$ & $r^{(1)}_2$  \\ 
	\hline 
0 & -5.6397e-4 & -6.9382e-05 & -20.2823 & 0.016165  \\ 
0 & -1.648e-4 & -4.72e-4 & 0.086621 & -10.0426  \\ 
	\hline 
\end{tabular} \label{tab:h1_row2} 
\end{table}

As discussed in Sect.~\ref{Sec_EMR}, the EMR has the structure of an MSM and it can be recast into an integro-differential equation. If one only considers the first added level, the EMR can be readily integrated giving the following equation for the evolution of the slow variables $\xx = (x_1,x_2)$:
\begin{equation}\label{intdiff equation}
\dot{\xx}(t) = \FFF (\xx(t)) + e^{-\mathrm{D}t}\yy (0) + \int _{0}^te^{-\mathrm{D}(t-s)} \Sigma \dd \mathbf{W}_s + \int _{0}^{t}e^{-\mathrm{D}(t-s)}\mathrm{C}\xx(s)\dd s.
\end{equation}
Here $\mathbf{W}_s$ is an independent two-dimensional Wiener process and
\begin{subequations} %% \label{eq:separate}
\begin{align}
	&\mathrm{D}=\begin{bmatrix}
	-19.9982 & 2.1122\cdot 10^{-3}  \\ 
	-0.77528 & -10.116
	\end{bmatrix}, \mathrm{C}=\begin{bmatrix}
	-5\cdot 10^{-3} & -5\cdot 10^{-4}\\
	-1\cdot 10^{-3} & -5\cdot 10^{-3} 
	\end{bmatrix}, \label{eq:D+C}
\\
	&\Sigma =\begin{bmatrix}
	0.2626 & -0.0014  \\ 
	-0.0014 & 0.5013
	\end{bmatrix}. \label{eq:covar}
\end{align}
\end{subequations}
First thing to note is that the matrix $\mathrm{C}$ has a small norm and, 
by virtue of Eq.~(\ref{intdiff equation}), it means that memory effects are going to be very small. On the other hand, the eigenvalues of the matrix $\mathrm{D}$ are $\lambda _1 \simeq -20, \lambda _2 \simeq -10$, which 
are approximatelly the drift coefficients of the uncoupled OU process driving the $\yy$-variables. This indicates that the exponential kernel is damping the effects of the $\xx$-variables in past times rather quickly. %% the further away in time, the smaller the contribution. 
Moreover, the covariance matrix $\Sigma$ corresponds to that obtained by integrating the $\yy$-dynamics independently, according to Eq.~({\ref{covariance matrix wl}}).

Regarding the WL approximation, we stress that the $\yy$-variables are no 
longer present, after taking the averages in its construction. The memory 
kernel $\mathcal{K}$ in this case differs from the matrix $\mathrm{D}$ above, although its dominant terms correspond to its eigenvalues. Note that, if $L_{13} = 0$, the coupling function $\CCC^{\xx}_{\yy}$ would project entirely onto the eigenfunction of the OU process associated with the eigenvalue $-(\gamma _1 + \gamma _2)$. The same statement would hold for $c_{134}=0$, where in this case $\CCC^{\xx}_{\yy}$ projects onto the eigenfunctions associated with the eigenvalue $-\gamma_1$.\\

%\subsubsection{Reduced Time Scale Separation}
\paragraph{Reduced Time Scale Separation.} The parameter $h$ controls the 
time scale separation in the evolution of the $\xx$- and $\yy$-variables. 
Here we set $h=1$ so that this separation is reduced by an order of magnitude as illustrated by the autocorrelation functions in Fig.~\ref{autocorrelations epsilon 1}. The question to be addressed in this subsection is the effect of such a reduction in the WL and EMR parametrizations and their respective performance.

In the WL parametrization, there is no need to sample the dynamics in order to construct it, since the formulas of Sect.~\ref{weak-coupling limit parametrization} are explicit and do not depend on $h$. In the case of model~\eqref{stochastic model 1}, the covariance matrix and time correlations of the WL noise correction are thus given by Eq.~(\ref{autocorrelation 
function wl}) with no reference to $h$.  The memory term, though, is expected to change as the kernel $\mathcal K$ will decay more slowly by a factor of $10$. Therefore, memory effects are more important, as expected.

The EMR approach, on the other hand, requires a new learning phase for this value of $h=1$. We used the same numerical integration parameters $\dd_{\ell} t = 10^{-3}$ and $T_{\ell} = 10^{4}$ time units as for the previous case. In the first level regression, one observes that the coefficient values listed in Table~\ref{table coeffs emr l1 epsilon 1} are essentially the same from those estimated in the previous case, for $h = 0.1$, and listed in Table~\ref{tab:h1_row1}, as well as being even more distant from the original ones in the table's first row. The second level coefficients, for $h=1$, are shown on Table~\ref{tab:second_level_h1}.

The covariance matrix $\Sigma$ of the noise correction is indicated in %% 
Eq.~(\ref{cov mat emr epsilon 1}) 
Eq.~\eqref{eq:covar_h2} and it agrees fairly well with the previous values, for $h = 0.1$, as given in Eq.~\eqref{eq:covar}. The matrix  $\mathrm{C}$ that indicates the strength of the memory effects also has a magnitude that is of the same order as that in the previous case of $h = 0.1$, which is rather suprising, given the factor of $10$ in time scale separation $h$; compare Eqs.~\eqref{eq:D+C} and \eqref{cov mat emr epsilon 1}. 
This observation tells us that the loss of Markovianity might be intrinsic to the nature of the coupling rather than being due to the time scale separation, even though, in the limit case of infinite scale separation, memory effects will dissapear entirely. The memory kernel, as determined by $\mathrm{D}$, scales almost exactly with the time scale separation and it is expected to change depending on how the coupling functions project onto the eigenspaces of the underlying Orstein-Uhlenbeck process, as discussed more generally earlier in Theorem~\ref{proposition: 1}.

\begin{figure}[H]
	\centering
	\begin{subfigure}[b]{0.32\textwidth}
		\centering
		\includegraphics[width=\textwidth]{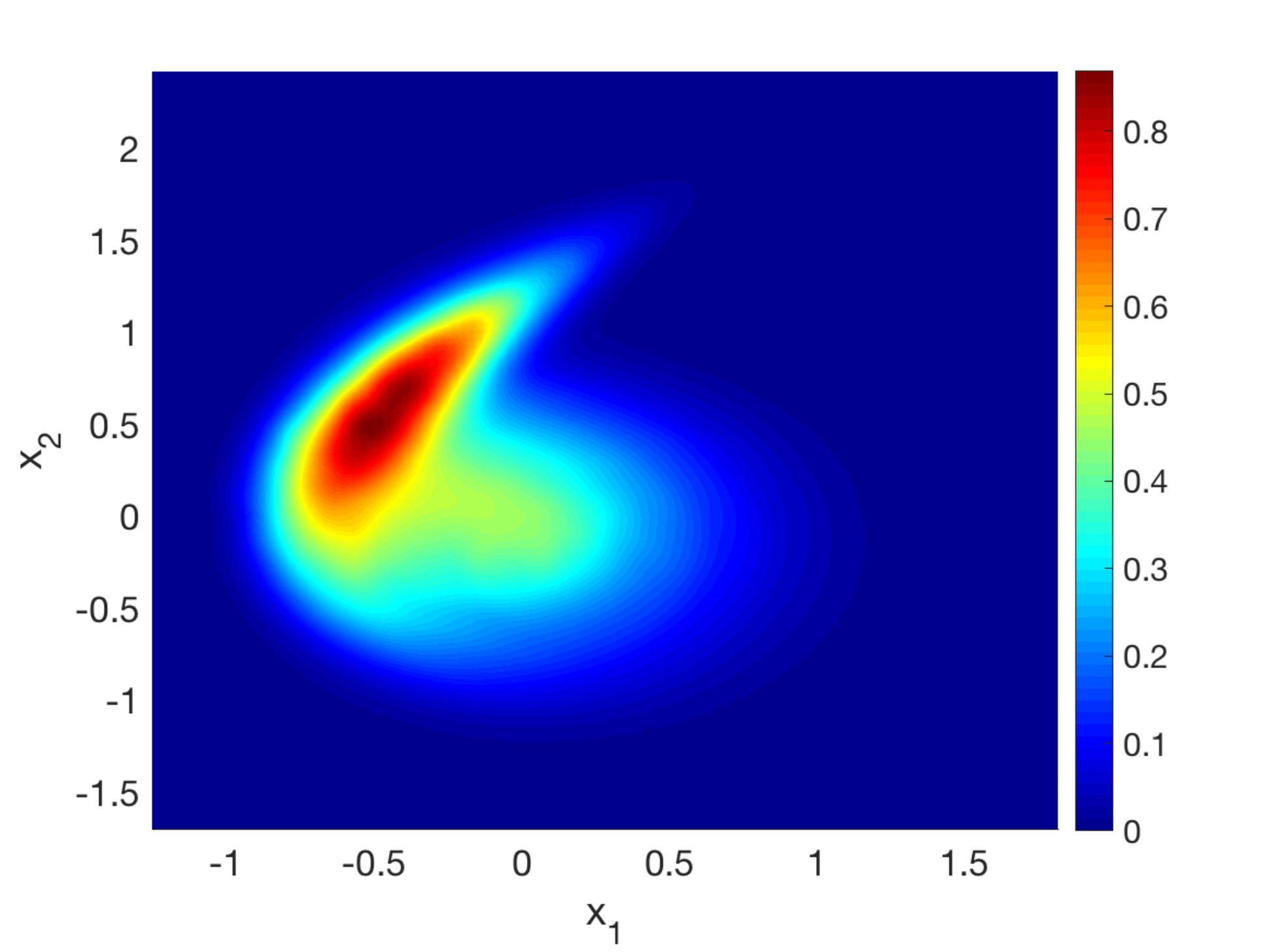}
		\caption{Full model}
	\end{subfigure}
	\hfill
	\begin{subfigure}[b]{0.32\textwidth}
		\centering
		\includegraphics[width=\textwidth]{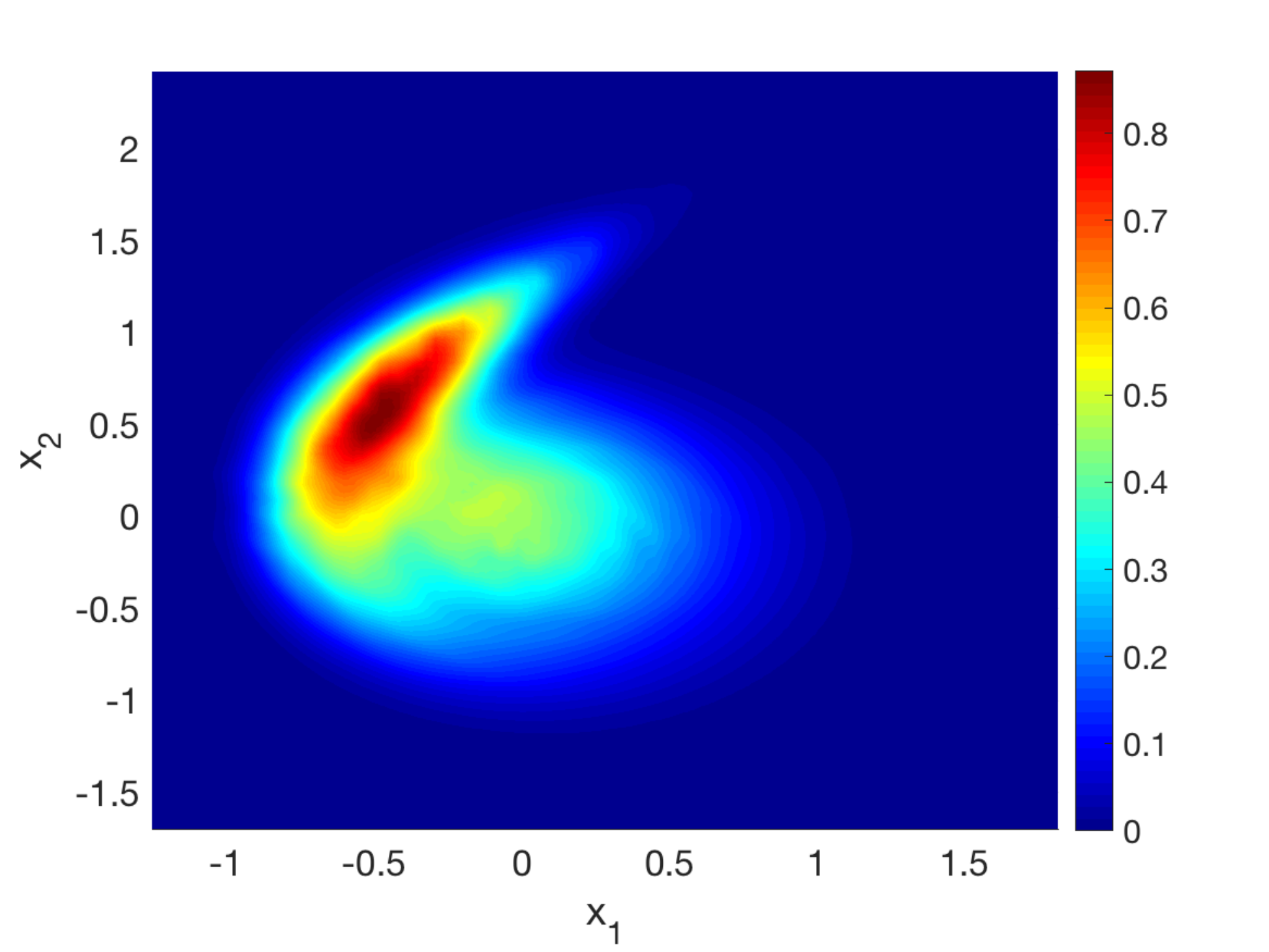}
		\caption{EMR model}
	\end{subfigure}
	\hfill
	\begin{subfigure}[b]{0.32\textwidth}
		\centering
		\includegraphics[width=\textwidth]{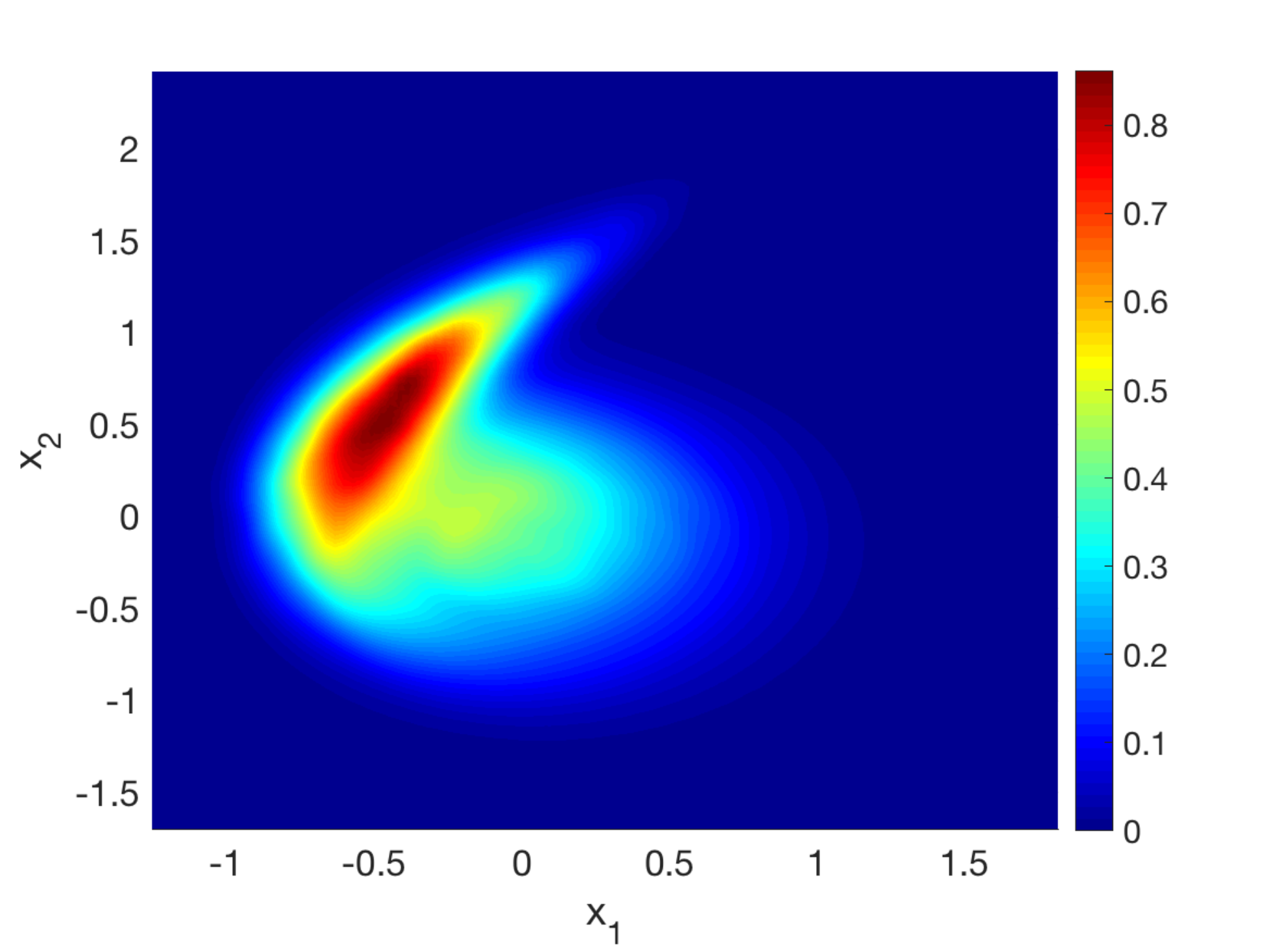}
		\caption{WL parametrization}
	\end{subfigure}
	\caption{\label{histograms 2d epsilon 1} 
	Two-dimensional smoothed PDFs of the stochastic model~\eqref{stochastic model 1},  but with a time scale separation of $h = 1$. Panels (a), (b) 
and (c) are calculated by integrating the full model, the EMR model and WL approximation, respectively,  as in Fig.~\ref{climatologies 1}.}
\end{figure}

\begin{figure}[H]
	\centering
	\begin{subfigure}[b]{0.49\textwidth}
		\centering
		\includegraphics[width=\textwidth]{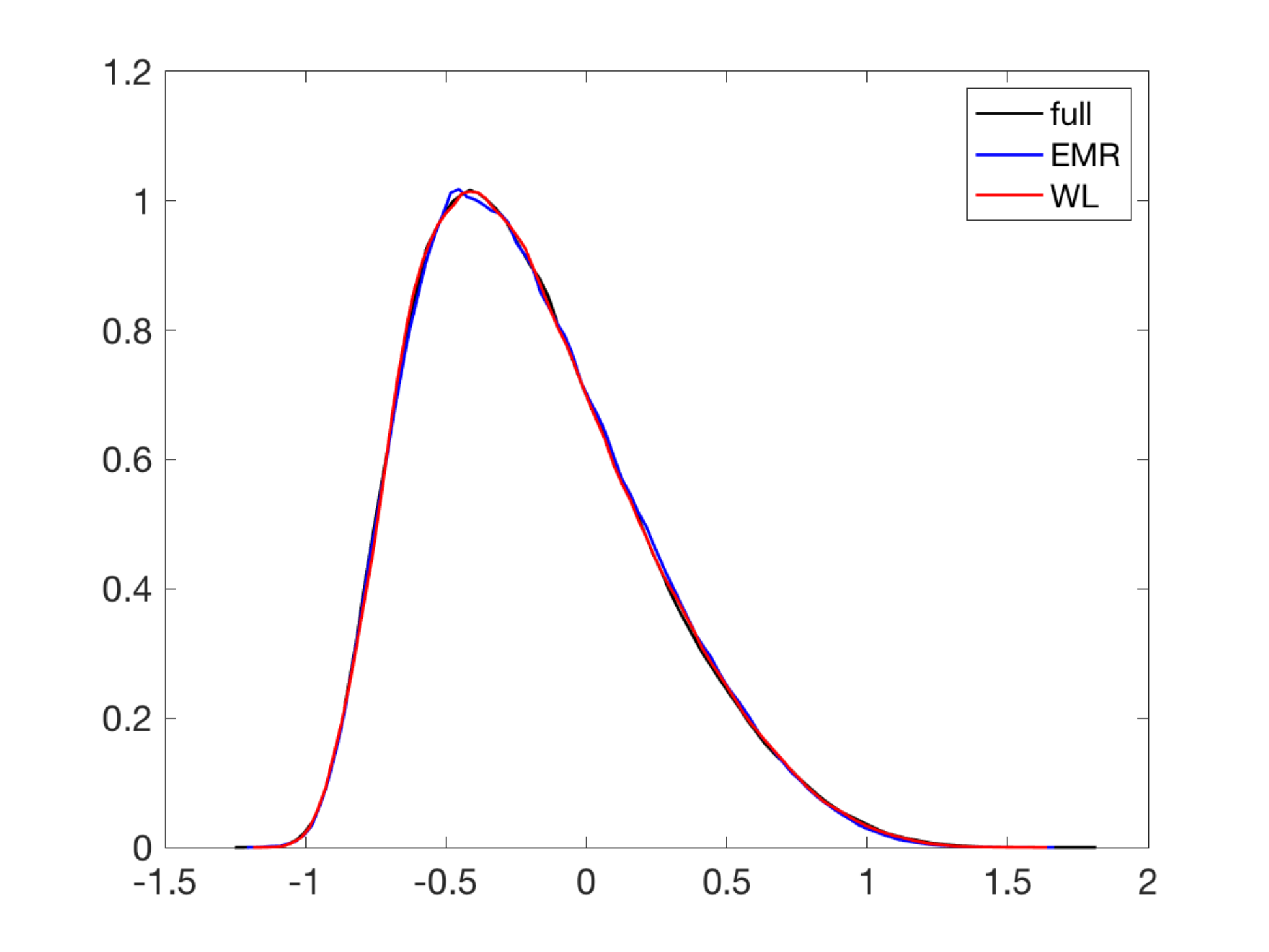}
		\caption{$x_1$-PDF}
	\end{subfigure}
	\begin{subfigure}[b]{0.49\textwidth}
		\centering
		\includegraphics[width=\textwidth]{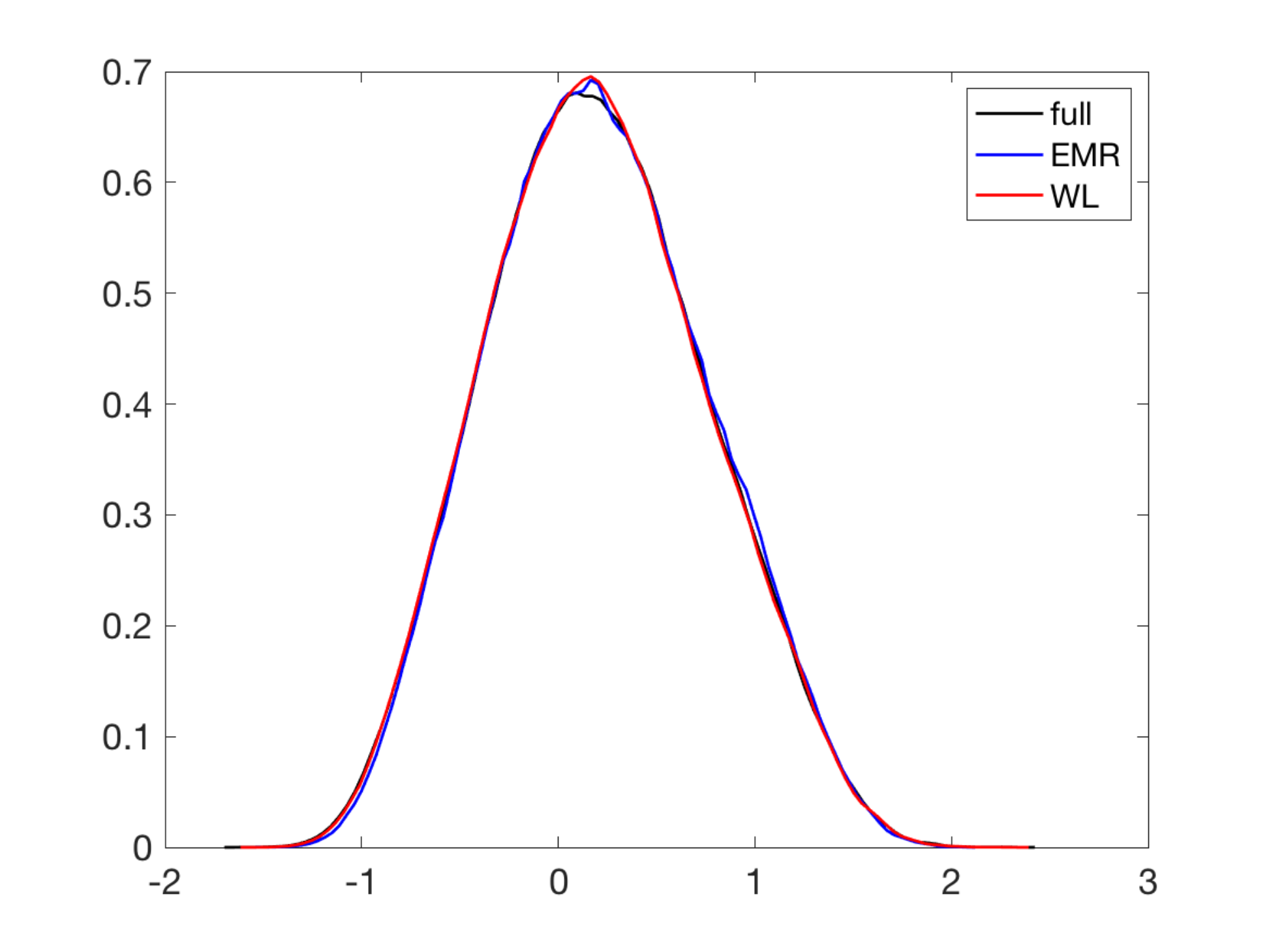}
		\caption{$x_2$-PDF}
	\end{subfigure}
	\caption{\label{pdf 2}PDFs of (a) the $x_1$ variable and (b) the $x_2$ variable, for a time scale separation of $h=1$; compare with Fig.~\ref{marginals 1}.
	}
\end{figure}

\begin{figure}[H]
	\centering
	\begin{subfigure}[b]{0.49\textwidth}
		\centering
		\includegraphics[width=\textwidth]{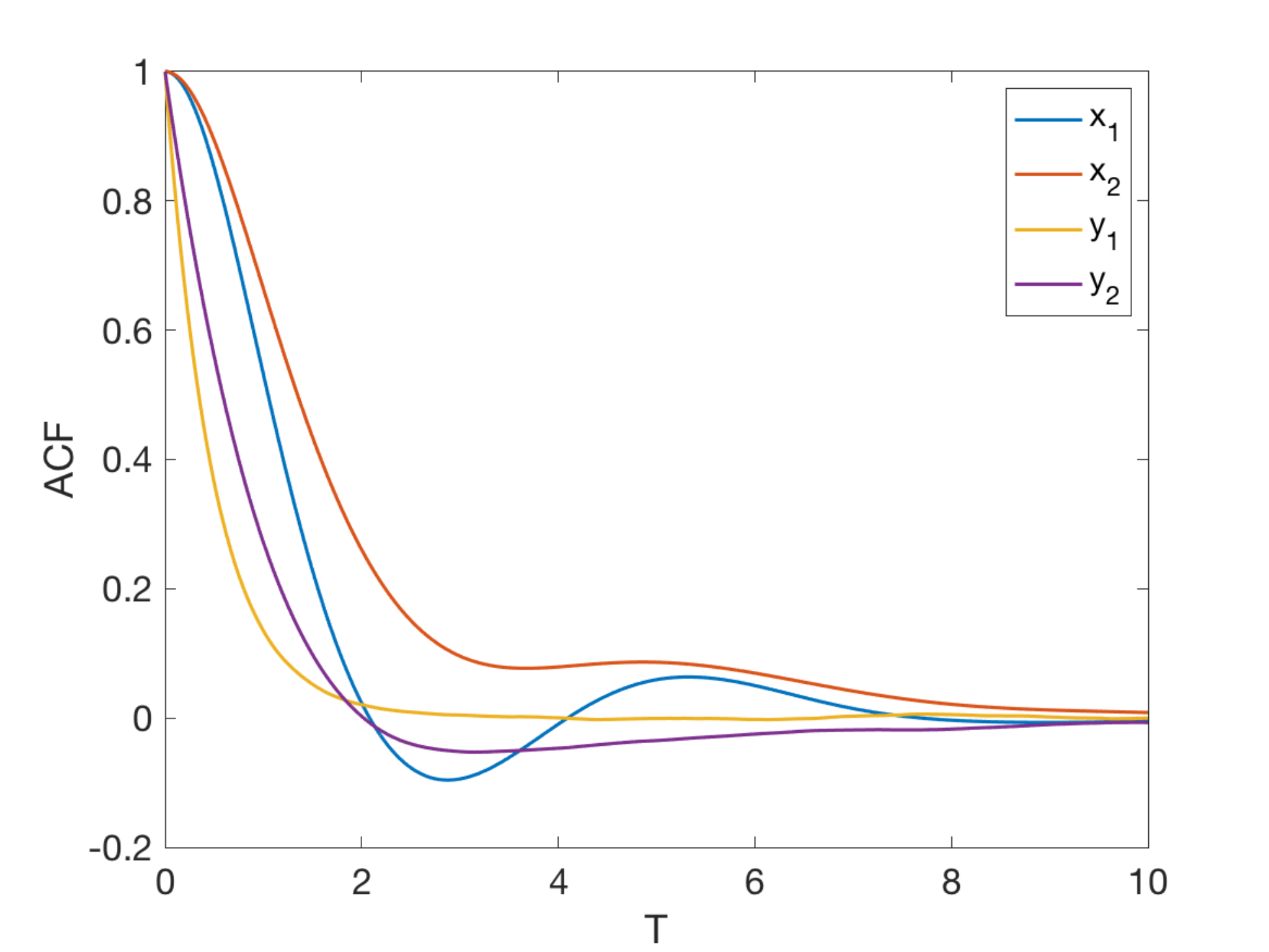}
		\caption{}
	\end{subfigure}
	\begin{subfigure}[b]{0.49\textwidth}
		\centering
		\includegraphics[width=\textwidth]{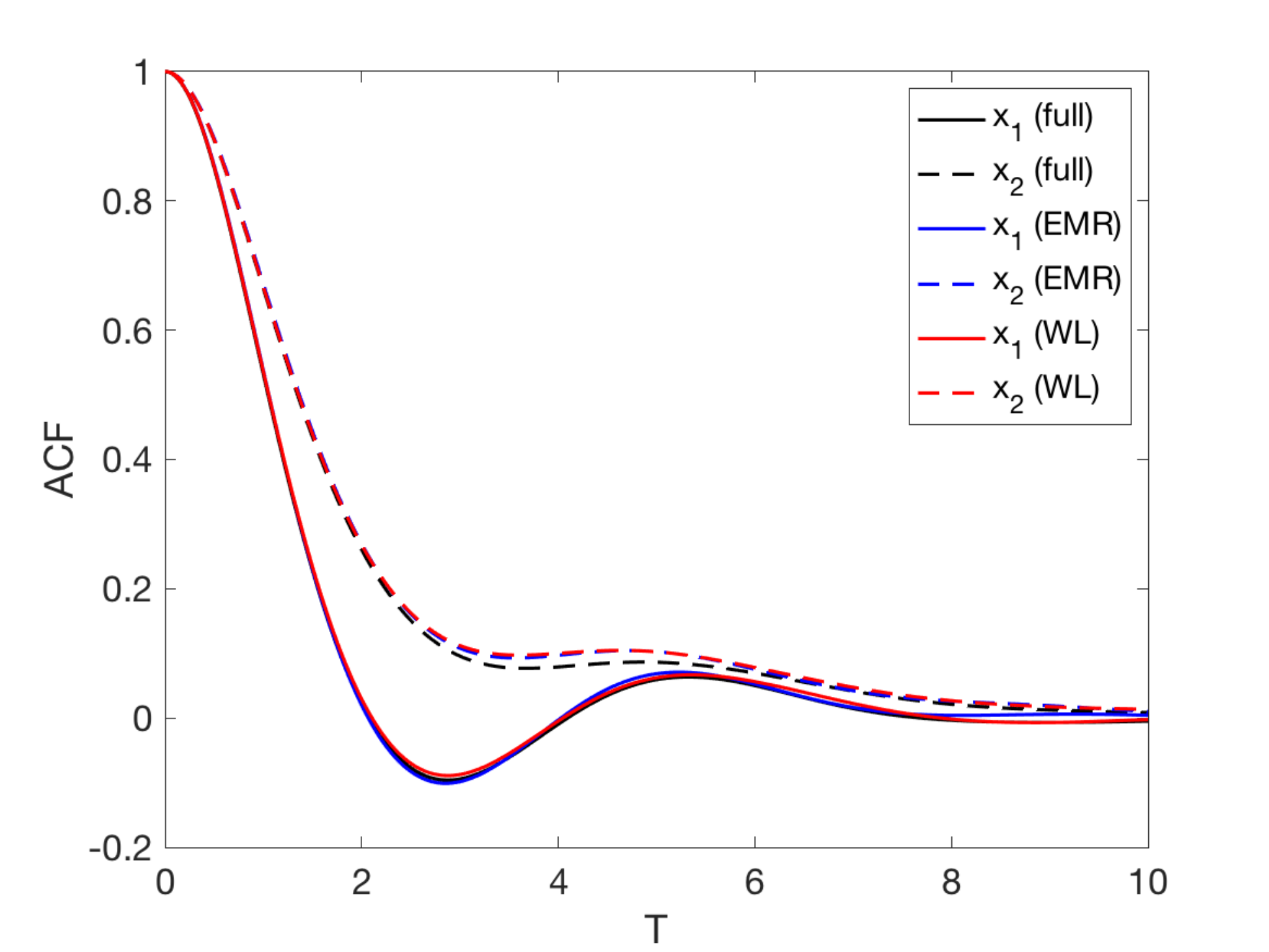}
		\caption{}
	\end{subfigure}
	\caption{\label{autocorrelations epsilon 1}Autocorrelation functions for 
the four variables 
		$x_1, x_2, y_1, y_2$ obtained (a) from the full model; %% (left); %% and the comparison of those obtained with
		and (b) the comparison of the corresponding results for $x_1, x_2$ with 
the full model, 
		the EMR model, and the WL parametrization. See legend for the choice of 
lines; $h = 1$.
	}
\end{figure}

The performance of both parametrization techniques is summarized in Figs.~\ref{histograms 2d epsilon 1}--\ref{autocorrelations epsilon 1}. These figures are the exact counterparts of Figs.~\ref{climatologies 1}--\ref{correlations 1} for the reduced time scale separation $h = 1$. First of all, we note from Fig.~\ref{autocorrelations epsilon 1}(a) that, for $h = 
1$  there is indeed no strict time scale separation, as indicated by the autocorrelation functions obtained from the full model. Secondly, consideration of Figs.~\ref{histograms 2d epsilon 1}(a)--(c), \ref{pdf 2}(a,b) and \ref{autocorrelations epsilon 1}(b) shows that neither the WL nor the EMR approach seems to be affected by the time scale reduction.

\begin{table}[H]\renewcommand{\arraystretch}{2}
	\centering
	\caption{\label{table coeffs emr l1 epsilon 1}Empirically estimated EMR model coefficients at the first level, for $h = 1.$}
	\begin{tabular}{cccccc}
		\hline 
		$f$ & $x_1$ & $x_2$ & $x_1^2$ & $x_1x_2$ & $x_2^2$ \\ 
		\hline 
-0.31181 & -0.339 & -0.42944 & 0 & -0.93439 & 0.97958 \\ 
-0.16925 & 0.46833 & 0.17513 & 0.93439 & -0.97958 & 0 \\ 
		\hline 
	\end{tabular}
\end{table}

\begin{table}[H]\renewcommand{\arraystretch}{2}
	\centering
	\caption{\label{tab:second_level_h1}Empirically estimated EMR model coefficients at the second level, for $h = 1$.}
	\begin{tabular}{ccccc}
		\hline 
		$f^{(1)}$ & $x_1$ & $x_2$ & $r^{(1)}_1$ & $r^{(1)}_2$  \\ 
		\hline 
0 & -3.9452e-3 & -1.5027e-4 & -2.1001 & 0.016509  \\ 
0 & -5.0312e-4 & -3.79e-3 & 0.054861 & -1.1214  \\ 
		\hline 
	\end{tabular}
\end{table}

\begin{subequations}
\begin{align}
	&\mathrm{D}=\begin{bmatrix}
-2.1001 & 0.016509  \\ 
0.054861 & -1.1214 
	\end{bmatrix} , \mathrm{C}=\begin{bmatrix}
-4\cdot 10^{-3} & -1\cdot 10^{-4} \\
-6\cdot 10^{-4} & -4\cdot 10^{-3}
	\end{bmatrix}, \label{cov mat emr epsilon 1} \\
	&\Sigma =\begin{bmatrix}
	0.2618 & 0.0011  \\ 
	0.0011 & 0.4554
	\end{bmatrix}. \label{eq:covar_h2}
\end{align}
\end{subequations}

\subsection{Memory Effects}

We would like to end this results section by analyzing the role of the memory effects when performing a reduction of the highly idealized model given by Eqs.~(\ref{stochastic model 1}). For this purpose, we apply the criterion~\eqref{criterion memory effects} discussed in the corresponding Appendix~\ref{Markovianity}. We thus spectrally approximate the autocorrelation functions of the variables $x_1$ and $x_2$ using Eq.~({\ref{criterion memory effects}}) with the Koopman operator $\mathcal{T}_{\tau}$ estimated using Ulam's method with a transition time of $\tau=0.5$ time units for the case where $h=0.1$ and $\tau=1$ time unit for $h=1$. 

The difference between the two $\tau$-values is due to the fact that we expect the coarse-grained phase space to be sensitive to the system's variability. Hence, if the time scale separation is large, a shorter transition time is requiered in order to capture the influence of the hidden processes. In fact, a range of transition times was tested in the case of $h=1$ to find that the optimal value was $\tau=1$. It follows that the methodology is robust in showing the effects of memory in the projected phase space.

\begin{figure}[H]
	\centering
	\begin{subfigure}[b]{0.49\textwidth}
		\centering
		\includegraphics[width=\textwidth]{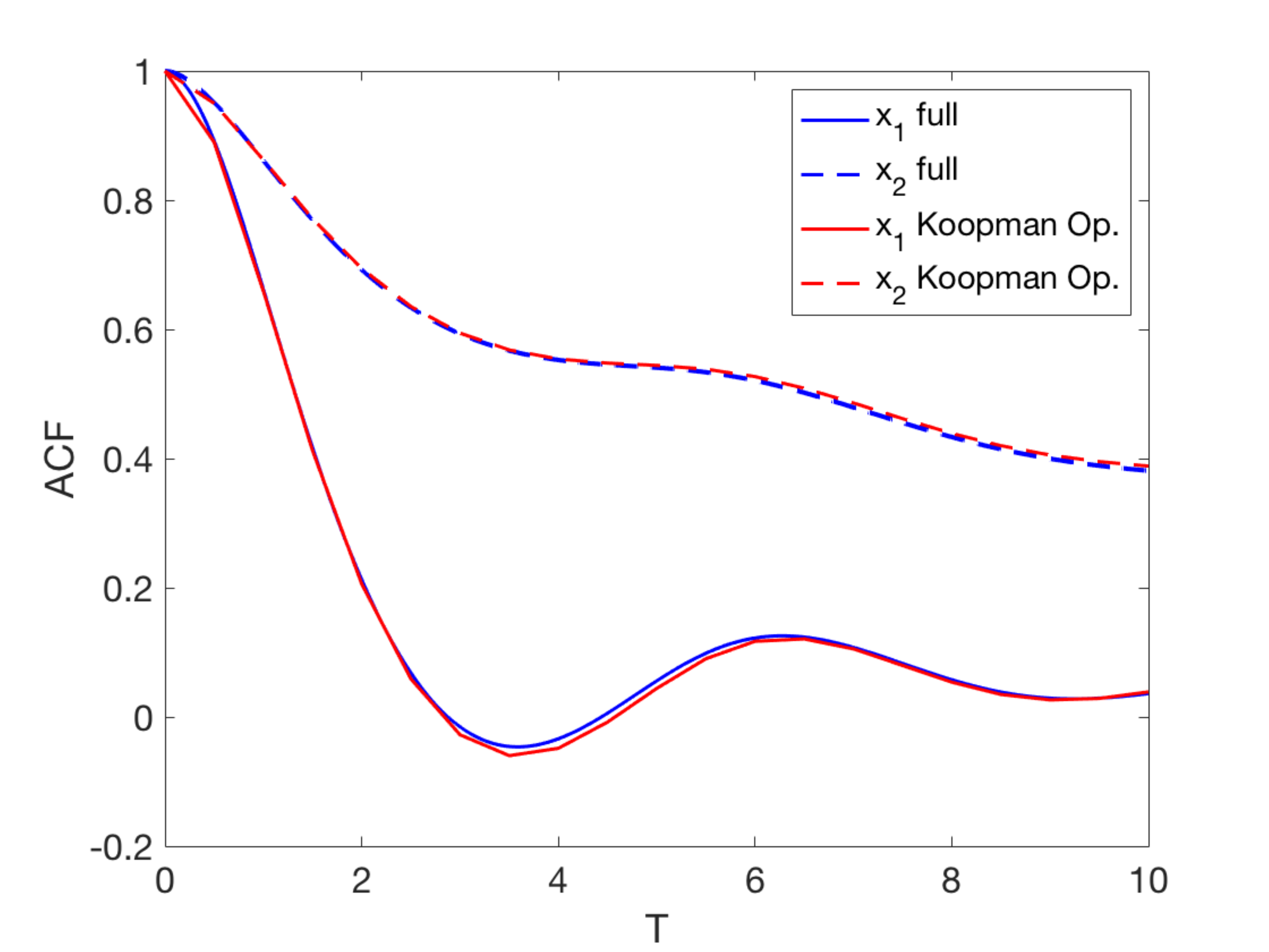}
		\caption{}
	\end{subfigure}
	\begin{subfigure}[b]{0.49\textwidth}
		\centering
		\includegraphics[width=\textwidth]{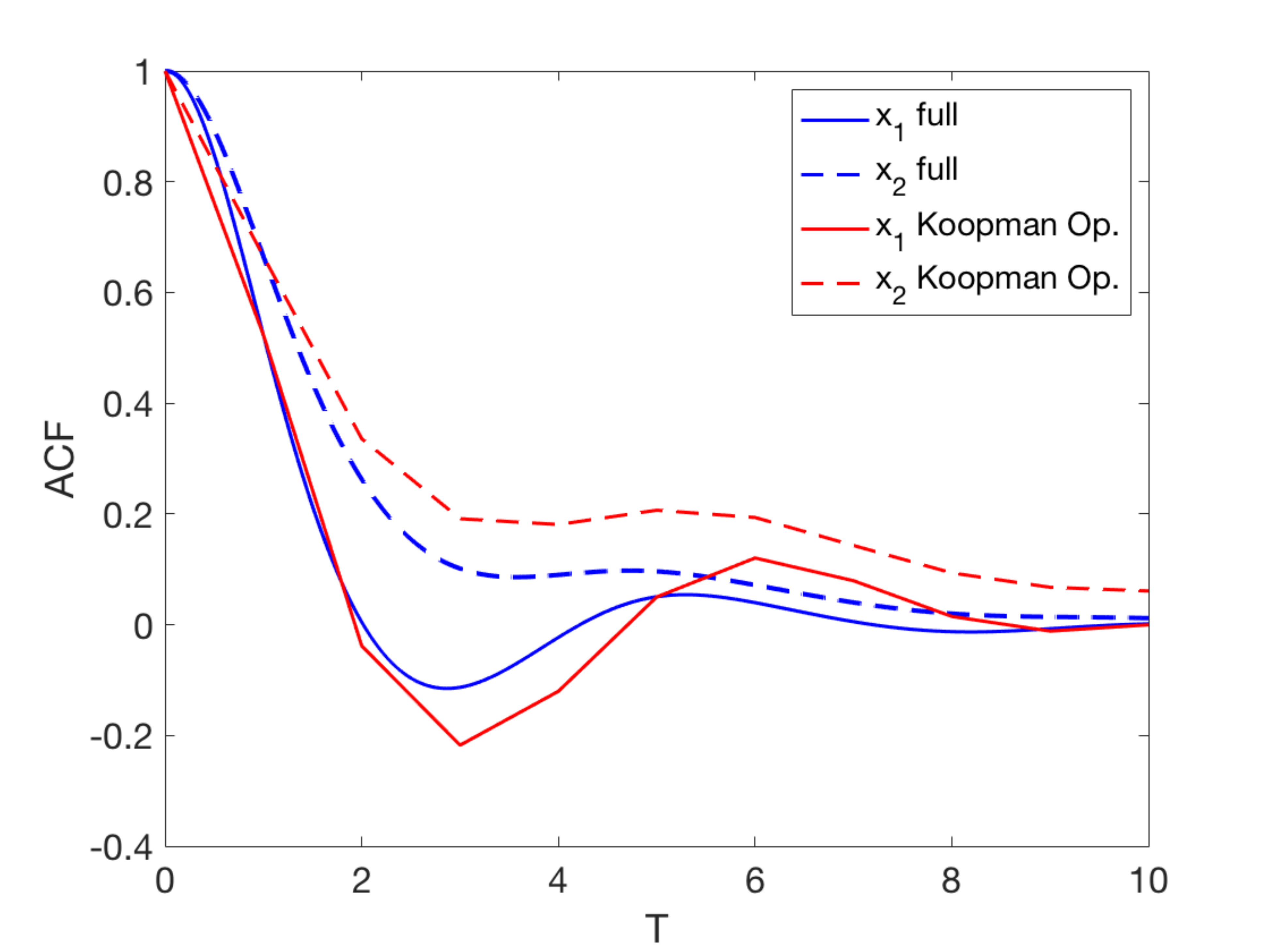}
		\caption{}
	\end{subfigure}
	\caption{\label{autocorrelations transfer operator}Autocorrelation functions for the 
	two $\xx$-variables obtained from the full model in the spectral reconstruction 
	using the Kooman operator: (a) $h=0.1$ and (b) $h=1$.}
\end{figure}

We clearly observe in Figure~\ref{autocorrelations transfer operator} that the correlation functions can be accurately reconstructed in the case of large time scale separation $h=0.1$ (figure (a)) but not so for $h=1$ (figure (b)). This indicates, naturally, that memory effects are negligible in the first case and relevant in the second.

\section{Conclusions}\label{CONCLUSIONS}
To formulate accurate and efficient parametrizations for multiscale processes is a crucial challenge in many areas of science and technology for one of two reasons: either the numerical simulation of all scales active in a given system is computationally unfeasible; or there is a mismatch between model resolution and the granularity and homogeneity of the observations, as in the case in geophysical flows and in the climate system. Moreover, the construction of parametrizations is instrumental to % are of the essence to 
help understand the nature of nonlinear fluxes across scales and the physical processes responsible for cascades, instabilities, and feedbacks.

There are two main approaches for constructing parametrizations:  top-down, by deriving the parametrizations directly from the evolution equations 
governing the system through the use of suitable approximations; and data-driven, in which the parametrizations are constructed through suitable optimization procedures, which are first tuned in a training phase and then actually used in the prediction phase. Both approaches aim to derive the effective dynamics for the variables of interest: formally, this is achieved by applying the Mori-Zwanzig projection operator \cite{mori_transport_1965, zwanzig_memory_1961} to the full dynamics. The result of doing so is to describe the impact of the hidden variables by formulating a generalized Langevin equation (GLE)~\eqref{mori zwanzig 2} for the variables of interest that includes a deterministic, a stochastic, and a non-Markovian component.  

Top-down and data-driven approaches are conceptually complemetary and have different practical advantages and disadvantages. In this paper, we have shown the fundamental equivalence between a top-down and a data-driven approach that have been formulated and applied in the recent literature. This equivalence was illustrated schematically in Fig.~\ref{schematic}. 

We first revisited in Sect.~\ref{weak-coupling limit parametrization} the 
WL parametrization of \cite{Wouters2012, Wouters2013}, which relies on an 
assumption of weak coupling between the hidden and observed variables, and have extended the previous results by considering more general coupling classes. We have also shown that the perturbative expansion that yields the WL parametrization is exact when the coupling between the hidden 
and resolved variables is additive.

The Dyson formalism~\eqref{dysonapprox} appears to be essential for computing the effects of the hidden processes on the dynamics of the observed variables, when working at the level of the system's observables. This methodology is explicit, in the sense that no information about the actual coupled process is needed, because the formal computations are performed by considering the limit in which no coupling is present. Other advantages of this approach are that it can be implemented without the need for any hypothesis on the time scale separation between the hidden variables and the observed ones, and that it is also scale adaptive \cite{Vissio2018a}.

We addressed systematically the problem of re-Markovianizing the WL memory equation, which was first pointed out in \cite{Wouters2016} and discussed further in Sect.~\ref{preliminary} here. In this example, a system~\eqref{eq:triad} with one observed and two hidden variables that yielded a scalar WL parametrization was re-Markovianized to a Markovian system with just two scalar differential equations. Throughout Sect.~\ref{weak-coupling limit parametrization}, we provided a broader framework for re-Markovianization. This framework was presented for a scalar equation in Theorem~\ref{proposition: 1} and described for higher dimensional systems in Remark~\ref{Rem_genedim}. The required assumptions for this treatment boil down to certain %% appropriate 
spectral properties of the Koopman operator for the hidden variables.

The multilevel structure of the re-Markovianization obtained in Sect.~\ref{weak-coupling limit parametrization} motivated the comparison with multilayer stochastic models (MSMs) in Sect.~\ref{MSMEMR}. Such MSMs arise naturally in data-driven reduction methods and they had been shown in \cite{kondrashovdata2015} to approximate the GLE predicted by Mori \cite{mori_transport_1965} and Zwanzig \cite{Zwanzig.2001}. 
	
We showed in Sect.~\ref{ssec:MSM} that a seemless application of the WL parametrization solves the MSM of Eq.~\eqref{Eq_MSM} and coincides with its It\^o integration, cf.~\cite{kondrashovdata2015}. Note that an MSM can be obtained from partial observations of the coupled system, which amounts to the special case of the data-driven empirical model reduction (EMR) methodology \cite{kondrashovdata2015, Kondrashov.Kravtsov.ea.2006, KKRG.ENSO.2005, Kravtsov2005, KKG.2010}. %% approach, 
	
The EMR methodology was revisited here in Sect.~\ref{Sec_EMR} and it is, in principle, dual to the WL parametrization, in the sense that only partial observations of the coupled system are required, without the need for 
knowing the actual equations of motion. Comparing the multilevel structure of Eq.~\eqref{eq: multilevel markovianization2} with that of Eq.~\eqref{eq:EMR} suggests that the Koopman eigenvalues $\lambda_j$ highlighted in 
Theorem~\ref{proposition: 1} may help provide insights into the number of 
levels needed for EMR to converge. This practical role of the $\lambda_j$'s deserves, therewith, a more careful examination %%that is aimed to be pursued elsewhere.
 in further work.

Additionally, we considered in Sect.~\ref{RESULTS} a conceptual climate model to which we applied both of the methodologies revisited herein. Since both the MSM and the WL parametrization yield a memory equation that involves integrals and stochastic noise, we were able to compare their structure, as well as their statistical outputs. We found that both methodologies produced equivalent numerical results and that the memory kernel and 
noise predicted in the WL parametrization agreed with what was found using the data-driven EMR approach.

Concluding, our view point is complementary to the dynamic mode decomposition \cite{Mezic2005,Rowley2009,Schmid2010} as it uses the basis of eigenvectors of the Koopman operator to construct the projected --- in the sense of Mori-Zwanzig --- dynamics of the observables of interest, which is then recast in Markovian form using the multilevel Markovian model framework, where the number of levels corresponds to the number of eigenvectors 
of the Koopman operators one considers in the reconstruction of the dynamics. In a nutshell, our findings support, on the one hand, the physical basis and robustness of the EMR methodology and, on the other hand, illustrate the practical relevance of the WL perturbative expansion used for deriving the parametrizations.

\section*{Acknowledgements}
It is a pleasure to acknowledge useful exchanges with Georg Gottwald, Jeroen Wouters and Gabriele Vissio. VL acknowledges the support received from the European Union's Horizon 2020 research and innovation program through the project CRESCENDO (Grant Agreement No. 641816). This article is TiPES contribution \# 56; the TiPES (Tipping Points in the Earth System) project has received funding from the European Union’s Horizon 2020 
research and innovation program under Grant Agreement No. 820970. This research was partially supported by the Israeli Council for Higher Education (CHE) via the Weizmann Data Science Research Center, by the European Research Council (ERC) under the European Union's Horizon 2020 research and innovation programme (grant agreement No. 810370), and by the Office of Naval Research (ONR) Multidisciplinary University Research Initiative (MURI) grant N00014-20-1-2023. This paper has also been partially supported by the EIT Climate-KIC; EIT Climate-KIC is supported by the European Institute of Innovation \& Technology (EIT), a body of the European Union, under Grant Agreement No. 190733 and by the Russian Science Foundation (Grant No. 20-62-46056).

\section*{Data Availability}
The data that support the findings of this study are available from the corresponding author
upon request.

\pagebreak

\appendix

%%%%%%%%%%%%%%%%%%%%%%%%%%%%%%%%%%%%

\section{Proof of Theorem \ref{proposition: 1}}\label{App:proof}

\begin{proof}
	The aim is to show that under the assumptions of this Theorem --- which require that the coupling function $\CCC^{\xx}:\mathbb{R}\longrightarrow\mathbb{R}$ projects entirely onto $\mathrm{span}\left\{\psi_{j}, \; j=1,\ldots,N \right\}$ --- the memory and noise terms of the WL equation \eqref{lw parametrization independent} 
	are obtained from the term $\epsilon \Lambda\cdot \mathbf{Z}(t)$ in Eq.~\eqref{remarkovianisation3}, after  integration of Eq.~\eqref{remarkovianisation2}. %% when the coupling function $\CCC^{\xx}:\mathbb{R}\longrightarrow\mathbb{R}$ projects entirely onto $\mathrm{span}\left\{\psi_{j}, \; 
%$j=1,\ldots,N \right\}$. 
	
	{\bf Step~1.} In this step, we expand the memory term and the lag-correlations of the noise in 
	the WL equation \eqref{lw parametrization independent}  in terms of the leading eigenelements of the uncoupled Koopman operator $\LLL_{0}^{\yy}$. 
These expansions will serve us in Step 2 below to compare the noise and memory terms of the WL-equation with those produced after integration of Eq.~\eqref{remarkovianisation2}.
	
	Let the $\lambda_j$ be sorted as in the hypotheses of the theorem. Hence, to distinguish real and complex conjugate eigenvalues and for notational convenience, we introduce the set of indices  $I_r$ and $I_{+}$ defined 
as:
	\begin{subequations}
		\begin{align}
			I_r&=\{j \in \{1,\ldots,N\} : \lambda_j \text{ is real} \},\\
			I_+&=\{j \in \{1,\ldots,N\} : \mathfrak{Im}\lambda_j>0\}.
		\end{align}
	\end{subequations}
	An immediate consequence that is used several times throughout the proof 
is that the sum of the eigenvalues is real, and may be split  as follows:
	\begin{equation}
		\sum_{j=1}^{N}\lambda_j=\sum_{j\in I_r}\lambda_j + \sum_{j\in I_+}\lambda_j+ \sum_{j\in I_+}\overline{\lambda_j}=\sum_{j=1}^{N}\mathfrak{Re}\lambda_j \in \mathbb{R}.
	\end{equation}		
	As previously stated, we expand the mean and correlation functions of the scalar noise term $\eta$ in the WL-equation \eqref{lw parametrization independent} in terms of the eigenpairs. The mean is zero by assumption, but the autocorrelation function can be expanded as follows, %% (see the notation in  Eq.~\eqref{spectral correlations} for clarity):
 based on Eq.~\eqref{spectral correlations},
	\begin{align}\label{proof wl noise 3}
		\bigg \langle \eta(t)\eta(0) \bigg \rangle = C_{\CCC^{\xx},\CCC^{\xx}}(t)   =\sum_{j=1}^{N}e^{\lambda_jt}\alpha_j\beta_j;
	\end{align}
	herein, $\alpha_j$ and $\beta_j$ are as defined in \eqref{eq:alpha} and \eqref{eq:beta}, respectively. 
	The expansion of the correlation function in Eq.~\eqref{proof wl noise 3} is a finite sum by virtue of the assumption that $\CCC^{\xx}$ lies in $\mathrm{span}\left\{\psi_{j}, \; j=1,\ldots,N \right\}$ and therefore there is no contribution from the essential spectrum.

	Regarding the complex scalars $\alpha_j$ and $\beta_j$ defined \eqref{eq:alpha} and \eqref{eq:beta}, it follows that for each $j$ such that $\lambda_j=\overline{\lambda_{j+1}}$, we get $\alpha_j=\overline{\alpha_{j+1}}$ and $\beta_j=\overline{\beta_{j+1}}$. Indeed,
	\begin{subequations}\label{eq: conjugacy relation alpha}
		\begin{align}
			\alpha_j &= \int \nu(\dd \yy)  \overline{\psi_j^{\ast}(\yy)}\CCC^{\xx}(\yy) = \overline{\int \nu(\dd \yy)  \psi_j^{\ast}(\yy)\CCC^{\xx}(\yy)} \\&= \overline{\int \nu(\dd \yy)  \overline{\psi_{j+1}^{\ast}(\yy)}\CCC^{\xx}(\yy)} = \overline{\alpha_{j+1}},
		\end{align}
	\end{subequations}
	in which we have exploited the fact that $\psi_j=\overline{\psi_{j+1}}$ when $\lambda_{j}$ is complex and $j$ in $\{1,\ldots, N\}$.

	The same proof can be repeated for $\beta_j$. Such a conjugacy relation also holds for the gradients of the eigenfunctions $\nabla  \psi_j$, for those $j$ in $\{1\ldots,N\}$ such that $\lambda_j=\overline{\lambda_{j+1}}$. This is observed by the following equality:
	\begin{align}\label{Eq_toto1}
		\nabla \psi_{j}(\yy)=\nabla \overline{\psi_{j +1}}(\yy)=\overline{\nabla\psi_{j+1}(\yy)},
	\end{align}
	since $\nabla$ is a differential operator that only involves here differentiation with respect to a real variable.

	Exploiting \eqref{Eq_toto1}, the memory kernel $\mathcal{K}$ then expands as (recalling that $\mathcal{R}(t)=0$):
	\bea\label{proof spectral kernel 2a}
	\mathcal{K}(t,s,x) &= \CCC^{\yy}(x(s))\cdot \bigg \langle \nabla\sum_{j=1}^Ne^{\lambda_j (t-s)}\alpha_{j} \psi _j (\yy) \bigg \rangle \\
	&=\CCC^{\yy}(x(s))\cdot  \sum_{j\in I_r}e^{\lambda_j (t-s)}\alpha_{j}\bigg \langle \nabla\psi _j (\yy)  \bigg \rangle 
	+\CCC^{\yy}(x(s))\cdot  \sum_{j\in I_+}e^{\lambda_j (t-s)}\alpha_{j}\bigg \langle \nabla\psi _j (\yy)  \bigg \rangle \\  
	&\hspace{1.2cm}+\CCC^{\yy}(x(s))\cdot  \sum_{j\in I_+}\overline{e^{\lambda_j (t-s)}\alpha_{j}\bigg \langle \nabla\psi _j (\yy)  \bigg \rangle} 
	\eea
	which leads to 
	\be\label{eq: last equality kernel proof} 
	\mathcal{K}(t,s,x)=\CCC^{\yy}(x(s))\cdot  \sum_{j=1}^N\mathfrak{Re}\bigg(e^{\lambda_j (t-s)}\alpha_{j}\bigg \langle \nabla\psi _j (\yy) \bigg 
\rangle\bigg).
	\ee
	
	Note that to go from \eqref{proof spectral kernel 2a} to \eqref{eq: last 
equality kernel proof},  we have made use of the aforementioned conjugacy 
relations, that led to a real-valued  memory kernel at the end. With \eqref{proof wl noise 3} and \eqref{eq: last equality kernel proof} at hand, the noise and memory terms in 
	the WL equation \eqref{lw parametrization independent} are thus characterized in terms of the leading eigenelements of the uncoupled Koopman operator $\LLL_{0}^{\yy}$.

	{\bf Step2.} The second step consists of analyzing the noise and memory terms produced by integration of  Eq.~\eqref{remarkovianisation2} and to compare these terms with those of the WL-equation.

	Performing an It\^o integration of Eq.~\eqref{remarkovianisation2} leads 
to:
	\begin{equation}\label{eq: ito integral proposition}
		\mathbf{Z}(t)=e^{\mathrm{D}t}\mathbf{Z}(0)+ \underbrace{ \int _{0}^{t}e^{\mathrm{D}(t-s)}\Sigma \dd W_s}_{\textrm{noise term}} + \epsilon\underbrace{\int _{0}^te^{\mathrm{D}(t-s)}\mathbf{R}(x(s))\dd s}_{\textrm{memory term}},
	\end{equation}
	where (for simplicity) we have assumed that the initial condition distributes normally with mean zero and variance equal to the identity matrix,
	and the function $\mathbf{R}:\mathbb{R}\longrightarrow \mathbb{C}^{N}$ is as defined  in \eqref{Eq_Rexpression}. 
	The noise and memory contributions of  $\mathbf{Z}(t)$ are as indicated by the brackets in \eqref{eq: ito integral proposition}.

	Let us denote the noisy component of Eq.~\eqref{eq: ito integral proposition} as $\mathbf{q}(t)$ in $\mathbb{R}^N$. Then, it is clear that $\mathbf{q}$ has zero mean and the lag cross-correlations read as:
	\begin{align}
		\mathbb{E}\left( \mathbf{q}(t)\mathbf{q}^{\top}(0) \right) =e^{\mathrm{D}t} 
		= \begin{bmatrix}
			e^{\lambda_1t} &&\\ &\ddots&\\&&e^{\lambda_Nt}
		\end{bmatrix}.
	\end{align}

	Let $\Lambda$ be defined as in \eqref{Eq_lambda} and let us calculate the mean and lag-correlations of the one-dimensional stochastic process $\Lambda\cdot\mathbf{q}$, aimed at approximating $\eta$ in the WL equation.	
	First, note that 
	$\Lambda\cdot \mathbf{q}$ is a zero-mean Gaussian process, with lag correlations given by:
	\begin{subequations}
		\begin{align}\label{eq: proof correlation A Z}
			\mathbb{E}\left( \left(\Lambda\cdot \mathbf{q}(t)\right)\left(\Lambda\cdot \mathbf{q}(0)\right)  \right) = \left(e^{\mathrm{D}t}\Lambda\right)\cdot \Lambda.
		\end{align}
	\end{subequations}
	Now, expanding  $\left(e^{\mathrm{D}t}\Lambda\right)\cdot \Lambda$ in \eqref{eq: proof correlation A Z} shows that  we recover 
	the right-hand side (RHS) of \eqref{proof wl noise 3}.

	However, the noise term $\eta$ in the WL-equation is a real-valued stochastic process, and we are dealing with complex scalars, so therefore
	we still have to show that $\Lambda\cdot \mathbf{q}(t)$ is real for every $t$ in $\mathbb{R}$.  To do so, let us denote by $\mathbf{w}^{\top}=[w_1,\ldots ,w_N]$ any arbitrary row vector in $\mathbb{R}^{N}$.  In other 
words,  $\mathbf{w}$ is an arbitrary column vector with real entries.
	
	Consider the following inner product:
	\begin{subequations}
		\begin{align}
			\Lambda\cdot e^{\mathrm{D}t}\Sigma \mathbf{w} &= \Lambda\cdot e^{\mathrm{D}t}\begin{bmatrix}\sqrt{-2\mathfrak{Re}\lambda_1}&& \\ &\ddots & \\ 
&&\sqrt{-\mathfrak{Re}\lambda_N}
			\end{bmatrix}\mathrm{H} \mathbf{w} \\
			&=\Lambda\cdot \begin{bmatrix}e^{\lambda_1t}\sqrt{-2\mathfrak{Re}\lambda_1}&& \\ &\ddots  & \\ &&e^{\overline{\lambda_2}t}\sqrt{-2\mathfrak{Re}\lambda_N}
			\end{bmatrix}\mathrm{H} \mathbf{w}. 
		\end{align}
	\end{subequations}
	By construction of the matrix $\mathrm{H}$ in Eq.~\eqref{eq: def covariance matrix proposition}-\eqref{Eq_H2}, the product $\mathrm{H}\mathbf{w}$ 
is given component-wise, for $j=2,\ldots,N$, as:
	\begin{equation}
		\left[\mathrm{H}\mathbf{w}\right]_j=\begin{cases}w_j, \text{ if } j\in I_r \text{ or, } j\in I_+, \\  w_{j-1}, \text{ if }\lambda_j=\overline{\lambda_{j-1}}, \end{cases}
	\end{equation}
	while $\left[\mathrm{H}\mathbf{w}\right]_1=w_1$. This implies in particular that $\left[\mathrm{H}\mathbf{w}\right]_j=\left[\mathrm{H}\mathbf{w}\right]_{j+1}$ whenever $\lambda_j=\overline{\lambda_{j+1}}$. As a consequence, we get 
	\bea	
	\Lambda\cdot e^{\mathrm{D}t}\Sigma \mathbf{w}=&\sum_{j\in I_r}\alpha_j^{1/2}\beta_j^{1/2}e^{\lambda_jt}\sqrt{-2\lambda_j}w_j+\sum_{j\in I_+}\alpha_j^{1/2}\beta_j^{1/2}e^{\lambda_jt}\sqrt{-2\mathfrak{Re}\lambda_j}w_j\\
	&\hspace{0.7cm}+\sum_{j\in I_+}\overline{\alpha_j^{1/2}\beta_j^{1/2}e^{\lambda_jt}}\sqrt{-2\mathfrak{Re}\lambda_j}w_j,
	\eea
	which shows that $\Lambda\cdot e^{\mathrm{D}t}\Sigma$ is a real-valued quantity, and thus  for any realization of the $N$-dimensional Wiener process $W_t$, the product $\Lambda \cdot e^{\mathrm{D}(t-s)}\Sigma \dd W_s$ is real and hence, $\Lambda \cdot \mathbf{q}(t)$ is also real for every $t$.

	Finally, we are left with showing that the memory kernel $\mathcal{K}$ of the WL-equation  coincides with that of the memory term in Eq.~\eqref{eq: ito integral proposition}, when multiplied by the vector $\Lambda$. To 
do so, we exploit the expansion \eqref{eq: last equality kernel proof} of 
$\mathcal{K}$ for this comparison, namely using the expression of $\mathbf{R}$ in \eqref{Eq_Rexpression}, we observe that 
	\bea
	\Lambda\cdot \int _{0}^te^{\mathrm{D}(t-s)}\mathbf{R}(x(s))\dd s &=\CCC^{\yy}(x(s))\cdot  \sum_{j\in I_r}e^{\lambda_j (t-s)}\alpha_{j}\bigg \langle \nabla\psi _j (\yy)  \bigg \rangle\\ 
	&\hspace{-1.9cm}+\CCC^{\yy}(x(s))\cdot  \sum_{j\in I_+}e^{\lambda_j (t-s)}\alpha_{j}\bigg \langle \nabla\psi _j (\yy)  \bigg \rangle +\CCC^{\yy}(x(s))\cdot  \sum_{j\in I_+}\overline{e^{\lambda_j (t-s)}\alpha_{j}\bigg \langle \nabla\psi _j (\yy)  \bigg \rangle} \\ 
	=& \CCC^{\yy}(x(s))\cdot \sum_{j=1}^{N}\mathfrak{Re}\bigg(e^{\lambda_j(t-s)}\alpha_j\bigg \langle \nabla\psi _j (\yy)  \bigg \rangle\bigg),
	\eea
	which indeed coincides with the expression of  $\mathcal{K}$, as desired.  The proof is complete. 
\end{proof}

\section{Semigroup Property of the Projected Koopman Operator Family}{\label{Markovianity}

It was shown in \cite[Theorem A]{Chekroun2014} that projection onto a reduced state space is closely related with a coarse graining of the (full) probability transitions on the original system's attractor, while \cite[Theorem 2]{chekroun2019c} dealt recently with the impact of such a projection in terms of reduction of the Koopman semigroup. 
In \cite{chekroun2019c}, the authors proposed a criterion based on the spectral theory of Markov semigroups to ascertain whether the reduced state 
space associated with a given projection can fully explain the statistics 
of the desired variables. This approach provides potential insights into the need for modeling non-Markovian effects by inspecting the loss of the 
semigroup property, as explained below. 

Moreover, it follows from \cite{chekroun2019c} that the analysis of correlation functions is not only of physical interest but also of methodological utility. Correlation functions can be defined by means of the Koopman 
operator or, dually, by means of the transfer operator's providing the solution of the Liouville equation. 

Let $\mu$ denote an ergodic invariant measure of the system and take two observables $\Phi_1,\Phi_2$ in the space $L^2_{\mu}$ of zero-mean functions that are square-integrable with respect to $\mu$. Assume furthermore that the spectrum of the operator $\LLL$ in  $L^2_{\mu}$ is a pure point spectrum, given by the eigenvalues $\{\lambda _j\}_{j=1}^{\infty}$ and their associated eigenfunctions $\{\psi_j\}_{j=1}^{\infty}$, where the eigenvalues are ordered by their decreasing real parts. 

Then, the correlation function $C_{\Phi_1,\Phi_2}(t)$ between the functions $\Phi_1$ and $\Phi_2$ is given by:
\begin{equation}\label{spectral correlations0}
C_{\Phi_1,\Phi_2}(t) = \int  \Phi_1\cdot e^{t\LLL }\Phi_2 \dd \mu = \int  e^{t\mathcal{L}^{\ast}}\Phi_1\cdot \Phi_2 \dd \mu
\end{equation}
and it can be expanded, formally, as
\be\label{spectral correlations}
C_{\Phi_1,\Phi_2}(t) = \sum_{j=1}^{\infty}e^{\lambda _jt}\left \langle \Phi_1,\psi_j \right \rangle _{\mu}\left \langle \psi_j^{\ast},\Phi_2 \right \rangle _{\mu}.
\ee
The dual operators in \eqref{spectral correlations0} and the adjoint eigenvectors  in \eqref{spectral correlations} are indicated by the superscript $(\cdot)^\ast$, while $\langle \cdot, \cdot \rangle _{\mu}$ denotes the inner product. We refer to \cite[Corollary 1]{chekroun2019c} for a proof of \eqref{spectral correlations} in the context of Markov semigroups. The proof actually applies to the case of the Koopman semigroups considered here as long as the Koopman semigroup $U_t$ defined by \eqref{Def_Koopman} is a strongly continuous semigroup in $L^2_\mu$. The  RHS of Eq.~\eqref{spectral correlations} consists of a linear combination of exponential 
terms whose coefficients are calculated by projecting $\Phi_1$ and $\Phi_2$ onto the corresponding eigenspaces. These coefficients weight each exponential function and they can become exceedingly large if the Koopman operator deviates very much from normality \cite{trefethen2005}.  Note also 
that the set of eigenvalues $\lambda_j$'s play a key role in defining the 
response of the system to perturbations \cite{Lucarini2018JSP,Tantet2018}.

The interactions between the resolved and hidden variables that are modeled by the Dyson expansion of the Koopman operator in Sec.~\ref{ssec:MZ} may introduce memory effects into the closed, reduced model for the $\xx$-variables, as given by Eqs.~\eqref{lw parametrization}--\eqref{wl parametrization additive}. In certain situations, such memory effects can be neglected, even in the absence of exact slaving relationships between the resolved and hidden variables \cite{CLM19_closure}. But the loss of slaving 
relationships requires, in general, an explicit representation of memory effects \cite{chekroun2016emergence} to achieve an efficient model reduction.

Furthermore it was shown in \cite{Chekroun2014, chekroun2019c} that the reduction of the Koopman semigroup to observables that act only on the reduced state space leads, in most circumstances, to a family of operators that, while Markovian, no longer satisfy the semigroup property. One might 
then ask to which extent this loss of the semigroup property arising from 
the reduction, and the related emergence of memory effects, are crucial for providing a faithful reduced model of the observed variables.

When considering reduced state spaces obtained by projection, along with observables $\Phi_1$ and $\Phi_2$ defined on them, Theorem~2 in \cite{chekroun2019c} shows the existence of a family of Markov operators $\{\mathcal{T}_t\}_{t\geq 0}$ that satisfies:
\begin{equation}\label{criterion memory}
\int \Phi_1 \cdot \mathcal{T}_t\Phi_2\dd \mu_{\xx} = \int _{0}^t[\Phi_1\circ \pi_{\xx}]\cdot e^{\LLL t}[\Phi_2\circ \pi _{\xx}] \dd \mu = C_{\displaystyle{(\Phi_1\circ \pi_{\xx},\Phi_2 \circ \pi_{\xx}})}(t),
\end{equation}
for every $t\geq0$, where ${\displaystyle{\pi_{\xx}}}$ is the canonical projection onto the reduced subspace and $\mu_{\xx}$ is the \emph{disintegrated} or {\em sample measure} associated with $\pi_{\xx}$; see \cite[Remark 3]{chekroun2019c}. However, due to the projection, the semigroup property is lost, namely, $\mathcal{T}_{s}\mathcal{T}_t\neq \mathcal{T}_{t+s}$ for some $t,s$. 

Following the reasoning above, one can establish a criterion for the need 
to model a memory contribution when performing the model reduction. Formally, if there exist $\tau > 0$ and $T\in \mathbb{N}$ such that, for every 
$t\in \{k\tau \in \mathbb{R} : 0\leq k \leq T \}$, we have
\begin{equation}\label{criterion memory effects}
C_{\displaystyle{(\Phi_1\circ \pi_{\xx},\Phi_2 \circ \pi_{\xx}})}(t) = \int \Phi_1\cdot \mathcal{T}_{t}\Phi_2\dd \mu = \int \Phi_1\cdot \left(\mathcal{T}_{\tau}\right)^k\Phi_2\dd \mu,
\end{equation}
and one can say that the semigroup is preserved, to some extent, depending on how large $T$ can be in Eq.~\eqref{criterion memory effects} above. Other such criteria are available in the context of mutually dual Koopman 
and transfer operators. Thus, A. Tantet and coauthors \cite{Tantet2015} had already considered empirical ways of quantifying the loss of the semigroup property in reduced dimensions.

The interpretation of $\tau$ comes from the practical implementation of the methodology and it is usually referred to as the \emph{transition time}. Indeed, numerically, the approximation of such Markov operators is done by Ulam's method, by projecting them onto a finite basis, typically using the characteristic functions of certain domains of phase space. Then, the transitions between domains --- after an adequate transition time $\tau$ --- are counted to obtain matrix estimates of the operator $\mathcal{T}_{\tau}$ acting on the reduced phase space. Hence, one seeks a suitably 
small, or large \cite{Tantet2018}, transition time to obtain the best candidate for applying Eq.~\eqref{criterion memory effects}, see also \cite[Sec.~3.3]{chekroun2019c}. Thus, Eq.~(\ref{criterion memory effects}) can be implemented in practice this way in order to (potentially) 
reconstruct correlation functions on the whole phase space. A very simple 
illustration of such a transfer operator calculation is given in \cite{Vissio.ea.2020}.

\section{It\^o Integration of the MSM}\label{ITO}

In the main text, we proposed a solution of the MSM given by Eqs.~\eqref{Eq_MSM} using the Dyson expansion for the linear operators involved in the backward Kolmogorov equation. %% Working there at the level of the observables, one can understand the action of the nonlinear flows as a linear 
advection acting on functions. Therefore, we substituted nonlinear ordinary differential equations for a partial differential equation, for the sake of having linear operators in hand.
 The same solution can be attained by direct integration of the MSM in the form \eqref{Eq_MSM}. We  convolute, in the It\^o sense, Eq.~(\ref{msm2}) to find an explicit solution for $\rr(t)$ when $d_1 = d_2$:
\begin{equation} \label{eq:Ito}
\rr (t) = e^{-\mathrm{D}t}\rr (0) + \int _{0}^te^{-\mathrm{D}(t-s)} \Sigma \dd W_s + \epsilon \int _{0}^{t}e^{-\mathrm{D}(t-s)}\mathrm{C}\xx(s)\dd s;
\end{equation}
here, $\rr(0)$ indicates an initial state that can be assumed to be distributed in a prescribed way. For the more general, nonlinear MSMs considered there, see \cite[Proposition 3.3]{kondrashovdata2015}. 

By substituting the expression \eqref{eq:Ito} into Eq.~(\ref{msm1}), we find an exact expression for the evolution of $\xx (t)$:
\begin{equation}\label{eq:ITO}
\dot{\xx}(t) =\FFF (\xx(t)) + \epsilon e^{-\mathrm{D}t}\rr (0) + \epsilon \int _{0}^te^{-\mathrm{D}(t-s)} \Sigma \dd W_s + \epsilon ^2 \int _{0}^{t}e^{-\mathrm{D}(t-s)}\mathrm{C}\xx(s)\dd s,
\end{equation}
in which the memory effects in the fourth term are of second order in $\epsilon$. Note that the $\epsilon$-order terms arise from a noise realization in the decoupled regime, whereas the memory term is exclusively due to the coupling of the main variables with the hidden ones. Hence, the only degree of freedom left is the distribution of the initial states $\rr(0)$.

\section{The coupled L84--L63 System}\label{L84L63}
The EMR methodology's ability to capture the statistics of low-dimensional dynamical systems was illustrated in \cite{Kravtsov2005}, where the authors considered the L63 system \cite{lorenzdeterministic1963} as a test case in which the phase space can be fully sampled. Moreover, provided that the integration time step is short enough, the parameters of the underlying model can be fully captured with a high degree of confidence. 

Here, we repeat the analysis of \cite{Kravtsov2005} to illustrate the effectiveness of EMR in capturing statistical and dynamical properties in an 
extended system. The model we consider is the result of coupling the $\X = (X, Y, Z)$ variables of the L84 system \cite{Lorenz1984a} with the $\x = (x,y,z)$ variables of the L63 system \cite{lorenzdeterministic1963}, namely:
\begin{subequations}\label{eq: l84l63}
\begin{align}
\dot X & = -Y^2-Z^2-aX+a(F_0 + hx),\\
\dot Y & = XY - bXZ - Y +G, \\
\dot Z & = XZ+bXY-Z,\\
\dot x & = \tau s (y-x),\\
\dot y & = \tau (\rho x - y -xz),\\
\dot z & = \tau (xy-\beta z);
\end{align}
\end{subequations}
the parameter values are: $a=0.25,b=4,$ $F_0=8, G=1,$ and $s=10,$ $\rho = 28, \beta =8/3$, respectively. The parameter $h$ measures the strength of the coupling, while $\tau$ scales the rate of change in the L63 system and, therewith, the time scale ratio between the two subsystems.

 This system is a skew-product, in the sense of \cite{Sell.1971}, since the coupling is one-way only, with the L63 system driving the L84 dynamics. Hence, one has --- as noted in \cite{Vissio2018b} --- a fully Markovian parametrization of the L63 variables. Furthermore, the correlation function that defines the stochastic noise $\eta(t)$ can be further expanded 
and simplified,with respect to Eq.~\eqref{eq:moments}.  One can, in fact, 
write explicitly:
\begin{equation}
C(\eta (0),\eta(t))=\left \langle (x(0),0,0)\cdot (x(t),0,0) \right \rangle,
\end{equation}
where the angular brackets $\langle \cdot \rangle$ indicate averages with 
respect to the physical measure associated with the L63 system. Since L84 
does not feed back into L63, the evolution of $\x(t)$ only obeys the dynamics of L63, and thus the decorrelation of the noise scales with $\tau$. %%  like that of $x(t)$. 

In most of the numerical experiments, the time scale separation between the two systems is $\tau=5$. The relevance of this time scale parameter was investigated in previous work \cite{Vissio2018b}. Here, we focus on the effects of the coupling strength $h$, and we shall study the cases of $h=0.25$ and $0.025$. Partial observations only will be used in these experiments, by sampling the three-dimensional outputs of the L84 system. Then, the observed tendencies are regressed and sequentially layered following the EMR approach, as explained in Sect.~\ref{MSMEMR}.

\subsection{EMR Outputs} \label{app:out}

The L84--L63 model is integrated for 730 time units that correspond in L84 to 10 natural years, with a time step of $5\cdot 10^{-3}$ time units; two separate runs are made for the coupling strengths $h = 0.25$ and $h = 0.025$. These two full-model runs are used to train the corresponding 
EMR model versions, both of which use only the slow $\X$-variables and eliminate the fast $\x$-variables. Then, two separate full-model simulations are run, for testing purposes, over 7~300 time units, and the EMR model's output is compared with it, for the two parameter values. Below we show the main statistical outputs of the EMR methodology compared to the two 
reference integrations of the full model. The results for the two separate $h$-values are shown in Figs.~\ref{trajectory h=0.25}--\ref{acf h=0.25} and Figs.~\ref{trajectory h=0.025}--\ref{acf h=0.025}, respectively.

The region of phase space explored by the EMR model clearly coincides with the one visited by the full model, as seen in Figs.~\ref{trajectory h=0.25} and \ref{trajectory h=0.025}, and the relative occupancies within 
this region---as indicated by the smoothed PDFs shown in Figs.~\ref{pdf h=0.25} and \ref{pdf h=0.025}, respectively---agree very well.  The time scales are also well captured, as indicated by the good approximation of the autocorrelation functions, cf.~Figs.~\ref{acf h=0.25} and \ref{acf h=0.025}.   

Notice that, while the original L84--L63 system is purely deterministic, the EMR model includes white noise acting on the hidden layers of the learned model. This fact could suggest that a smoothing of the invariant measure is inevitable and that the EMR methodology may not be able to capture fractal geometries in phase space, since the EMR model does not satisfy the H\"ormander's hypoellipticity condition \cite{csg11, Hoermander.1967}. The numerical evidence in Figs.~\ref{trajectory h=0.25} and \ref{trajectory h=0.025}, however, illustrates a strikingly good approximation of the full model's attractor, including its very fine, and presumably fractal, structure.
	
Actually, \cite[Theorem~3.1 and Corollary~3.2]{kondrashovdata2015} provided sufficient conditions for the existence of a random attractor for a broad class of MSMs that are not subject to a non-degeneracy condition of H\"ormander type. In other words, one can have an MSM that possesses a random attractor and is thus dynamically quite stable, while exhibiting in a forward sense an invariant measure of the associated Fokker-Planck equation that is singular with respect to the Lebesgue measure. This mathematically rigorous result helps explain what is observed numerically not only in the present paper for the EMR of the L84--L63 model, but also in the case of the EMR model of the Lotka-Volterra example in \cite[Fig.~7]{kondrashovdata2015}.

Ulam's method was used on the  projection of the full $(\X, \x)$ phase space onto the $\X$ subspace to approximate the spectrum of the Koopman operator, since it can provide further information on the characteristics of 
the time series, beyond PDFs and correlation functions. The observed spectra  (red $\times$ symbols) using a coarse partition of phase space into $512$ nonintersecting boxes showed good agreement with the spectra based on the full model (blue open circles); see Fig.~\ref{ergodicity spectrum L84-L63}. This agreement confirms further that, at this level of coarse graining, the EMR model captures well the the characteristics of the full model's solutions.

\begin{figure}[H]
	\centering
	\begin{subfigure}[b]{0.49\textwidth}
		\centering
		\includegraphics[width=\textwidth]{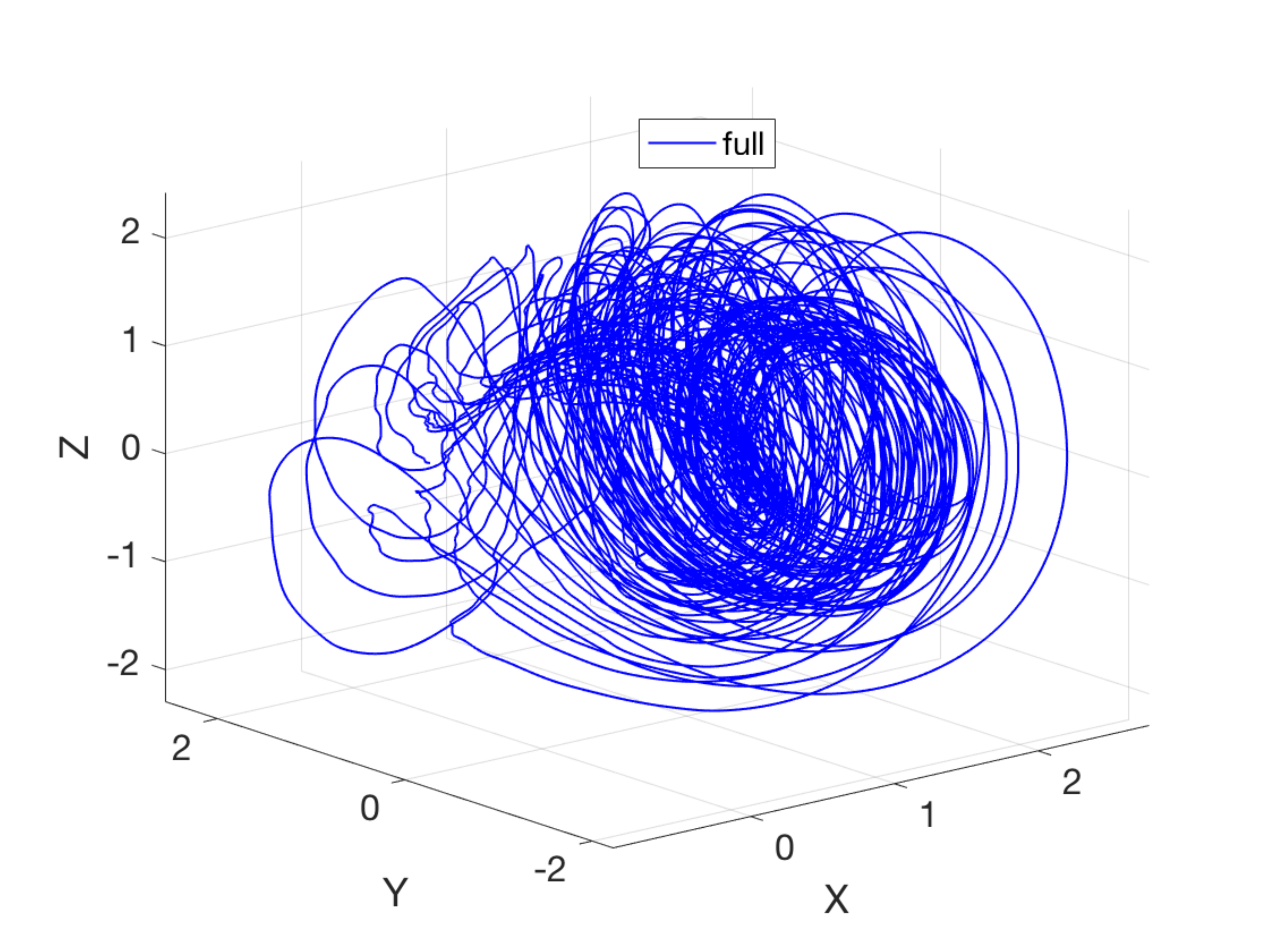}
		\caption{Full model}
	\end{subfigure}
	\hfill
	\begin{subfigure}[b]{0.49\textwidth}
		\centering
		\includegraphics[width=\textwidth]{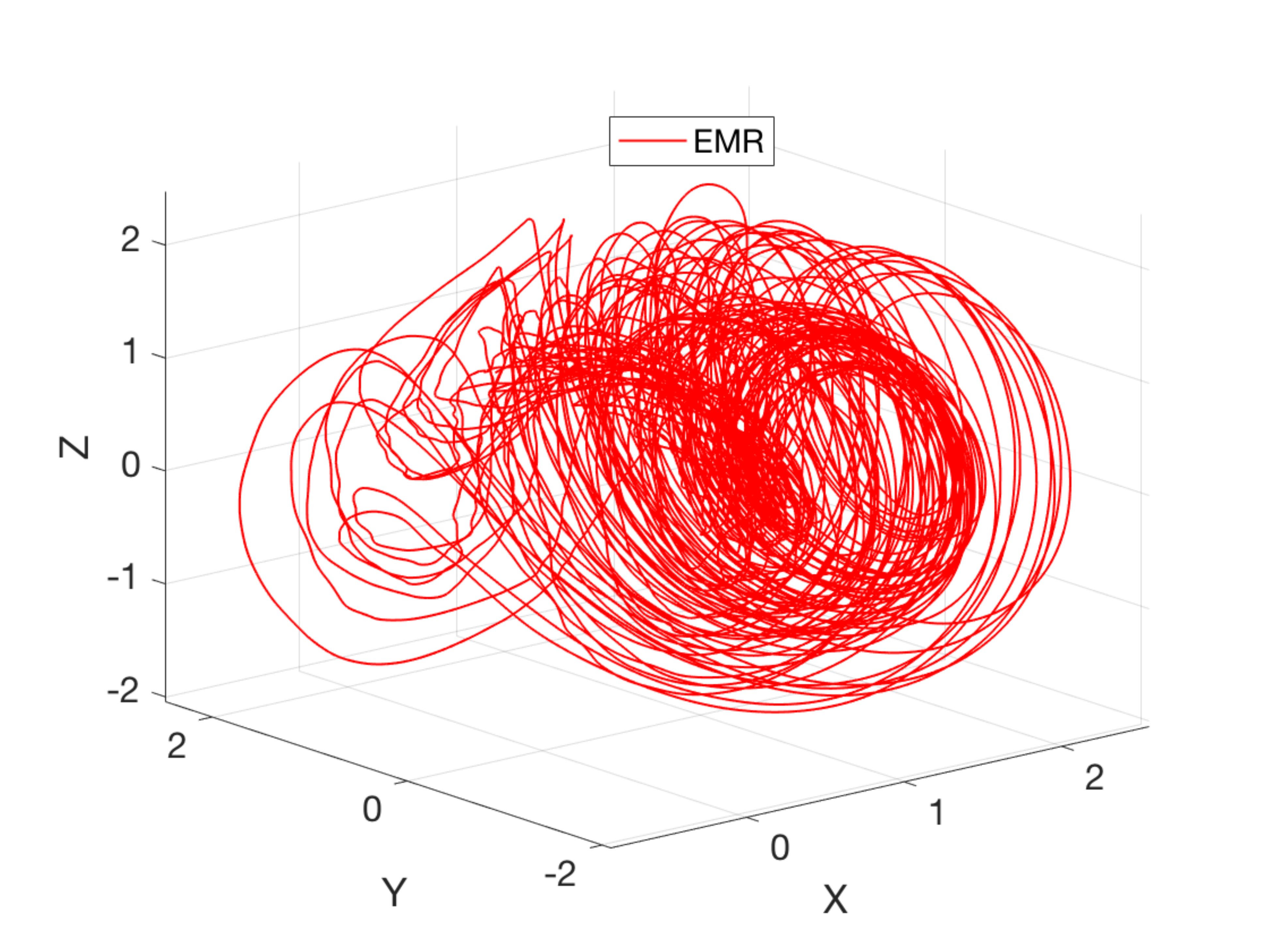}
		\caption{EMR model}
	\end{subfigure}
	\hfill
	\caption{\label{trajectory h=0.25} Trajectories of the L84--L63 model in the three-dimensional $(X,Y,Z)$ 
	phase space of the L84 model, for $h=0.25$ and 200 time units:
	(a) for the full L84--L63 model governed by Eqs.~\eqref{eq: l84l63} (blue); (b) for the EMR model (red).}
\end{figure}

\begin{figure}[H]
	\centering
	\begin{subfigure}[b]{0.3\textwidth}
		\centering
		\includegraphics[width=\textwidth]{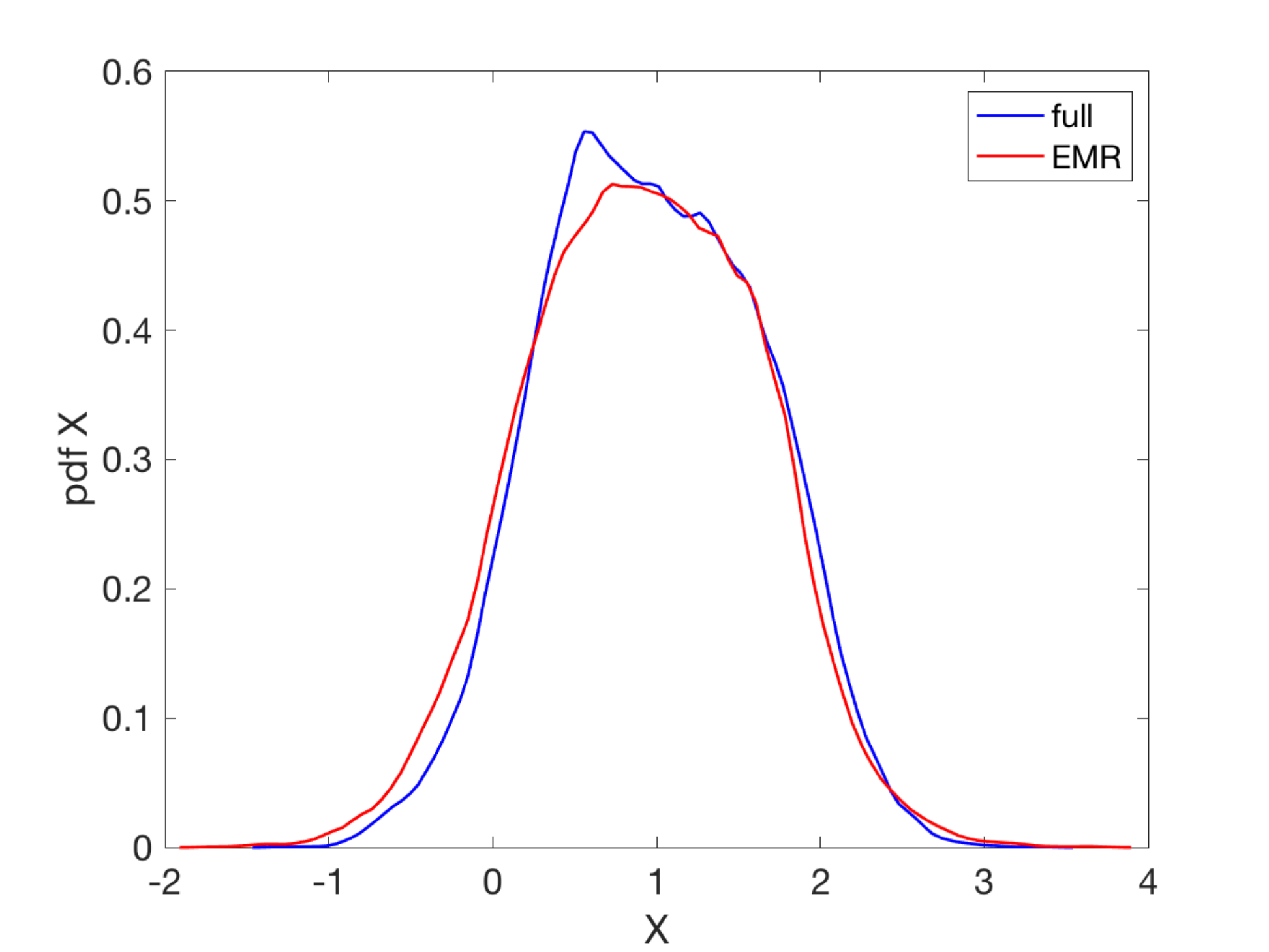}
		\caption{$X$-PDF}
	\end{subfigure}
	\hfill
	\begin{subfigure}[b]{0.3\textwidth}
	\centering
		\includegraphics[width=\textwidth]{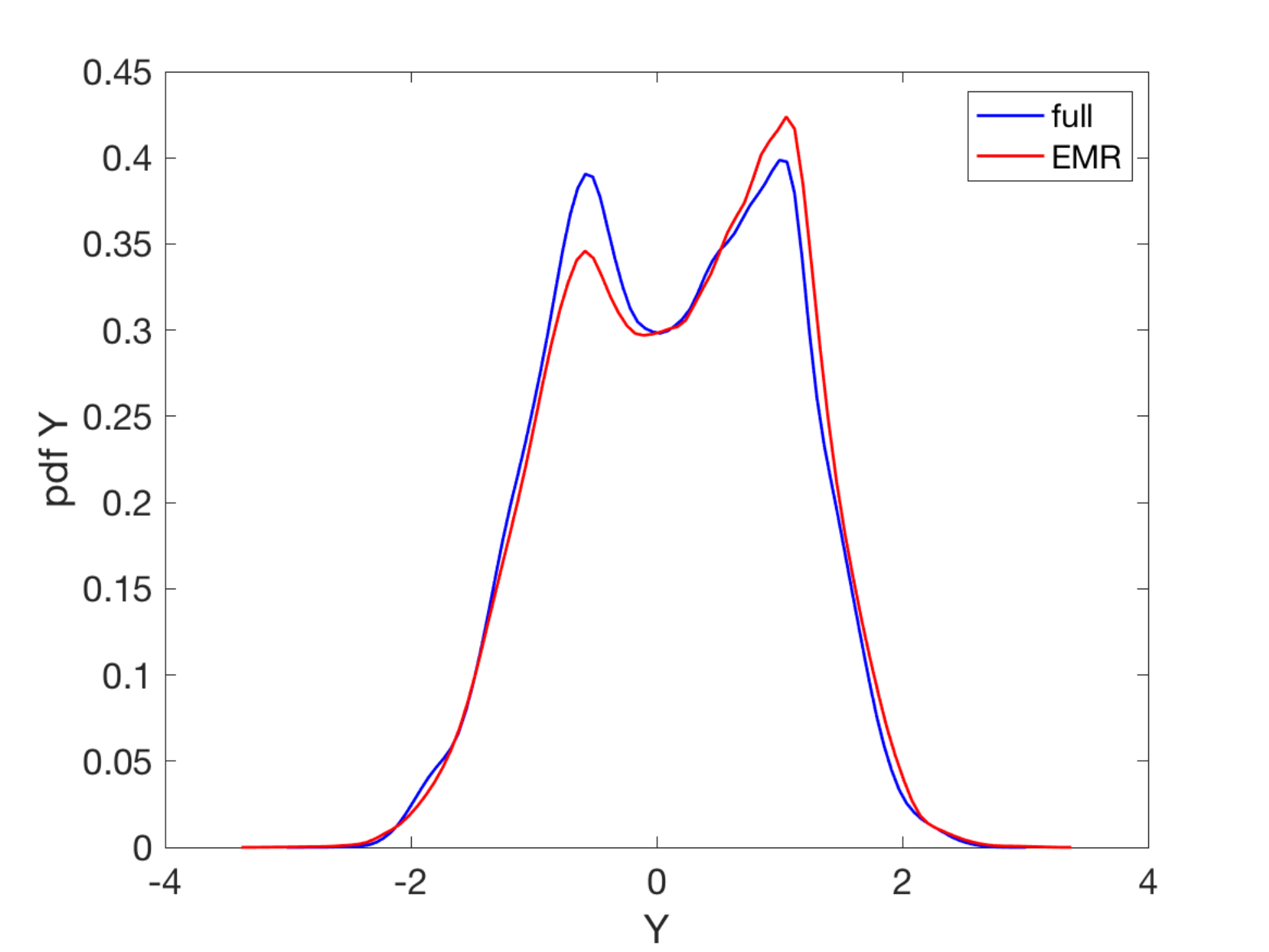}
		\caption{$Y$-PDF}
	\end{subfigure}
	\hfill
	\begin{subfigure}[b]{0.3\textwidth}
		\centering
		\includegraphics[width=\textwidth]{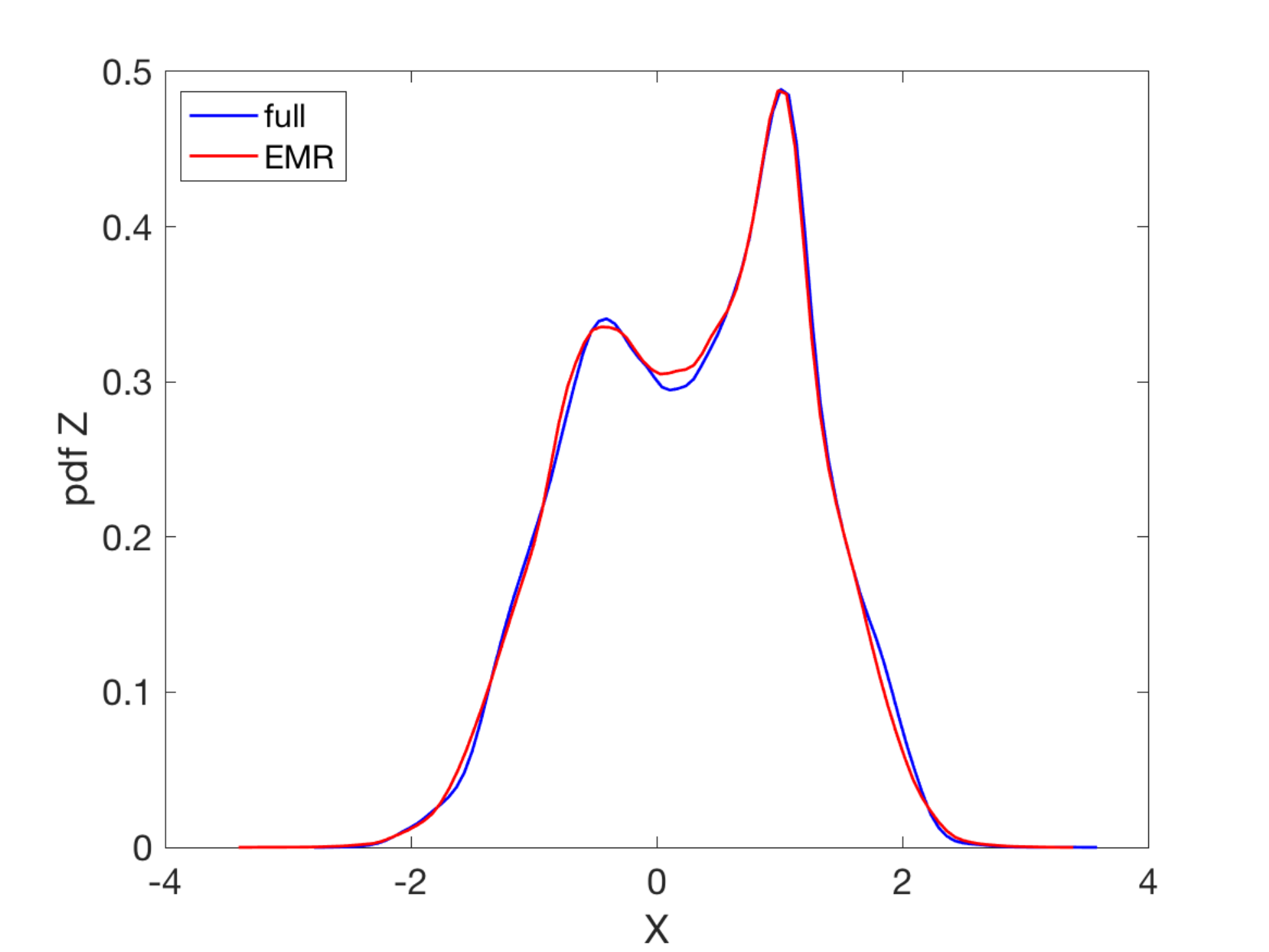}
		\caption{$Z$-PDF}
	\end{subfigure}
	\caption{\label{pdf h=0.25} Smoothed PDFs of the L84--L63 variables (a) $X,$  (b) $Y,$ and (c) $Z$ with a coupling strength of $h=0.25$. The blue curve corresponds to the full model; the red curve corresponds to the EMR model. These PDFs and those in Fig.~\ref{pdf h=0.025} were obtained by using the Matlab R2019a kernel smoothing function \textit{ksdensity}.
	} 
\end{figure}

\begin{figure}[H]
	\centering
	\begin{subfigure}[b]{0.3\textwidth}
		\centering
		\includegraphics[width=\textwidth]{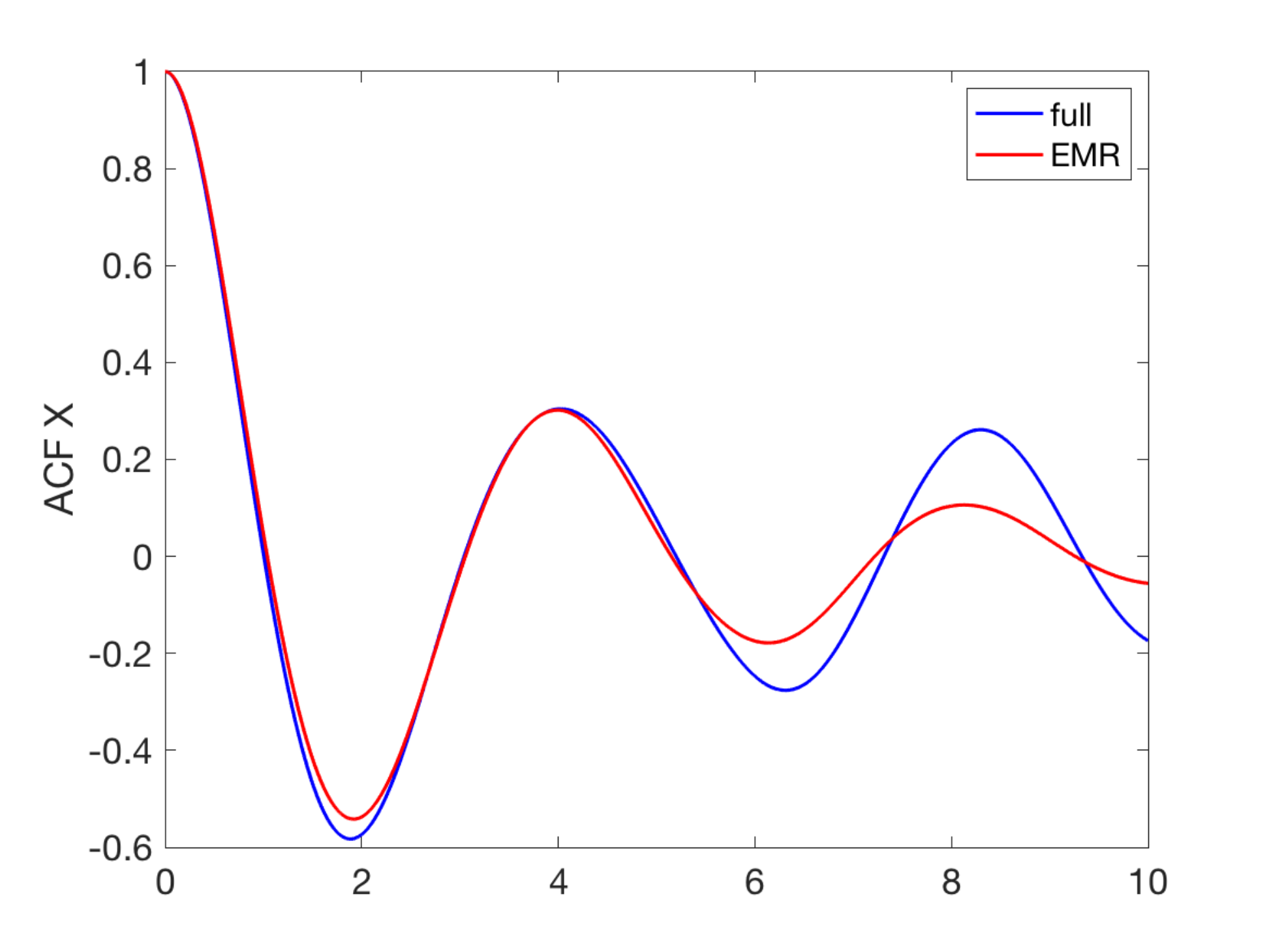}
		\caption{$X$-ACF}
	\end{subfigure}
	\hfill
	\begin{subfigure}[b]{0.3\textwidth}
		\centering
		\includegraphics[width=\textwidth]{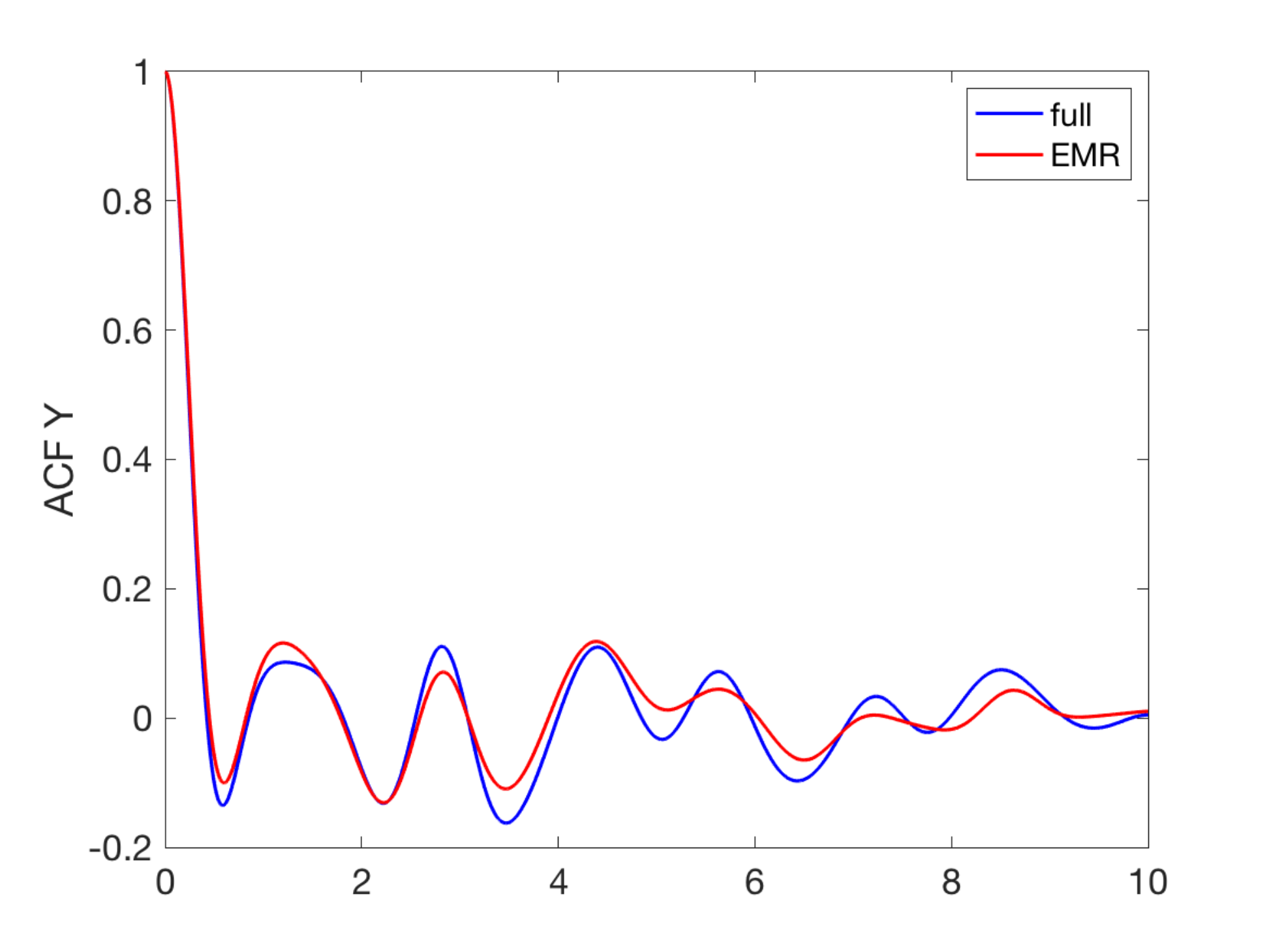}
		\caption{$Y$-ACF}
	\end{subfigure}
	\hfill
	\begin{subfigure}[b]{0.3\textwidth}
		\centering
		\includegraphics[width=\textwidth]{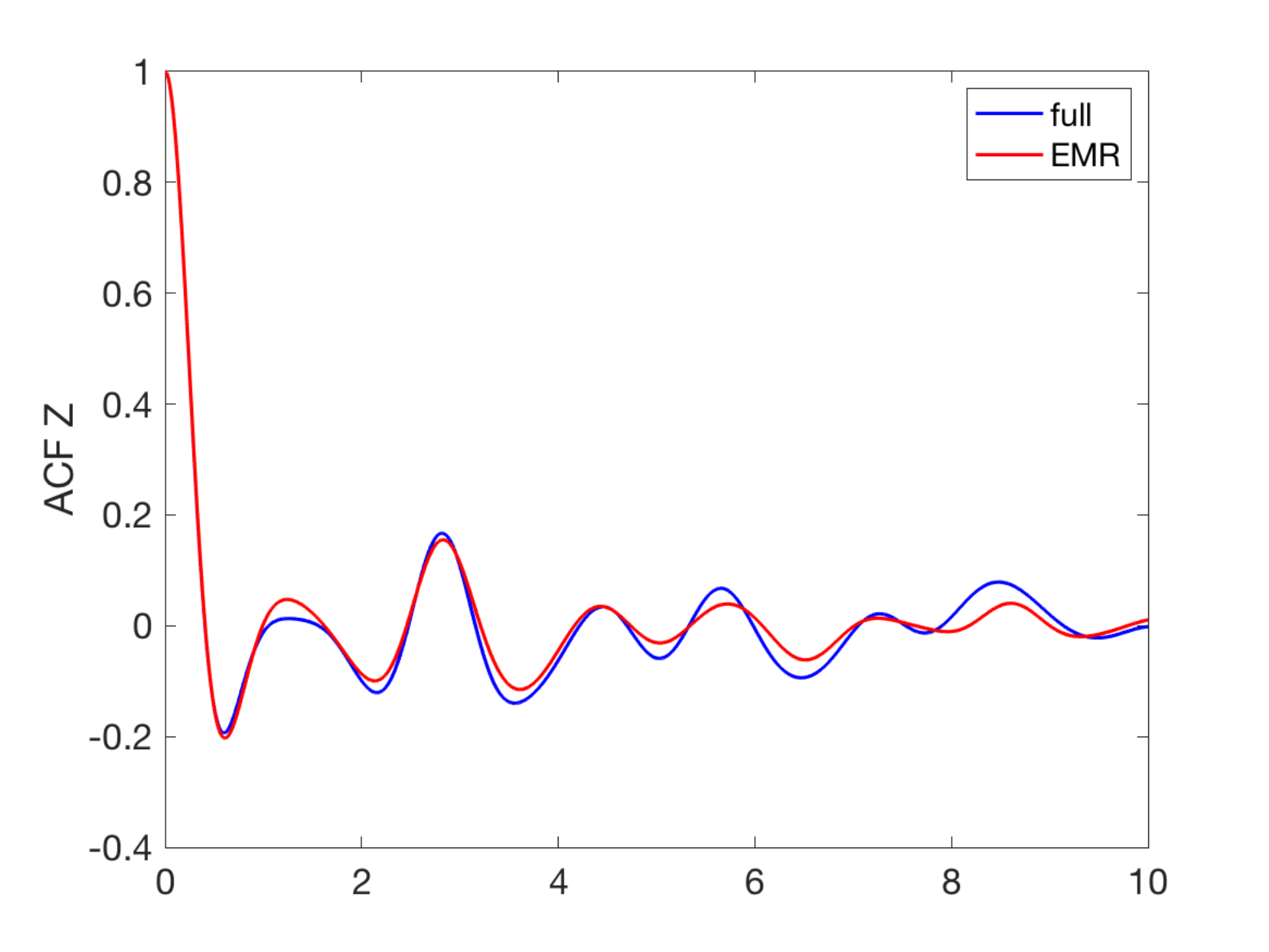}
		\caption{$Z$-ACF}
	\end{subfigure}
	\caption{\label{acf h=0.25} Autocorrelation functions (ACFs) of the L84--L63 variables 
	(a) $X,$  (b) $Y,$ and (c) $Z$ for a coupling strength of $h=0.25$.
	%%The blue curve corresponds to the data obtained by integrating Eq.~\eqref{eq: l84l63}, 
	%% the red curve corresponds to the EMR model.
	}
\end{figure}

\begin{figure}[H]
	\centering
	\begin{subfigure}[b]{0.49\textwidth}
		\centering
		\includegraphics[width=\textwidth]{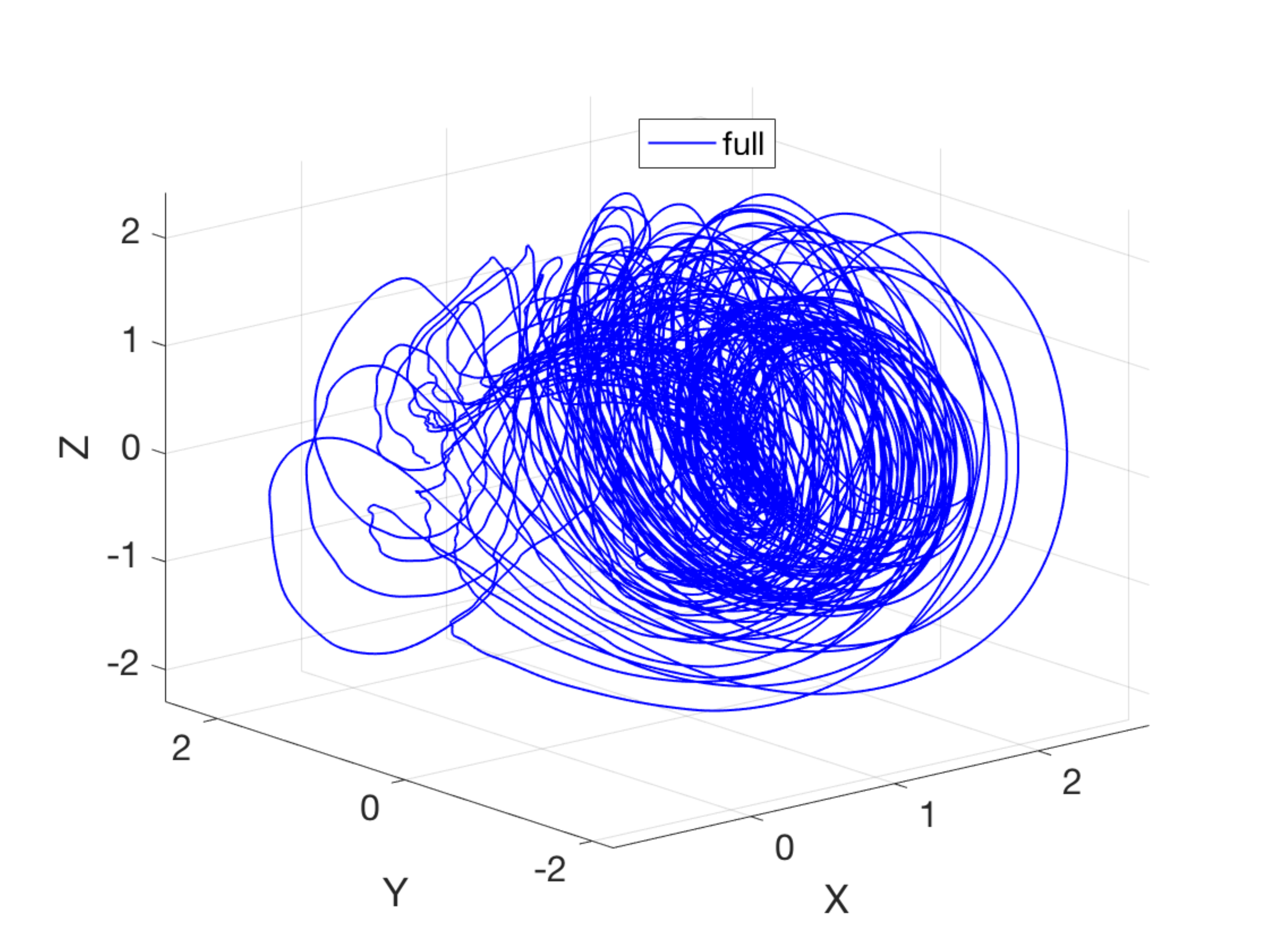}
		\caption{Full model}
	\end{subfigure}
	\hfill
	\begin{subfigure}[b]{0.49\textwidth}
		\centering
		\includegraphics[width=\textwidth]{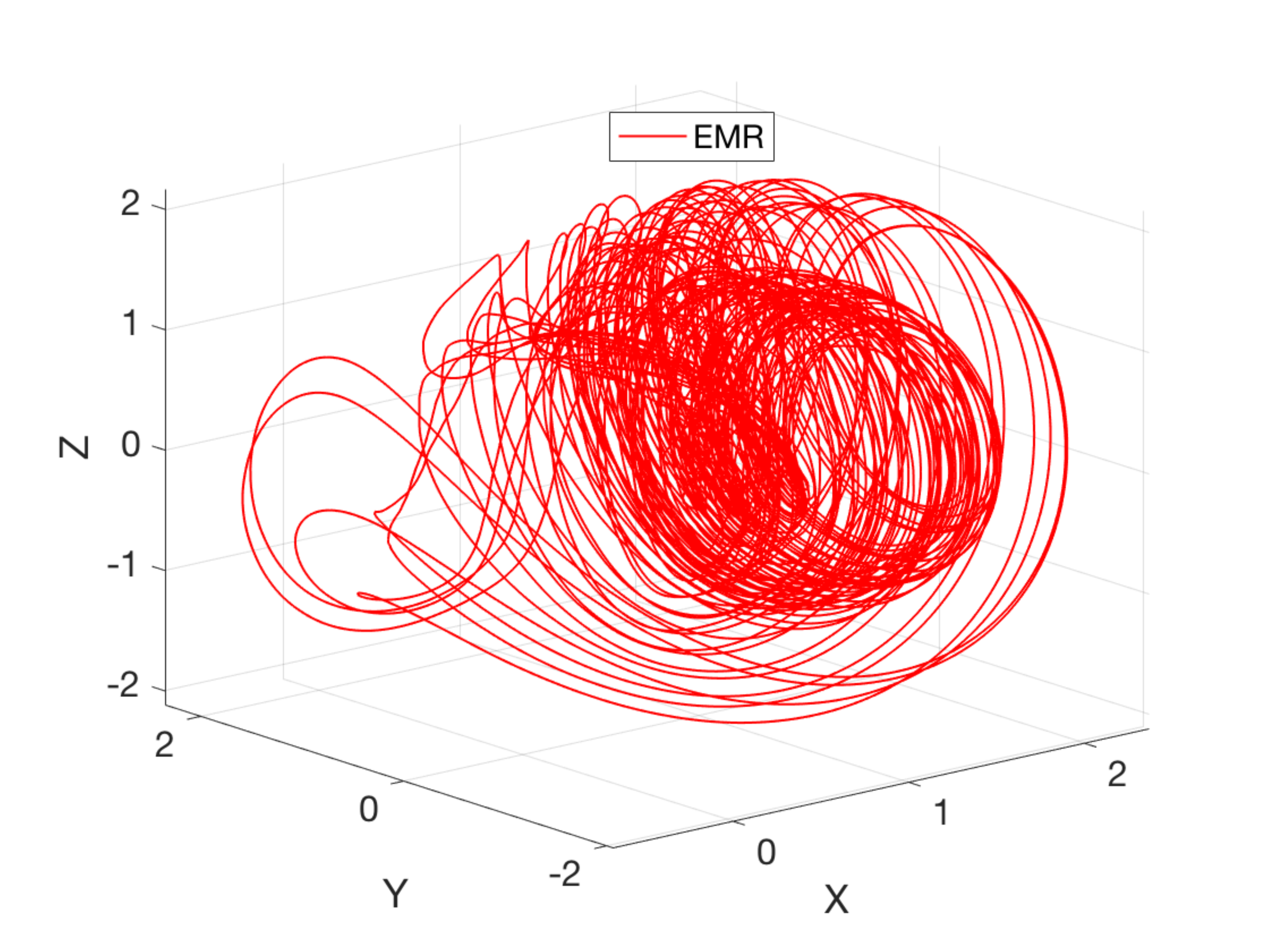}
		\caption{EMR model}
	\end{subfigure}
	\hfill
	\caption{\label{trajectory h=0.025} Example trajectories of the L84-L63 model on the $(X,Y,Z)$ domain with a coupling strength of $h = 0.025$ 
integrated for 200 time units. Subfigure (a) corresponds to the full model \eqref{eq: l84l63} (blue) and subfigure (b) refers to the EMR model (red).}
\end{figure}

\begin{figure}[H]
	\centering
	\begin{subfigure}[b]{0.3\textwidth}
		\centering
		\includegraphics[width=\textwidth]{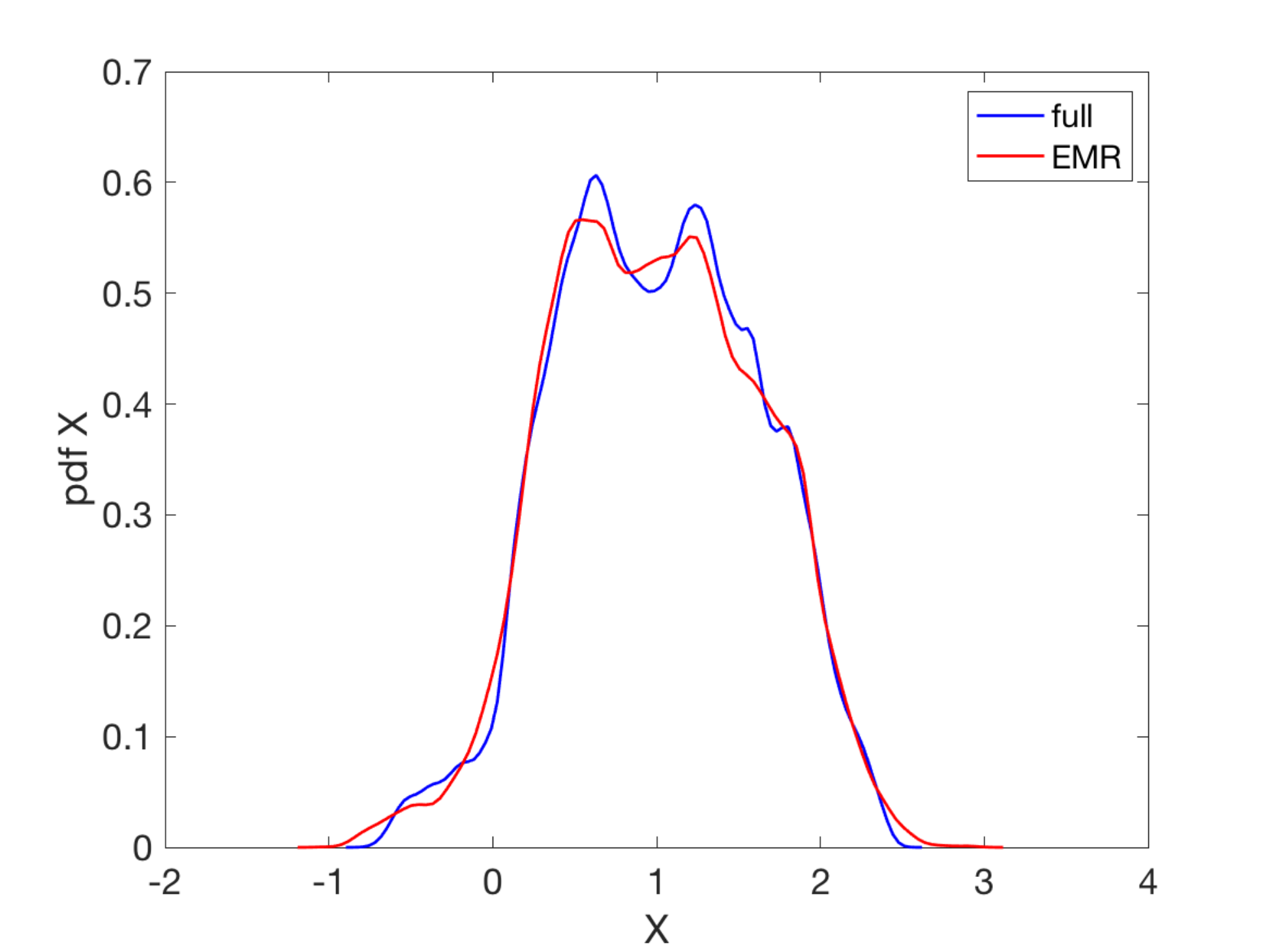}
		\caption{$X$-PDF}
	\end{subfigure}
	\hfill
	\begin{subfigure}[b]{0.3\textwidth}
		\centering
		\includegraphics[width=\textwidth]{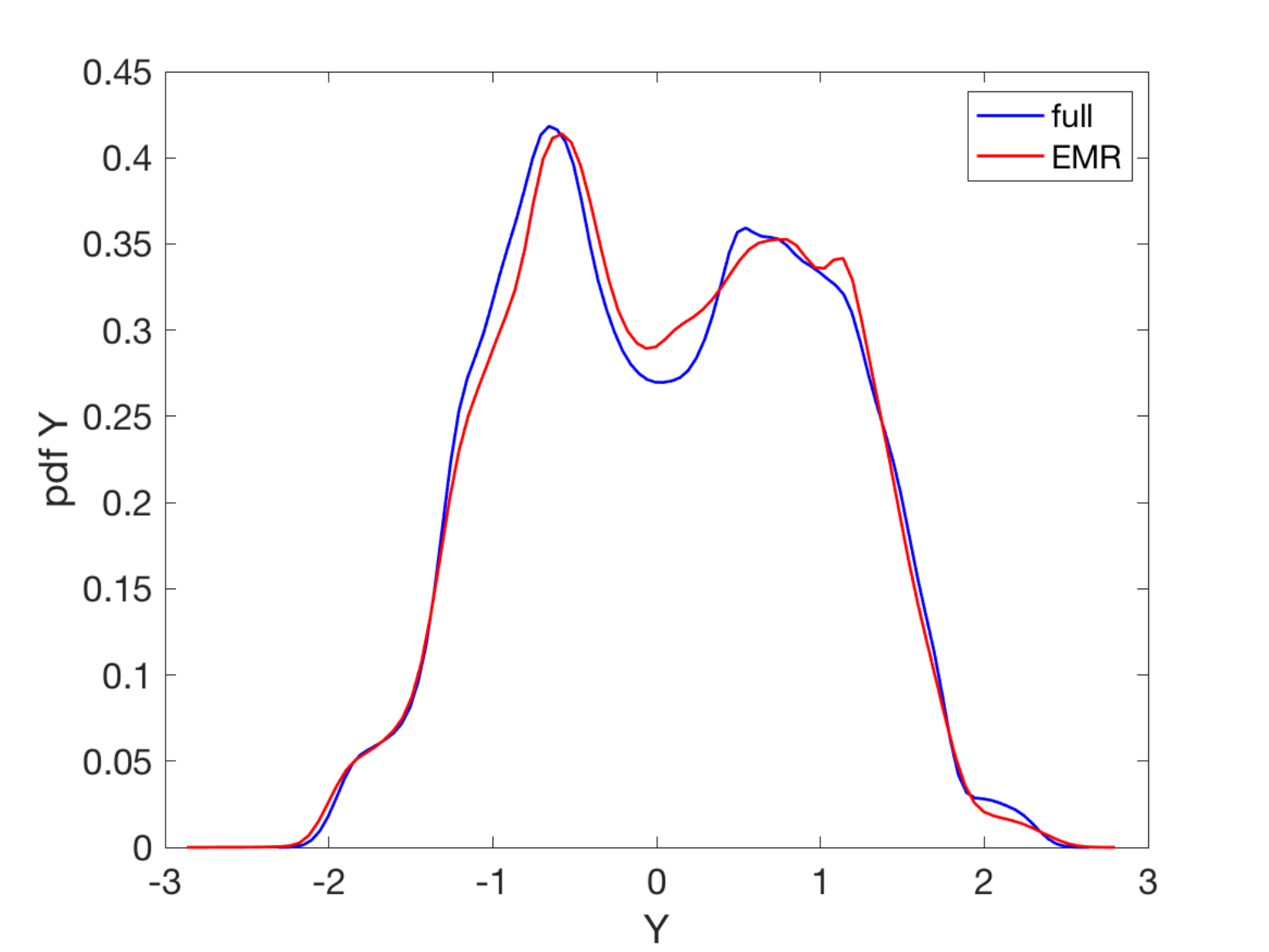}
		\caption{$Y$-PDF}
	\end{subfigure}
	\hfill
	\begin{subfigure}[b]{0.3\textwidth}
		\centering
		\includegraphics[width=\textwidth]{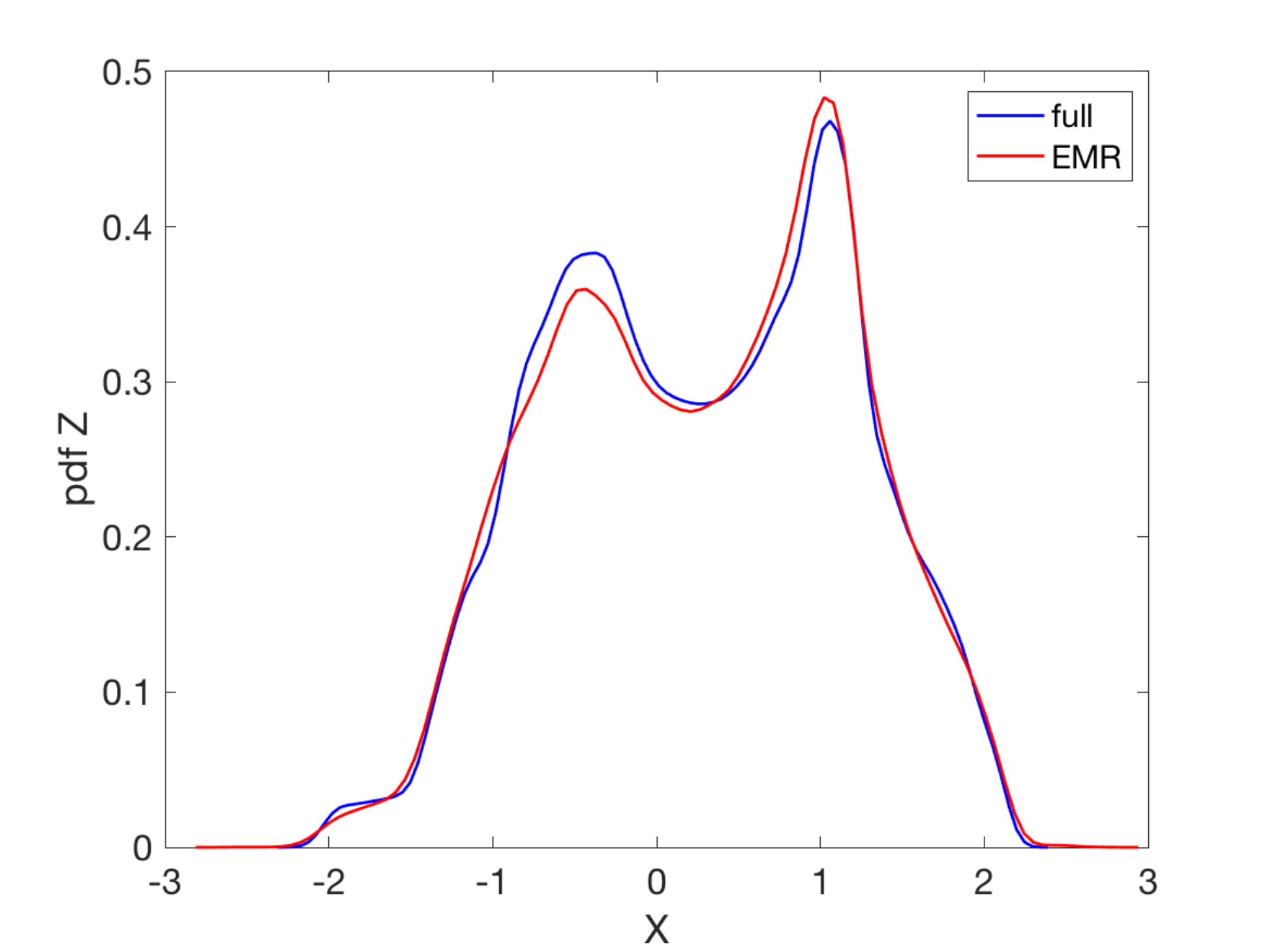}
		\caption{$Z$-PDF}
	\end{subfigure}
	\caption{\label{pdf h=0.025} Smoothed PDFs of the L84-L63 variables (a) $X$, (b) $Y$, and (c)   $Z$, with a coupling strength of $h=0.025$. The blue curve corresponds to the full model; the red curve corresponds to 
the EMR model.}
\end{figure}

\begin{figure}[H]
	\centering
	\begin{subfigure}[b]{0.3\textwidth}
		\centering
		\includegraphics[width=\textwidth]{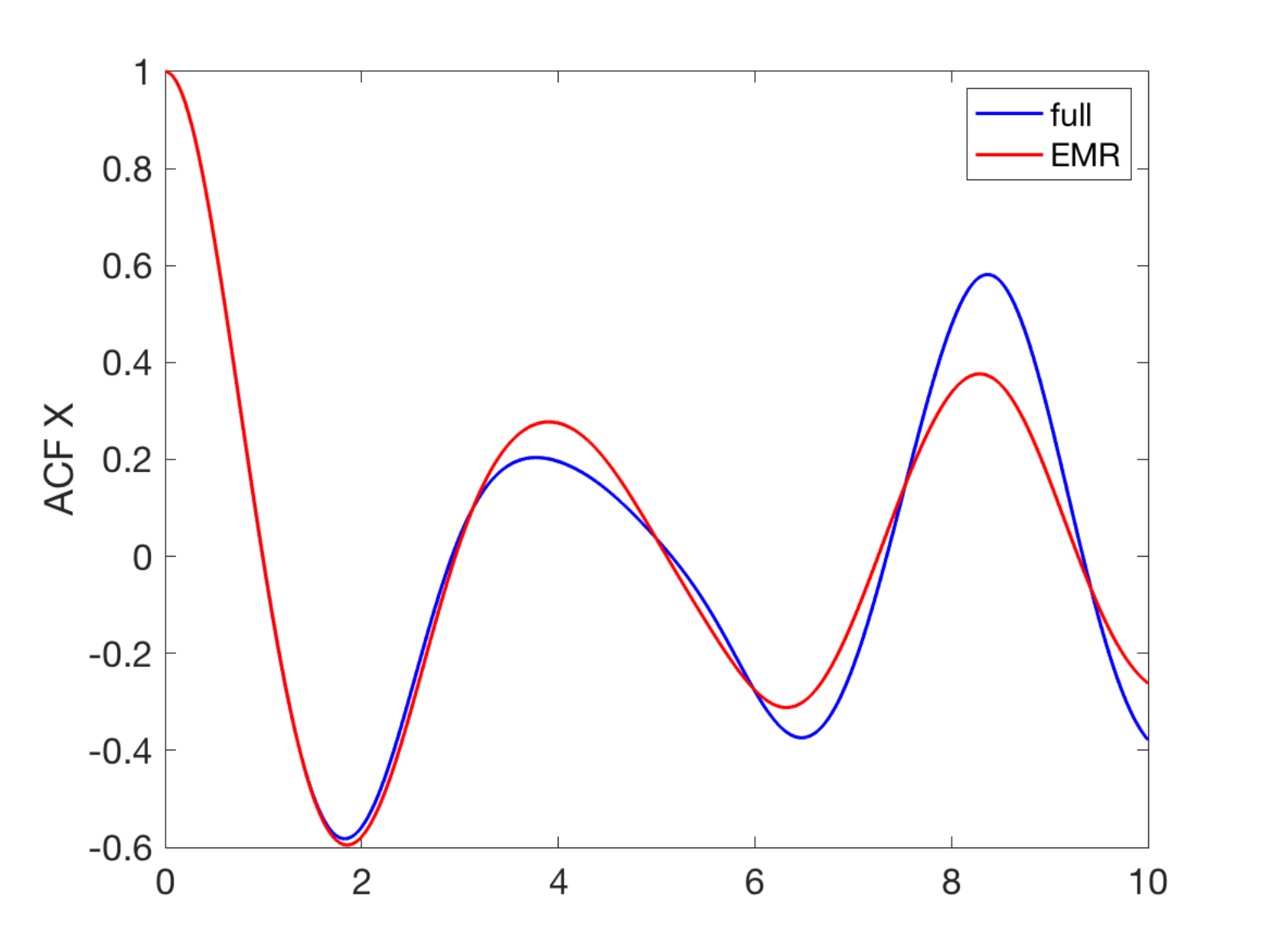}
		\caption{$X$-ACF}
	\end{subfigure}
	\hfill
	\begin{subfigure}[b]{0.3\textwidth}
		\centering
		\includegraphics[width=\textwidth]{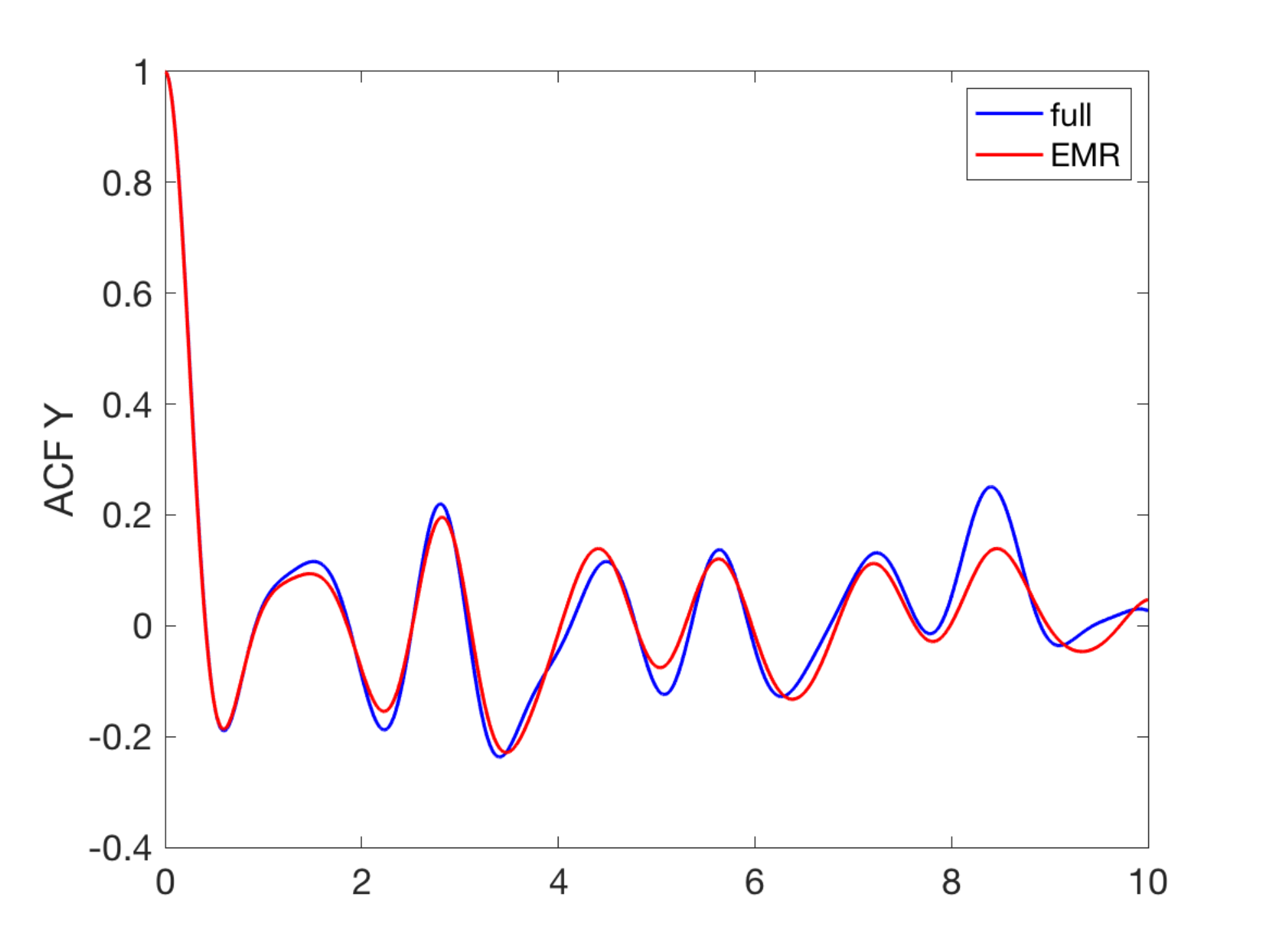}
		\caption{$Y$-ACF}
	\end{subfigure}
	\hfill
	\begin{subfigure}[b]{0.3\textwidth}
		\centering
		\includegraphics[width=\textwidth]{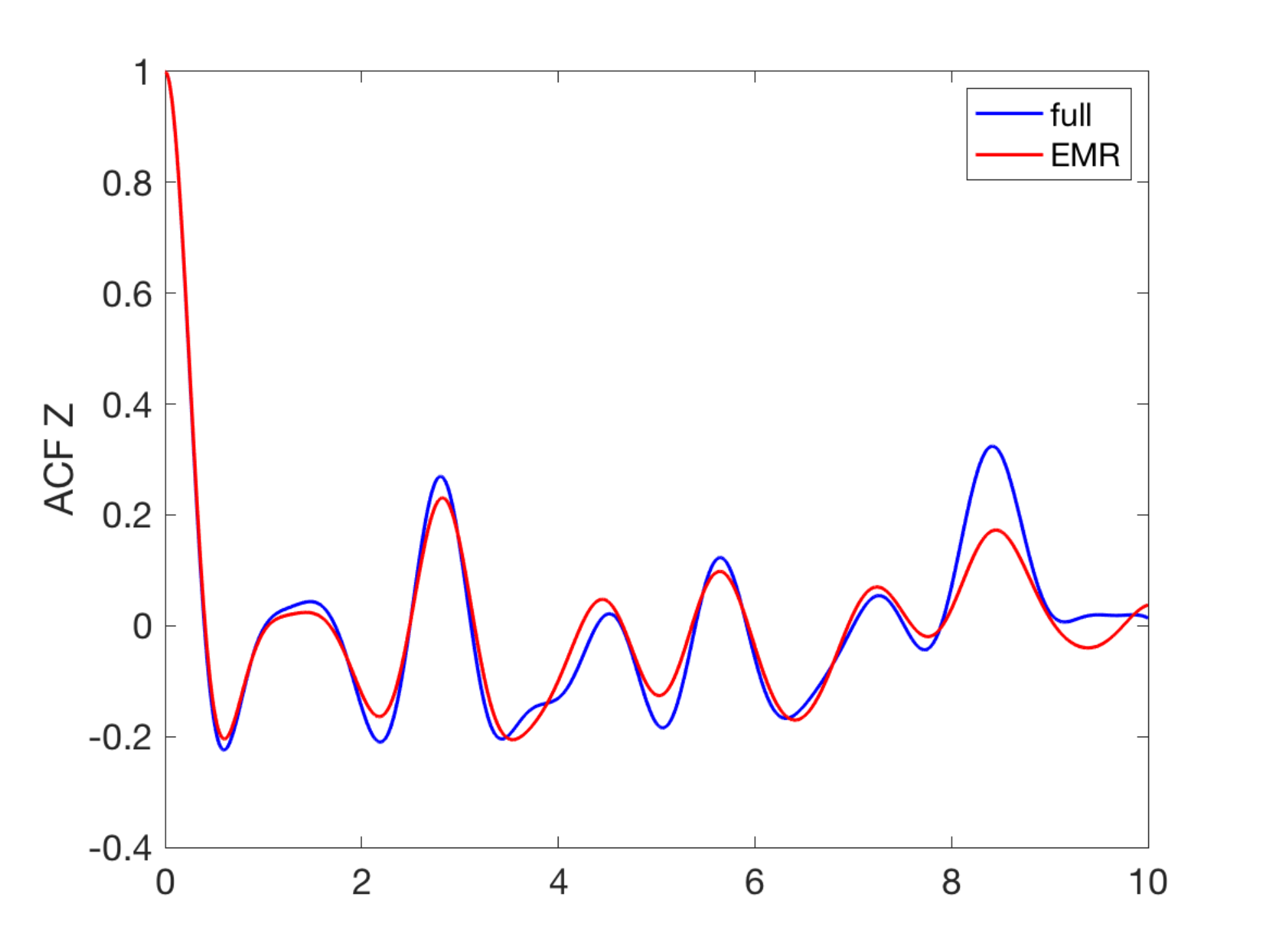}
		\caption{$Z$-ACF}
	\end{subfigure}
	\caption{\label{acf h=0.025} ACFs of the L84-L63 variables (a) $X$, (b) $Y$, and (c) $Z$ for $h = 0.025$. The blue curve corresponds to the full model; the red curve corresponds to the EMR model.}  
\end{figure}

\begin{figure}[H]
	\centering
	\begin{subfigure}[b]{0.49\textwidth}
		\centering
		\includegraphics[width=\textwidth]{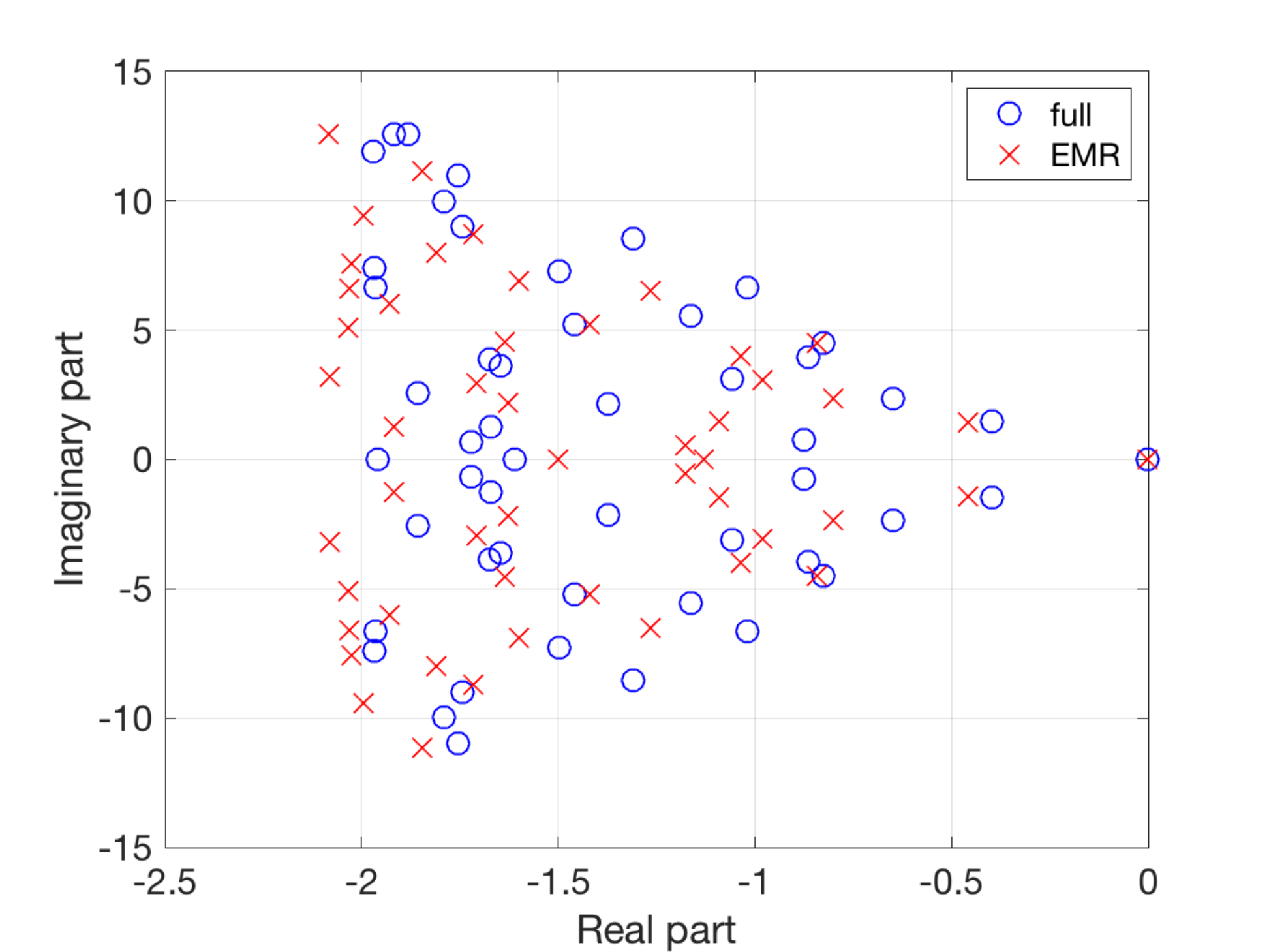}
		\caption{$h=0.25$}
	\end{subfigure}
	\hfill
	\begin{subfigure}[b]{0.49\textwidth}
		\centering
		\includegraphics[width=\textwidth]{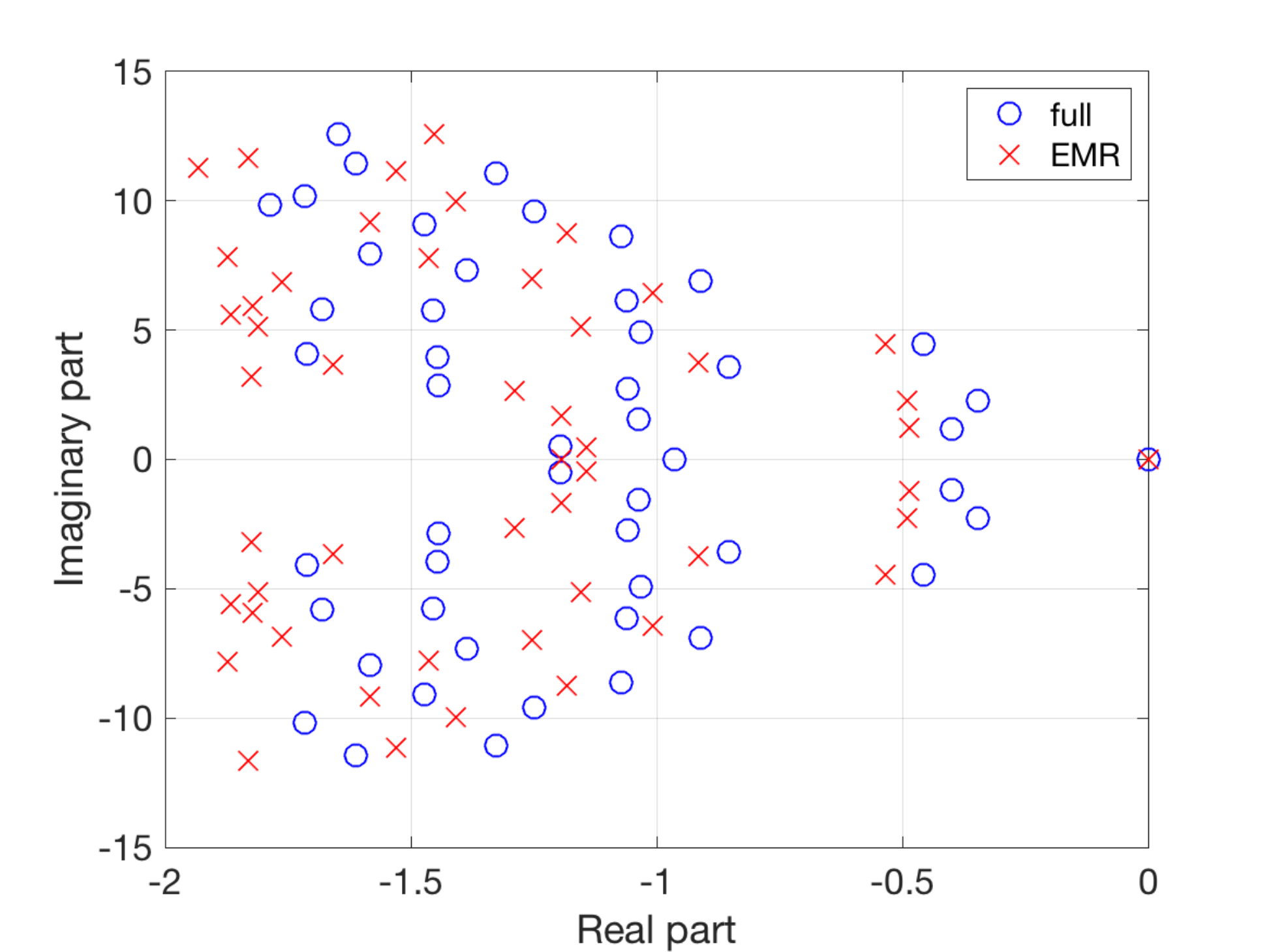}
		\caption{$h=0.025$}
	\end{subfigure}
	\caption{\label{ergodicity spectrum L84-L63} Leading eigenvalues of the discretized Koopman operator 
	in the L84 model's phase space. The blue open circles correspond to the data obtained by 
	integrating the full model's Eqs.~\eqref{eq: l84l63}, the red $\times$ symbols correspond 
	to the EMR model. (a) $h=0.25$; and (b) $h=0.025$.
	}
\end{figure}

\subsection{Convergence}

Convergence in the EMR approach is determined by the "whiteness" of the last-level residual, as explained in Sect.~\ref{Sec_EMR}; see Eq.~\eqref{emrdiagnostic} and discussion thereof. In Fig.~\ref{convergence emr}, we plotted the mean of the determination coefficients $R^2$ for the three $\X$-variables and we show that its convergence in the EMR approach depends only mildly on the coupling parameter $h$. Indeed, for $h=0.25$ we observe in panel (a) that around 18 levels are necessary before achieving the 
optimal level, whereas for weaker coupling with $h=0.025$ convergence is attained in panel (b) already with 15 levels, as one might expect.  

Furthermore, as already pointed out in \cite[Sec.~7]{kondrashovdata2015} on a different example, the results in Fig.~\ref{convergence emr}(c) illustrate that a smaller time scale separation $\tau$ can require a higher number of levels for EMR to attain convergence: in the case at hand, around 25 levels are needed. For completeness, Fig.~\ref{convergence emr}(d) shows that including additive white noise in the L63 system can, in fact, accelerate the convergence of the method, with convergence achieved at $\ell = 7$.

\begin{figure}[H]
	\centering
	\begin{subfigure}[b]{0.49\textwidth}
		\centering
		\includegraphics[width=\textwidth]{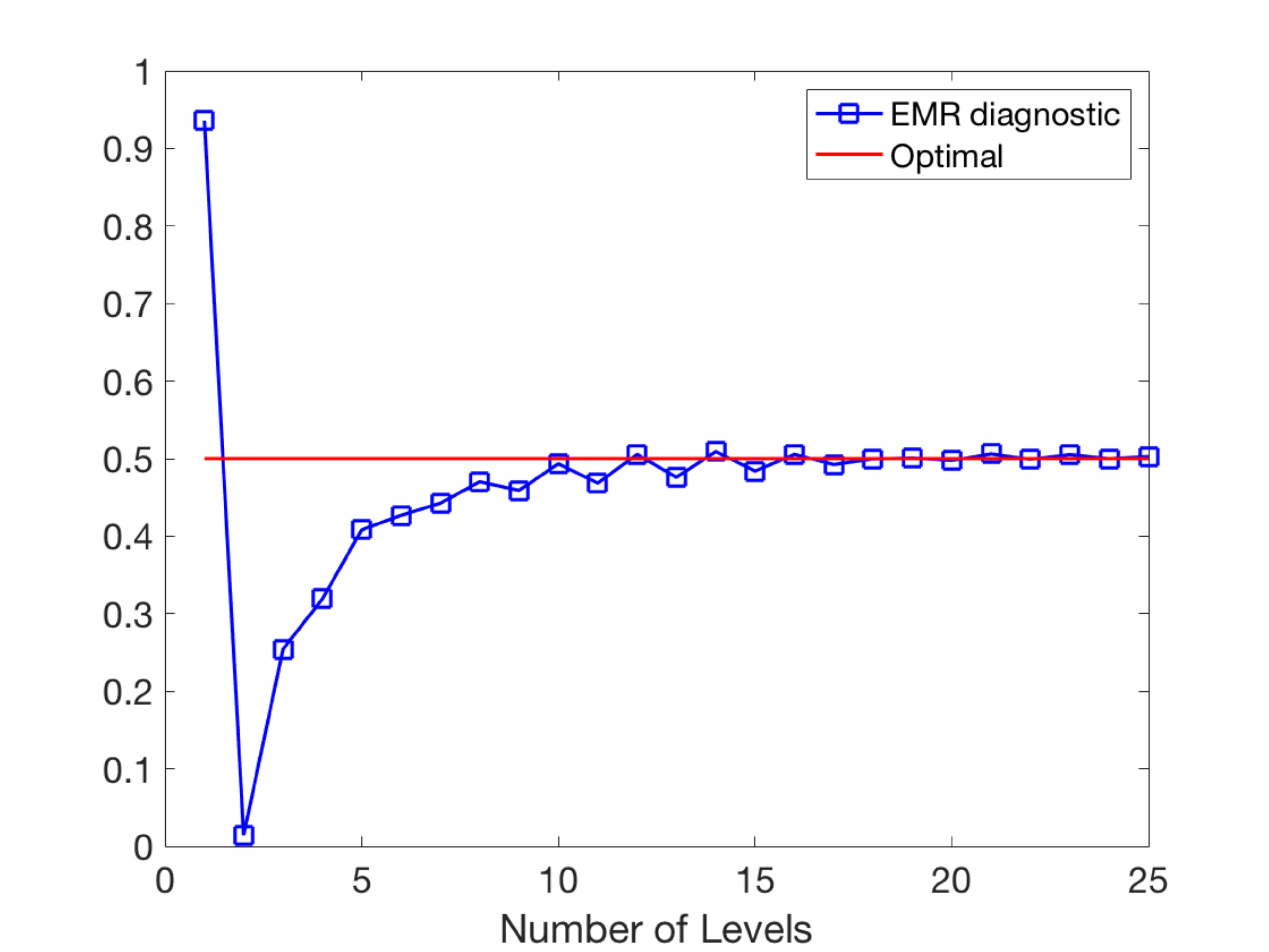}
		\caption{$h=0.25$ and $\tau = 5$}
	\end{subfigure}
	\hfill
	\begin{subfigure}[b]{0.49\textwidth}
		\centering
		\includegraphics[width=\textwidth]{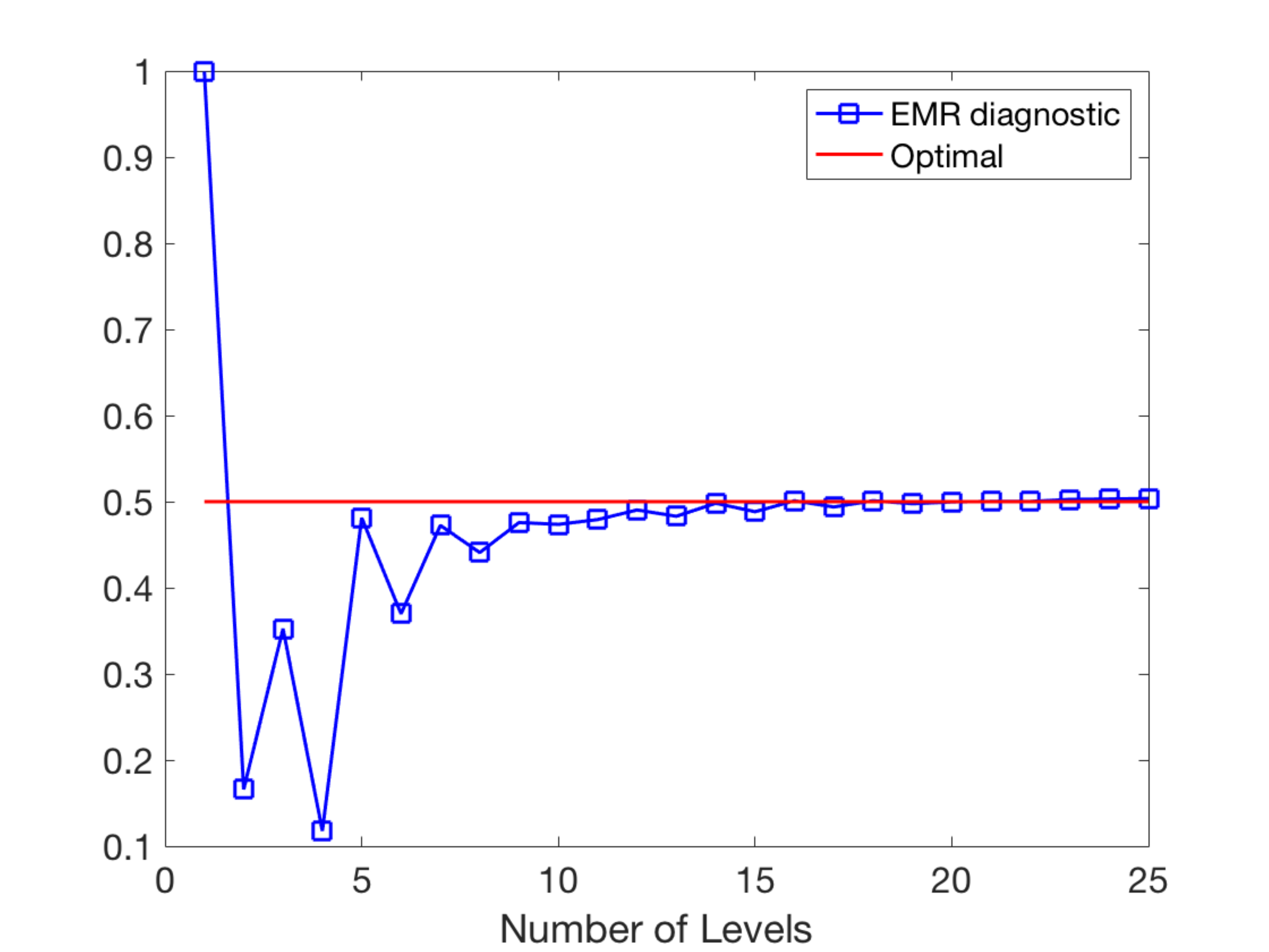}
		\caption{$h=0.025$ and $\tau = 5$}
	\end{subfigure}
	\hfill
	\begin{subfigure}[b]{0.49\textwidth}
		\centering
		\includegraphics[width=\textwidth]{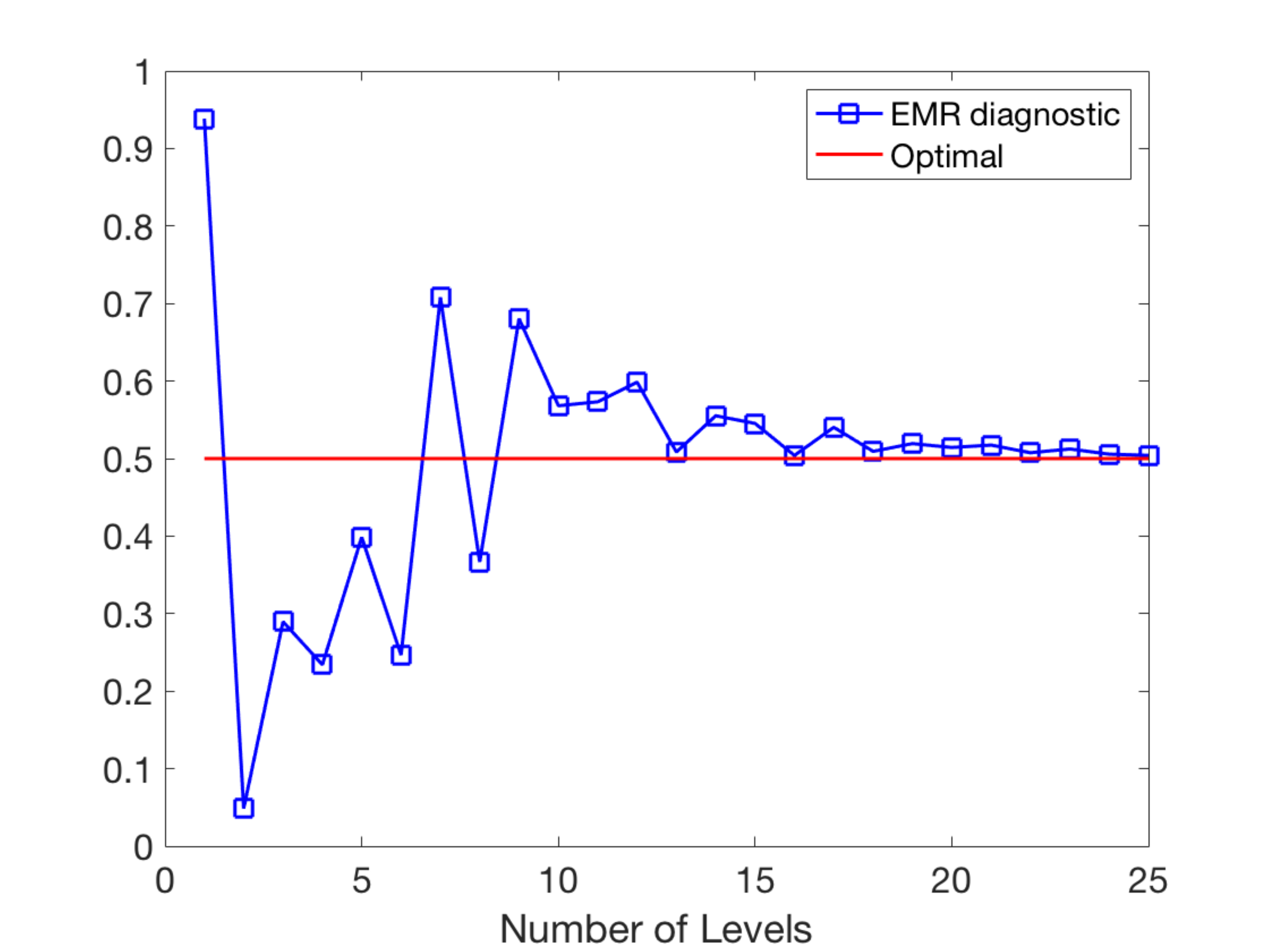}
		\caption{$h=0.25$ and $\tau=2$}
	\end{subfigure}
	\begin{subfigure}[b]{0.49\textwidth}
	\centering
	\includegraphics[width=\textwidth]{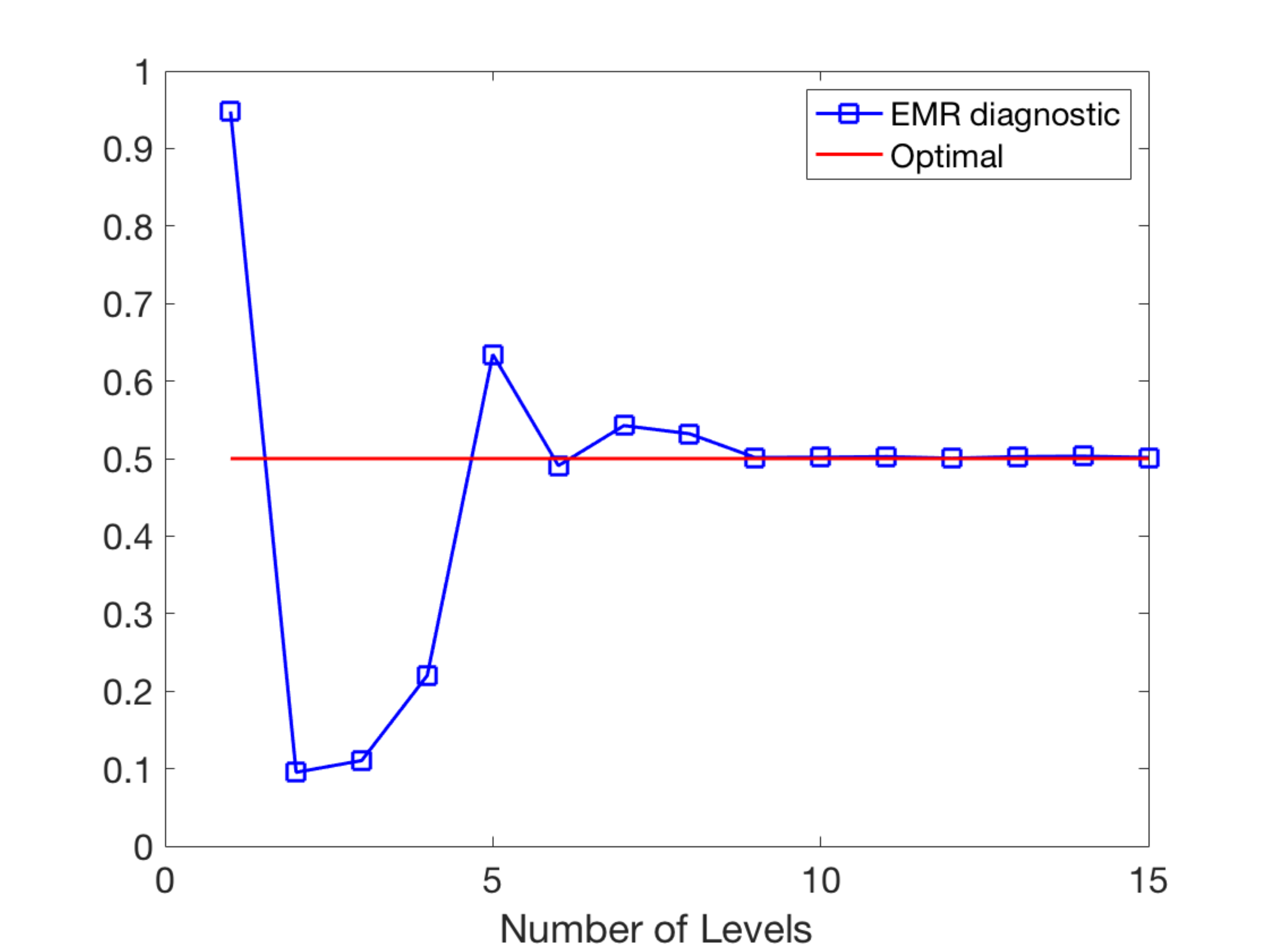}
	\caption{Noisy L63 model with $h=0.25$ and $\tau = 5$}
	\end{subfigure}
	\caption{\label{convergence emr}Determination coefficients $R^2$ of the EMR method 
	as a function of the number $\ell$ of levels. (a) $h=0.25$; (b) $h=0.025$; and (d) $h=0.25$ but with the L63 model including additive noise. Panels (a,b,d) all have the time scale separation $\tau = 5$,
	while in panel (c) $h=0.25$ and $\tau=2$.
}
\end{figure}

\subsection{Model Coefficients}

We show here that the EMR model coefficients can be efficiently approximated when phase space subsampling is carried out. Here, regressions are performed over $50$ short time series of $10$ time units each, with a time step of $5\cdot 10^{-3}$, as in Sect.~\ref{app:out}. The reason for taking this sample length here is that $10$ time units is visually enough for the slow variable $\X$ to go through a cycle, as illustrated in Fig.~\ref{oned traj}, for both $h=0.25$ and $0.025$.

\begin{figure}[H]
	\centering
	\begin{subfigure}[b]{0.49\textwidth}
		\centering
		\includegraphics[width=\textwidth]{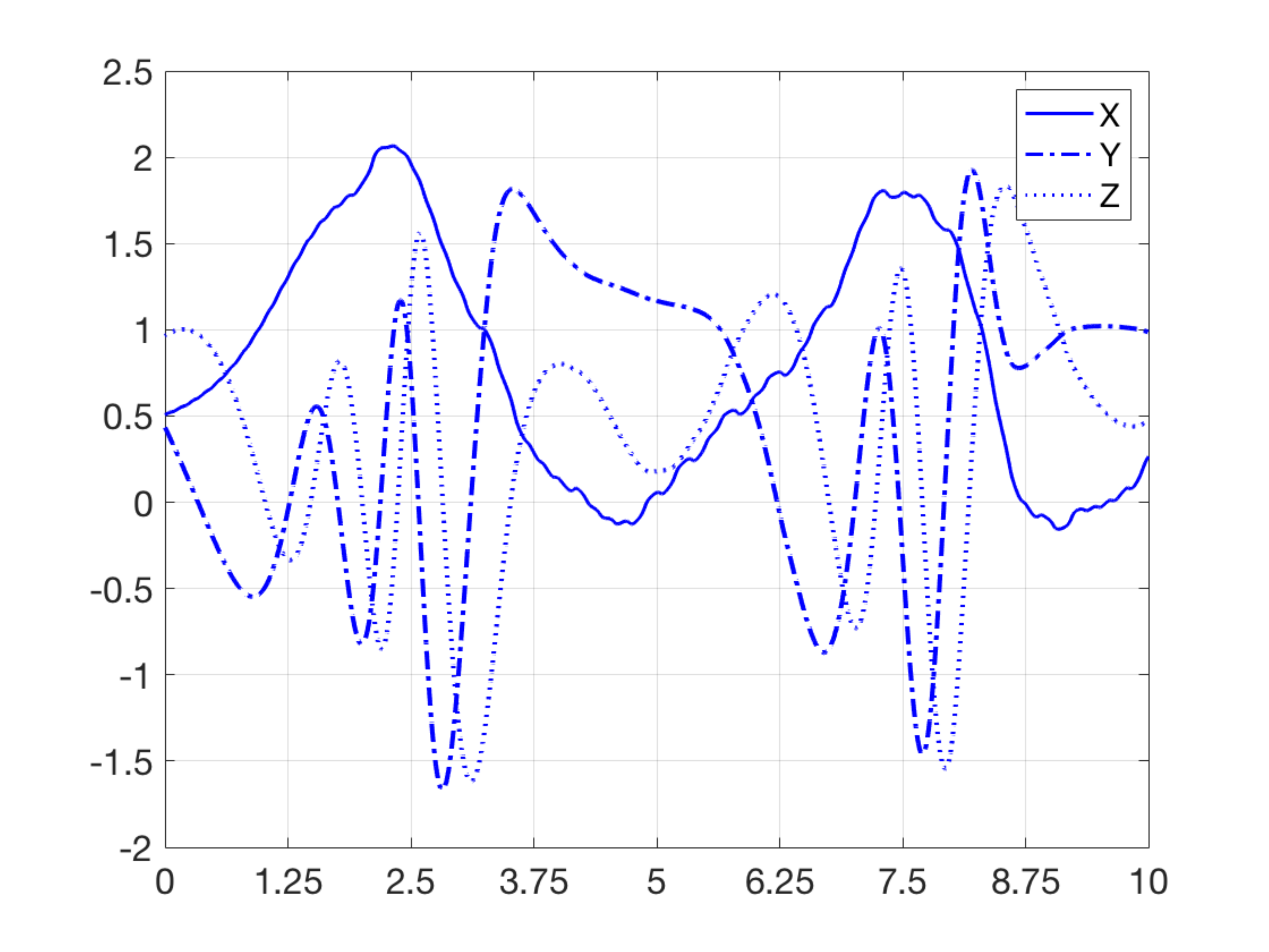}
		\caption{$h=0.25$}
	\end{subfigure}
	\hfill
	\begin{subfigure}[b]{0.49\textwidth}
		\centering
		\includegraphics[width=\textwidth]{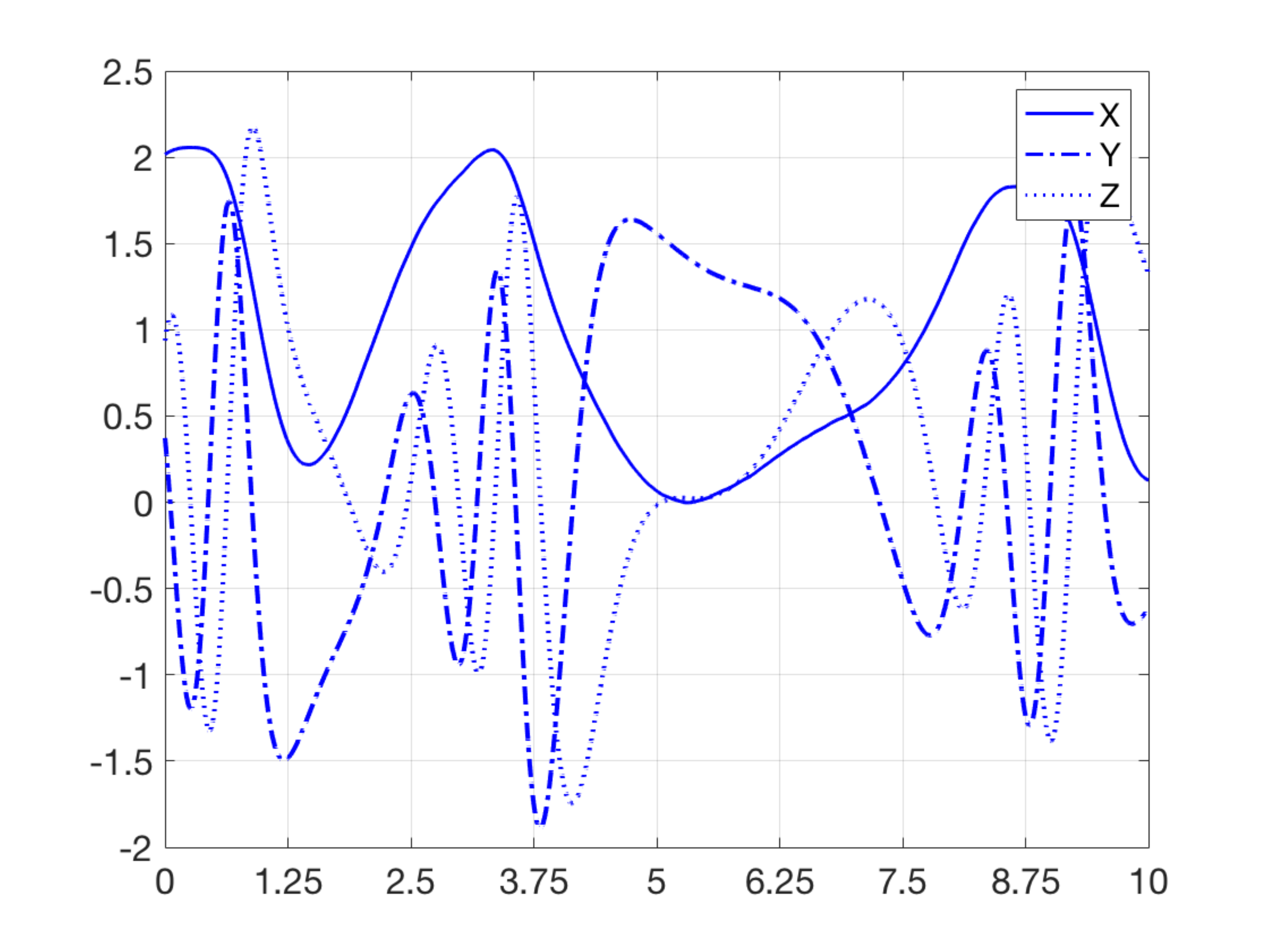}
		\caption{$h=0.025$}
	\end{subfigure}
	\caption{\label{oned traj}Time-series of the L84 variables $ (X, Y, Z)$
	over $10$ time units: (a) $h=0.25$, and (b) $0.025$.}
\end{figure}

The estimated  coefficients and their standard deviations using the EMR regressions are listed below in Tables~\ref{tab: coef l84l63 h025 m} and \ref{tab: coef l84l63 h025 std} for $h = 0.25$ and in Tables~\ref{tab: coef l84l63 h0025 m} and \ref{tab: coef l84l63 h0025 std} for $h = 0.025$. The tables show the coefficients of the linear and quadratic forms at first level in the EMR regressions: see Eq.~\eqref{EMR_0}. 

As expected, a stronger coupling of $h=0.25$ leads to greater uncertainty in the estimation, as indicated by the corresponding standard error. For the fairly complex and chaotic system at hand, we note that no memory effects are artificially introduced in the regressions at the second level. Indeed, we found that the coupling of the main level with the subsequent ones was $0$ to the fourth decimal place. 

\begin{table}[H]\renewcommand{\arraystretch}{2}
	\centering
	\begin{tabular}{c | cccccccccc}
		\hline
		EMR & $1$ & $x$ & $y$ & $z$ & $x^2$ & $xy$ & $y^2$ & $xz$ & $yz$ & $z^2$  \\ \hline
		$f_X$ & 1.949 & -0.352 & -0.002 & -0.104 & -0.015 & 0.01 & 0.052 & -0.946 & -0.001 & -0.921 \\
		$f_Y$ & 0.999 & 0.001 & -1.001 & 0.002 & 0 & 1.001 & -4.003 & 0 & 0 & 0 
\\
		$f_Z$ & 0.002 & -0.003 & -0.002 & -1.001 & 0.001 & 4.003 & 1.001 & 0 & 0 & 0 \\  
	\end{tabular}
	\caption{\label{tab: coef l84l63 h025 m}Means of the EMR coefficients 
	of the L84--L63 model, estimated from an ensemble of 50 runs over 10 time units 
	for $h=0.25$.}
\end{table}

\begin{table}[H]\renewcommand{\arraystretch}{2}
	\centering
	\begin{tabular}{c | cccccccccc}
		\hline
		EMR & $1$ & $x$ & $y$ & $z$ & $x^2$ & $xy$ & $y^2$ & $xz$ & $yz$ & $z^2$  \\ \hline
		$f_X$ & 0.543 & 1.005 & 0.246 & 0.291 & 0.428 & 0.152 & 0.188 & 0.104 & 
0.076 & 0.099 \\
		$f_Y$ & 0.001 & 0.002 & 0 & 0.001 & 0.001 & 0 & 0 & 0 & 0 & 0 \\
		$f_Z$ & 0.001 & 0.002 & 0.001 & 0.001 & 0.001 & 0.001 & 0 & 0 & 0 & 0 \\
	\end{tabular}
	\caption{\label{tab: coef l84l63 h025 std} Standard deviations of the EMR coefficients 
		of the L84--L63 model, estimated from an ensemble of 50 runs over 10 time units 
		for $h=0.25$.}
\end{table}

\begin{table}[H]\renewcommand{\arraystretch}{2}
	\centering
	\begin{tabular}{c | cccccccccc}
		\hline
		EMR & $1$ & $x$ & $y$ & $z$ & $x^2$ & $xy$ & $y^2$ & $xz$ & $yz$ & $z^2$  \\ \hline
		$f_X$ & 2.006 & -0.253 & -0.001 & -0.003 & -0.004 & 0 & 0.001 & -1.002 & 0.001 & -1.003 \\
		$f_Y$ & 1 & 0 & -1.001 & 0.002 & 0 & 1.001 & -4.003 & 0 & 0 & 0 \\
		$f_Z$ & 0.001 & -0.002 & -0.002 & -1.001 & 0.001 & 4.003 & 1.001 & 0 & 0 & 0 \\
	\end{tabular}
	\caption{\label{tab: coef l84l63 h0025 m} Means of the EMR coefficients 
		of the L84--L63 model, estimated from an ensemble of 50 runs over 10 time units 
		for $h=0.025$.}
\end{table}

\begin{table}[H]\renewcommand{\arraystretch}{2}
	\centering
	\begin{tabular}{c | cccccccccc}
		\hline
		EMR & $1$ & $x$ & $y$ & $z$ & $x^2$ & $xy$ & $y^2$ & $xz$ & $yz$ & $z^2$  \\ \hline
		$f_X$ & 0.034 & 0.063 & 0.019 & 0.021 & 0.026 & 0.01 & 0.014 & 0.007 & 0.008 & 0.006 \\
		$f_Y$ & 0.001 & 0.002 & 0 & 0.001 & 0.001 & 0 & 0 & 0 & 0 & 0 \\
		$f_Z$ & 0.001 & 0.002 & 0.001 & 0.001 & 0.001 & 0.001 & 0 & 0 & 0 & 0 \\ 
	\end{tabular}
	\caption{\label{tab: coef l84l63 h0025 std} Standard deviations of the EMR coefficients 
		of the L84--L63 model, estimated from an ensemble of 50 runs over 10 time units 
		for $h=0.025$.}
\end{table}

\bibliographystyle{abbrv}
\bibliography{MSG+VL+MDC-MG2} 

\end{document}